\documentclass[a4paper,11pt]{article}
\pdfoutput=1 

\usepackage{jheppub} 

\usepackage[T1]{fontenc} 

\usepackage{lmodern}

\usepackage{enumitem} 	

\usepackage{amsmath}		
\usepackage{amssymb}		
\usepackage{bbold}			
\usepackage{breqn}			
\usepackage{mathtools}		

\usepackage{multirow,ctable,colortbl} 

\usepackage{xcolor}			

\newcommand{\IZ}{\mathbb{Z}}
\newcommand{\IC}{\mathbb{C}}

\newcommand{\IR}{\mathbb{R}}
\newcommand{\IQ}{\mathbb{Q}}
\newcommand{\IK}{\mathbb{K}}
\newcommand{\IH}{\mathbb{H}}

\newcommand{\SL}{{\rm SL}}
\newcommand{\Sp}{{\rm Sp}}
\newcommand{\Mod}{{\rm Mod}}
\newcommand{\Gr}{{\rm Gr}}
\newcommand{\U}{{\rm U}}

\newcommand{\SU}{{\rm SU}}
\newcommand{\OO}{{\rm O}}

\newcommand{\dd}{{\rm d}}

\newcommand{\tp}{{{}^\intercal}}
\newcommand{\mtp}{{{}^{-\intercal}}}

\newcommand{\Zuntw}{\mathsf{Z}^{\mathrm{untw}}}
\newcommand{\Ztw}{\mathsf{Z}^{\mathrm{tw}}}

\newcommand{\DT}{\mathsf{DT}}
\newcommand{\Shet}{S_{\mathrm{het}}}

\newcommand{\SQ}{\mathcal{S}_{\mathcal{Q}}}
\newcommand{\PQ}{\Phi_{\mathcal{Q}}}
\newcommand{\ZQ}{\mathsf{Z}_{\mathcal{Q}}}
\newcommand{\cP}{\Phi}
\newcommand{\cO}{Z}
\newcommand{\crho}{\tau}
\newcommand{\cv}{z}
\newcommand{\csig}{\sigma}

\newcommand{\ar}[2]{{\textstyle \big[{#1\atop #2}\big]}}

\newcommand{\Zo}{\mathsf{Z}^{(0)}}
\newcommand{\Zp}{\mathsf{Z}^{(+)}}
\newcommand{\Zm}{\mathsf{Z}^{(-)}}

\title{\boldmath Lost Chapters in CHL Black Holes:\\ Untwisted Quarter-BPS Dyons in the $\IZ_2$ Model}

\author[a]{Fabian Fischbach}
\author[a,b]{Albrecht Klemm}
\author[a]{Christoph Nega}

\affiliation[a]{Bethe Center for Theoretical Physics,\\Universit\"at Bonn, D-53115, Germany}
\affiliation[b]{Hausdorff Center for Mathematics,\\Universit\"at Bonn, D-53115, Germany}

\emailAdd{fischbach@physik.uni-bonn.de}
\emailAdd{aklemm@th.physik.uni-bonn.de}
\emailAdd{cnega@th.physik.uni-bonn.de}

\abstract{%
Motivated by recent advances in Donaldson-Thomas theory, four-dimensional $\mathcal{N}=4$ string-string duality is examined in a reduced rank theory on a less studied BPS sector. In particular we identify candidate partition functions of ``untwisted'' quarter-BPS dyons in the heterotic $\IZ_2$ CHL model by studying the associated chiral genus two partition function, based on the M-theory lift of string webs argument by Dabholkar and Gaiotto. This yields meromorphic Siegel mo\-dular forms for the Iwahori subgroup $B(2) \subset \Sp_4 (\IZ) $ which generate BPS indices for dyons with untwisted sector electric charge, in contrast to twisted sector dyons counted by a multiplicative lift of twisted-twining elliptic genera known from Mathieu moonshine. The new partition functions are shown to satisfy the expected constraints coming from wall-crossing and S-duality symmetry as well as the black hole entropy based on the Gauss-Bonnet term in the effective action. In these aspects our analysis confirms and extends work of Banerjee, Sen and Srivastava, which only addressed a subset of the untwisted sector dyons considered here. Our results are also compared with recently conjectured formulae of Bryan and Oberdieck for the partition functions of primitive DT invariants of the CHL orbifold $X=( \text{K3} \times T^2 )/ \IZ_2$, as suggested by string duality with type IIA theory on $X$.

}

\begin{document} 

\rightline{BONN-TH-2020-02}
\maketitle
\flushbottom


\section{Introduction}
\label{sec:intro}
\noindent \textit{Extremal black hole entropy from counting quarter-BPS dyons in $\mathcal{N}=4$ string theory}\\
\noindent

\noindent A benchmark test of any theory of quantum gravity is to provide a microscopic, statistical explanation of the entropy carried by a black hole, which is semi-classically given by the Bekenstein-Hawking entropy~\cite{Bekenstein:1974ax,Hawking:1976de}.
Within string theory such an explanation was provided in~\cite{STROMINGER199699,Breckenridge:1996is}, where the statistical entropy of a D-brane system indeed matches the Bekenstein-Hawking entropy of the corresponding five-dimensional extremal black hole in the large charge limit.
 As shown in~\cite{Dijkgraaf1997d}, the entropy can also be derived in four-dimensional $\mathcal{N}=4$ string theory (heterotic strings on $T^6$ or type IIA strings on $\text{K3} \times T^2$) from an exact microscopic index formula for the quarter-BPS dyons of the theory. According to~\cite{Dijkgraaf1997d} the dyon degeneracies\footnote{These degeneracies are actually indices, i.e., sixth helicity supertraces~\cite{Bachas_1997,Gregori_1998,Kiritsis:1997hj}, making them invariant under (small) variations of the moduli. An interpretation of this microscopic BPS \textit{index} as a macroscopic black hole \textit{degeneracy} was justified in~\cite{Sen_2009arithmetic,Sen_2014}.\label{fn:indicesdeg}} are given by a contour integral of the reciprocal Igusa cusp form $\chi_{10}^{-1}$, extracting Fourier coefficients of this $\Sp_4(\IZ)$-Siegel modular form. Their growth, as an asymptotic expansion in large charges, is estimated by a saddle-point approximation~\cite{Dijkgraaf1997d,Cardoso_2005} that picks up the dominant pole of the integrand, again reproducing the macroscopic entropy formula of the extremal dyonic black hole~\cite{Cvetic:1995uj,Cvetic:1995bj} along with a series of power suppressed and exponentially suppressed contributions (from the dominant, respectively, sub-dominant pole)~\cite{Banerjee:2008ky}.
 Here the leading correction to the entropy can, on the macroscopic supergravity side, be attributed to the inclusion of the Gauss-Bonnet term in the effective action~\cite{Harvey_1998,Gregori_1998} by following Wald's generalization~\cite{Wald_1993} of the black hole entropy~\cite{Sen:2005higher,Sen_2006entrophet}. Apart from such higher-derivative corrections there are also quantum corrections to the dyonic extremal black hole entropy. Based on the AdS/CFT correspondence the proposed quantum entropy function of~\cite{Sen:2008EF,Sen:2008QEF} captures both kinds of corrections and accounts for exponentially suppressed contributions\footnote{Logarithmic corrections consistently vanish micro- and macroscopically for $\mathcal{N}=4$ theories~\cite{Banerjee_2011,Keeler_2014,Gupta_2014,Sen_2014,Gupta_2014B}.} as demanded by the microscopic index formula~\cite{Banerjee:2008ky,Sen_2009arithmetic,Murthy:2009dq}. See~\cite{Mandal_2010,Gomes:2011zzc,Sen_2014} for reviews and~\cite{Murthy_2016,Gomes:2015ExHol,Gomes2017_stringyexcl,Ferrari:2017mixed,Gomes:2017GenKlo,Gomes:2017Uduality} for more recent studies of the (quarter-BPS) quantum entropy that rely on localization of the supergravity path integral.\\ 

\noindent
\textit{Dyon counting in CHL models from dual perspectives}\\
\noindent

\noindent The quarter-BPS index formula of~\cite{Dijkgraaf1997d} was generalized to dyons in $\mathcal{N}=4$ CHL orbifolds~\cite{Chaudhuri:1995fk,Chaudhuri:1995bf,Chaudhuri:1995dj,Schwarz:1995bj,Aspinwall_1996} in~\cite{Jatkar:2005bh,David:2006ji,Dabholkar07Borcherds,Dabholkar_2007} with an appropriate Siegel modular form taking the role of $\chi_{10}$.%
 \footnote{As was pointed out in~\cite{Dabholkar:2007vk}, the dyon partition functions of~\cite{Dijkgraaf1997d,Jatkar:2005bh} actually only capture dyons satisfying a primitivity constraint, namely that the discrete invariant $I=\gcd(Q\wedge P)$ built from the dyon charge $(Q,P)$ is unity. Partition functions for $I>1$ have subsequently been worked out in~\cite{Banerjee:2008pv,Banerjee:2008pu,Dabholkar:2008zy,Sen:2008ht,Dabholkar_2010}.}
These theories are obtained upon orbifolding heterotic strings on $T^2 \times T^4$ by a $1/N$ shift along a circle in $T^2$ and a supersymmetry-preserving order $N$ action on the internal CFT describing heterotic strings on $T^4$. By string-string duality~\cite{Hull:1994ys,Duff:1994zt,Witten:1995ex,Sen:1995cj,Harvey:1995rn} these are dual to type IIA theory on orbifolds $(\text{K3} \times T^2)/\IZ_N$, where the $\IZ_N$ action is a shift on $T^2$ combined with an order $N$ holomorphic-symplectic automorphism of the K3. 

Clearly, apart from its eminent role in understanding black hole microstate entropy  versus its macroscopic counterpart, the counting of quarter-BPS states provides a non-perturbative window to various string dualities. This manifests also in the approaches taken to physically derive the proposed counting functions.
In the type IIB frame these are computed at weak coupling as partition functions of a rotating D1-D5-sytem in a Taub-NUT background~\cite{David_2006}, similar to the proposal of~\cite{StromingerSY_0505094}.\footnote{See~\cite{Sen:2007qy,Dabholkar:2012zz,Mandal_2010} for reviews.}
Another derivation was given in~\cite{gaiotto2005rerecounting,Dabholkar_2007}, representing the dyons as string webs lifted to M-theory where the dyon partition function can eventually be related to a genus two partition function of a heterotic string.
More recently the BPS indices have also been extracted from four- and six-derivative couplings in the low energy effective action of three-dimensional heterotic CHL vacua with 16 supercharges. \label{pg:3dcouplings} Upon circle decompactification to four dimensions~\cite{Bossard_2016,Bossard_2017,Bossard_2019} the index formula is found with the correct choice of moduli-dependent integration contour proposed in~\cite{Cheng_2007}. This contour prescription renders the quarter-BPS index duality-invariant and captures the (dis-)appearance of two-centered bound states at walls of marginal stability~\cite{Sen:2007vb,Sen:2007twocenter,Cheng_2007,Sen:2008ht,Cheng:2008fc}. In the configurations contributing to the quarter-BPS index the two centers are each half-BPS~\cite{Dabholkar:2009dq}.\footnote{Three-center BPS bound states  are conjecturally enumerated by a degree three Siegel modular form~\cite{Denef:2017zxm}.} \\

\noindent
\textit{(Mock) Jacobi forms, moonshine and dyon counting }\\

\noindent At fixed magnetic charge invariant, quarter-BPS dyons are captured by meromorphic Jacobi forms arising in the Fourier-Jacobi decomposition of the respective Siegel modular form.\footnote{For an introduction to the theory of Jacobi forms and their connection to Siegel modular forms see~\cite{EichlerZagier}.} A decomposition of the former into a finite part (a mock Jacobi form) and a polar part (an Appell-Lerch sum) separates the counting of single-centered black holes, which are stable across walls of marginal stability, from the two-centered black hole bound states~\cite{dabholkar2012quantum}. Modular invariance can be restored upon addition of a non-holomorphic completion, and the  completion of the polar part has recently been interpreted physically as the continuum contribution in the supersymmetric index of the quantum mechanical bound states~\cite{Murthy:2018bzs}. Understanding mock modularity from the physics perspective is an active research area \nolinebreak~\cite{Murthy_2016,Ferrari:2017mixed,Chattopadhyaya:2018xvg,Wrase1912}.

Recent interest in strings on K3, $\mathcal{N}=4$ CHL models and their BPS counting has also arisen in moonshine contexts and we shall briefly sketch some connections. Quarter-BPS partition functions can be constructed for more general CHL compactifications involving orbifolds by any -- not necessarily geometric -- symmetry of the K3 non-linear sigma model (NLSM)~\cite{Aspinwall:1996mn,Nahm:1999ps} that commutes with the worldsheet $\mathcal{N}=(4,4)$ superconformal algebra and the half-integral spectral flows. Such K3 NLSM symmetries\footnote{K3 NLSM symmetries can also be interpreted in terms of derived equivalences of K3 surfaces~\cite{2013arXiv1309.6528H}.\nolinebreak } belong to elements of the largest Conway group $Co_0$ and were classified in~\cite{Gaberdiel:2011fg}, partially extended in~\cite{Cheng:2016org} to full type IIA theory at singular loci in moduli space, while~\cite{Persson:2015jka} classify the resulting CHL models and observe that the Fricke involution acts as S-duality in the  self-dual models (i.e., when the symmetry has balanced Frame-shape, as is the case with geometric symmetries).\label{cor:B1}\footnote{A yet broader notion of ``CHL models'' was proposed in~\cite{Persson:2017lkn}, generically exhibiting Atkin-Lehner dualities.} The dyon partition functions are obtained by multiplicative lifts~\cite{BorcherdsInfProd,BorcherdsSingOnGras,GritsenkoNikulinI,GritsenkoNikulinII} of the twining genera  associated to symmetry conjugacy classes~\cite{Cheng:2016org}, which map (vector valued) weak Jacobi forms to Siegel modular forms for some congruence subgroup of $\Sp_4(\IZ)$~\cite{Cheng:2010pq,Cheng:2012uy,Raum1208}.\footnote{Similarly quarter-BPS partition functions for twisted BPS indices in the sense of~\cite{Sen:2009Atwist,Sen:2010Discrete,Govindarajan:2010fu} can be constructed from twisted-twining genera, see~\cite{Persson:2013xpa} for an overview.} This includes the original case of type IIA theory on $\text{K3} \times T^2$ and $\chi_{10}$~\cite{Dijkgraaf1997d,DMVVellgenus,GritsenkoEllCY} as a special case. In combination with constraints from modularity and wall-crossing this construction has recently~\cite{Paquette:2017gmb} been used to explicitly derive (almost) all of the twining genera. These K3 twining genera often coincide with those of Mathieu~\cite{Eguchi:2010ej,Cheng:2010pq,Gaberdiel:2010ch,Gaberdiel:2010ca,Eguchi:2010fg,Gannon:2012ck}, Umbral~\cite{Cheng:2012tq,Cheng:2013wca,Cheng:2014zpa,MR3433373,Kachru:2016ttg,Zimet:2018dev} and Conway~\cite{MR2352133,MR3376736,MR3465528} moonshine phenomena and hence establish a direct link to the dyon partition function. An overview of moonshine phenomena from the physics perspective can be found in~\cite{Kachru:2016nty,Anagiannis:2018jqf}.\\
		
\noindent
\textit{Quarter-BPS indices, Donaldson-Thomas invariants and the lost chapters}\\

\noindent Finally, the problem of counting quarter-BPS states has an avatar in the enumerative geometry of holomorphic curves in the  Calabi-Yau threefold $X = \text{K3} \times T^2$ 
as was first pointed out for the type IIA theory on $X$ in~\cite{Katz:1999xq}. The reduced Gopakumar-Vafa invariants on $X$ were given an interpretation in terms of the cohomology 
of the moduli space ${\cal M}^n_{\beta}$  associated to D0-D2-brane bound states inside $X$. Given a D2-brane wrapping a holomophic genus $g$ curve $C_g$ in the class $\beta\in H_2(X,\mathbb{Z})$, the 
moduli space ${\cal M}^n_{\beta}$  was constructed in~\cite{Katz:1999xq}  as a singular Jacobian fibration ${\rm Jac}(g)$ over the  deformation space of the curve $C_g$ in $X$. 
Roughly the integer $n$ can be thought of as the number of D0-branes and its relation to the degenerations  of ${\rm Jac}(g)$ maps it  to the genus counting parameter $g$. 
The decomposition of  the cohomology of the moduli  space ${\cal M}^n_{\beta}$ with respect to an $\mathfrak {sl}(2)_r\times \mathfrak {sl}(2)_l$ Lefshetz action constructed  
using the Abel-Jacobi map allows explicit computations, if the singularities of the Jacobian fibration are not too bad, and was used to conjecturally 
identify the quarter-BPS states of the heterotic string  with the  reduced Gopakumar-Vafa invariants.  These symplectic invariants have been 
mathematically rigorously defined in  terms of stable pair invariants and Donaldson-Thomas (DT) invariants in~\cite{Pandharipande:2007qu}. For unit-torsion dyons 
the proposal of~\cite{Katz:1999xq} leads to the Igusa cusp form conjecture~\cite{Oberdieck2016,Bryan:2015uva} for the primitive DT invariants, proven in~\cite{MR3827207,Oberdieck_2019}. 

Given the success of this highly non-trivial physics prediction, one is immediately lead to the question of how this generalizes to the CHL orbifolds. This question has recently been addressed in~\cite{Bryan:2018nlv}, focussing on orbifolds by a symplectic automorphism $g$ of the K3 (i.e., geometric K3 NLSM symmetries in the above terminology) and in particular the $\langle g \rangle = \IZ_2$ orbifold. Conjecture A of~\cite{Bryan:2018nlv} proposes that the primitive DT partition function of the $\IZ_2$ CHL orbifold $(\text{K3} \times T^2)/\IZ_2$ is given by the multiplicative lift $\Ztw$ of the twisted-twined elliptic genera belonging to the $[g]=2A$ conjugacy class as in~\cite{David:2006ji} --- but only for DT invariants coming from ``twisted'' curve classes. There is a binary distinction between twisted and untwisted curve classes and for the latter Conjecture B of~\cite{Bryan:2018nlv} proposes an alternative primitive DT partition function $\Zuntw$. 
So far this new Siegel modular form, somewhat surprisingly, does not seem to have made any appearance in physics, where the twisted partition function has (almost) exclusively been considered. Does it have a physical (dyon counting) interpretation? Regarding string-string duality, we should be able to provide a derivation from the heterotic perspective. Apart from that, there are stringent constraints coming from wall-crossing, S-duality invariance and the black hole entropy. 
  Addressing these questions will be the content of the lost chapters.

Before proceeding, we shall explain where to possibly fit them in the CHL story, so let us comment on the distinction between twistedness and untwistedness on the physics side.
 It is known (though mentioned less frequently) that the partition function of~\cite{David:2006ji} counts unit-torsion dyons whose electric charge $Q$ in the heterotic frame belongs to the ``twisted sector'', i.e., for perturbative half-BPS states of charge $(Q,0)$ the component corresponding to the string winding number around the CHL circle is half-integral. This is in contrast to untwisted sector charges corresponding to integral winding along the CHL circle. It has been argued~\cite{Sen:2007qy} that S- and T- transformations (i.e., those inherited from the parent theory compatible with the orbifolding procedure) do not mix dyon charges with twisted and untwisted sector charges in the winding number sense.\footnote{Physically this distinction does not apply for the unorbifolded case, for which the electric and magnetic charge lattices are isomorphic and the U-duality group acts transitively on the unit-torsion dyon charges. Due to duality invariance of the BPS index we hence expect only one quarter-BPS partition function.\label{fn:nosectorsinhett6}}
However, the definition of twisted curve classes in DT theory a priori only concerns the $E_8(-\frac{1}{2})$-part of the (co-)homology lattice of $(\text{K3} \times T^2)/\IZ_2$, which is a sublattice of the electric charge lattice $\Lambda_e$, while the twisted and untwisted charge sectors in physics (independently of the $E_8$ components) refer to the electric components along cosets $U+\frac{\delta}{2} \subset U(\frac{1}{2}) \subset \Lambda_e $ and $U \subset U(\frac{1}{2}) \subset \Lambda_e$, respectively.
Both notions basically specify appropriate components of the ``residue'' \cite[app. B]{Bryan:2018nlv}\label{cor:typo:appB1} of a dyon charge, which is the class in the discriminant group $\Lambda_e/\Lambda_e^*$.
Although the two notions of twistedness are not equivalent, one clearly needs to understand both the BPS counting in the twisted and untwisted charge sectors, keeping track also of the remaining residue components.

Motivated by these observations, we will independently derive (re-derive) quarter-BPS partition functions for unit-torsion dyons with untwisted (twisted) sector electric charge, distinguishing subsectors specified by the remaining charge residue components.
 Following the M-theory lift of string webs approach of~\cite{Dabholkar_2007}, the appropriate candidate BPS partition functions are deduced from a chiral genus two orbifold partition function in the heterotic $\IZ_2$ CHL model.  
For each subsector we check the above mentioned constraints coming from charge quantization, $\Gamma_1(2)$ S-duality, wall-crossing and black hole entropy. As we will see, these modular and polar constraints\footnote{Modular constraints mean that the partition function is expected to transform as a Siegel modular form under certain $\Sp_4(\IZ)$ elements, while polar constraints give the singular behaviour near certain divisors associated with walls of marginal stability.} are strong enough to (almost) guess the partition functions once the appropriate ring of Siegel modular forms has been identified. In the cases where generators for the latter are explicitly known, namely cases leading to Siegel modular forms for the Iwahori subgroup $B(2) \subset \Sp_4(\IZ)$, these constraints indeed fix the BPS partition function. First steps in that direction have been presented in~\cite{Banerjee:2008pv} quite some time ago, though the analysis in~\cite{Banerjee:2008pv} was not carried through and remains limited to a small subsector of the untwisted sector addressed here.%
\footnote{It is worth mentioning that in~\cite{Banerjee:2008pv} these constraints were also used to propose the correct partition function of dyons with torsion greater than one for the unorbifolded theory.}
Also, as it is the case for the twisted sector partition function of ~\cite{David:2006ji}, for large charge invariants the asymptotic growth of the Fourier coefficients reproduces the correct black hole entropy. In accordance with expectations from quarter-BPS black holes in four dimensions, we not only see the leading Bekenstein-Hawking term, but also a subleading term that can be associated with the (model-dependent) Gauss-Bonnet term in the effective action.

Our results relying on the string web argument of \cite{Dabholkar_2007} also agree with those relying on an analysis of suitable 3D protected couplings in~\cite{Bossard_2019}. Indeed, eq. (2.14) of~\cite{Bossard_2019} gives an expression for the quarter-BPS index in (conjecturally) arbitrary charge sector in terms of combinations of Fourier coefficients of the same Siegel modular forms and reproduces the result of~\cite{David:2006ji} as a special case.\footnote{The twisted/untwisted nomenclature employed here is not fully equivalent to the one of ~\cite{Bossard_2017,Bossard_2019}, but the results are.} \label{pg:3dcouplings2}
 However, apart from supporting the results in~\cite{Bossard_2019} from an independent perspective and demonstrating an extended range of applicability of the approach in~\cite{Dabholkar_2007}, we also provide a physical explanation for the $B(2)$ Iwahori modular symmetry of certain quarter-BPS partition functions (and show that this in turn fixes them via the above constraints) and further compare with the DT conjectures of \cite{Bryan:2018nlv}.

Now regarding the DT conjecture B, we first give an alternative (but equivalent) expression for $\Zuntw$ in terms of the multiplicative lift of twining elliptic genera and two of its modular transforms.
The dyon counting interpretation of $\Zuntw$ is, however, subtle. There are several subsectors of the (untwisted) charge sector and for none of these the dyon counting functions exactly matches $\Zuntw$. Rather, $\Zuntw$ is --- at least formally --- a sum or average of two such functions, denoted $\Zp$ and $\Zo$ below. Hence, one possible interpretation is that the DT invariants in $\Zuntw$ are sums of quarter-BPS indices of two representative states with charge in the union of two charge subsectors, instead of giving the BPS index of a unique charge configuration (or its orbit).
Clearly, it would be desirable to further improve the physical understanding of the DT formulae and see whether the given interpretation is really the right one.

As an independent minor point discussed in an appendix, the ``twisted'' and ``untwisted'' helicity supertraces considered in~\cite{Sen:2009Atwist,Sen:2010Discrete,Govindarajan:2010fu} are not in one-to-one correspondence with (standard) helicity supertraces for states with charge respectively belonging to the twisted or untwisted charge sector in the sense above.

The paper is organized as follows:
In section \ref{sec:countingbpsN4} we review $\mathcal{N}=4$ CHL models with a focus on quarter-BPS dyon counting. 
 Section \ref{sec:DHstatesZ2} reviews the half-BPS counting functions specific to the $\IZ_2$ model, which appear as wall-crossing data.
The genus two derivation of our candidate partition function in section \ref{sec:genustwo} then proceeds in a similar fashion.
 Modular and polar constraints on the latter are checked in section \ref{sec:sdwcconstraints}, %
 with the black hole entropy being treated separately in section \ref{sec:bhentropy}. 
We compare our results to the DT results in section \ref{sec:DTtheory} and conclude in section \ref{sec:conclusion}. 
Background material on Siegel modular forms and twisted helicity traces is collected in the appendices \ref{app:siegelmodforms} and \ref{sec:twistedIndex}. 


\section{Counting BPS dyons in four-dimensional $\mathcal{N}=4$ theories}
\label{sec:countingbpsN4}

This section provides a brief review of four-dimensional $\mathcal{N}=4$ CHL models~\cite{Chaudhuri:1995fk,Chaudhuri:1995bf,Chaudhuri:1995dj}, the focus lying on BPS state counting via automorphic forms. Our presentation follows~\cite{Sen:2007qy,Paquette:2017gmb,Bossard_2017,Bossard_2019,Dabholkar:2012zz,Banerjee:2008pv,Bryan:2018nlv}. More on CHL models and their duals can be found, for instance, in~\cite{Schwarz:1995bj,Mikhailov_1998,Vafa_1996,Chaudhuri:1995dj,Aspinwall_1996,Witten_1998wv,Lerche_1998,Bershadsky:1998vn,Berglund_1999,Park_1998,Clingher:2018fgx,Persson:2015jka}, see also~\cite{Kachru_1998,Bianchi_1998,de_Boer_2000}.

\subsection{Construction of $\mathcal{N}=4$ CHL models}
\label{ssc:ConstructionCHLmodels}

By virtue of $\mathcal{N}=4$ string duality these models have dual descriptions as freely acting orbifolds of heterotic string theory on $T^6$ or IIA string theory on $\text{K3} \times T^2$. Such models have been classified in~\cite{Persson:2015jka}. We will mainly be interested in the simplest and most studied case where the orbifolding group $G_{\mathrm{orb}}=\IZ_N$ is a cyclic group of order $N \in\lbrace 1,2,3,5,7 \rbrace$ and the rank of the resulting gauge group in the four non-compact spacetime dimensions is $r=2k+4$, the integer $k=24/(N+1)$ being determined by $N$.

 As is well-known, the \textit{maximal} rank case (i.e., the trivial orbifold) gives a gauge group $\U(1)^{28}$ at a generic point of the moduli space 
\begin{equation}
[\OO(22,6;\IZ) \backslash \OO(22,6) / (\OO(6) \times \OO(22))] \ \times \ [ \SL_2(\IZ) \backslash \SL_2(\IR)/\U(1)] \ ,
\end{equation}
corresponding to 22 vector multiplets, six graviphotons and the heterotic axio-dilaton. Here, the first factor can be understood as the heterotic Narain moduli space~\cite{NARAIN198641} and the quotient is taken by the discrete automorphism group $\OO(\Lambda_{22,6})\cong \OO(22,6;\IZ)$ of the Narain lattice $\Lambda_{22,6}$ of momentum-winding modes.%
\footnote{Recall that locally the Narain moduli space is parametrized by the metric and the antisymmetric B-field on $T^6$ as well as by the 16 Wilson lines for the Cartan-torus of the $E_8 \times E_8$ or $\mathrm{Spin}(32)/\IZ_2$ gauge group. This provides an embedding of the abstract lattice $E_8^{\oplus 2} \oplus U^{\oplus 6}\cong \Lambda_{22,6}$, which is the unique even unimodular lattice of signature (22,6) up to isomorphism, into the pseudo-Riemannian space $\IR^{22,6}$. Here we denoted by $U = \binom{0\ 1}{1\ 0}$ the hyperbolic lattice of signature $(1,1)$. Furthermore, the Grassmannian $\Gr_{r,s}\coloneqq \OO(r,s)/(\OO(r)\times \OO(s))$ parametrizes splittings $\IR^{r,s} \cong  \IR^{r,0} \oplus \IR^{0,s} $. In our notation $\OO(r)=\OO(r,\IR)$ etc. } %
  The discrete groups acting from the left are the T- and S-duality group of that theory.

For a given factorization $T^6 = T^4 \times S^1 \times \hat{S}^1$, the dual type II description is IIA[$\text{K3} \times S^1 \times \hat{S}^1$], or via T-duality on the last circle IIB[$\text{K3}   \times S^1 \times \tilde{S}^1$]. The complex structure modulus of $ S^1 \times \tilde{S}^1$ in type IIB, the complexified K\"{a}hler modulus of $ S^1 \times \hat{S}^1$ in type IIA and the heterotic axio-dilaton are dual to each other. Also the Narain lattice can be reinterpreted in the type IIA theory as
\begin{equation}
\Lambda_{22,6} \cong \Lambda_{20,4} \oplus \Lambda_{2,2} \ ,
\end{equation}
where $\Lambda_{20,4} \cong H^*(\text{K3},\IZ)$ is the integral cohomology lattice of the K3 surface, while  $ \Lambda_{2,2} $ is the winding-momentum lattice for $S^1 \times \hat{S}^1$. As an abstract lattice, the latter is given by the direct sum of two hyperbolic lattices, i.e., $ \Lambda_{2,2}\cong U^{\oplus 2}$.%
\footnote{On some subspaces of the Narain moduli space the generic gauge group will be enhanced, with non-Abelian gauge bosons arising from additional root vectors in the Narain lattice. Enhanced gauge symmetry occurs in the type IIA duality frame for degenerations of the K3 surface (see, for instance,~\cite{Aspinwall:1996mn}).} 

Let us turn to the \textit{reduced} rank theories. In the type IIA theory the cyclic orbifold group is generated by a pair $(g,\delta)$, consisting of an order $N$ action $g$ on the $\mathcal{N}=(4,4)$ K3 non-linear sigma model (NLSM) and a simultaneous order $N \lambda$ shift in the direction $\delta$ on $S^1$, where $\delta \in \Lambda_{2,2}$ has square zero in order to satisfy level matching. The condition on $g$ is to fix the superconformal algebra on the worldsheet and the spectral flow generators, see~\cite{Gaberdiel:2011fg} for a precise characterization. Indeed, one can choose $\lambda=1$ for symmetries $g$ that are geometric in the sense that $g$ describes an automorphism of the K3 surface that fixes the holomorphic-symplectic (2,0)-form (and thus keeps the $\SU(2)$ holonomy). Such symmetries are uniquely determined by their induced action on the lattice $H^2(\text{K3}, \IZ)$. They are in fact, up to lattice automorphisms, already determined by the order $1 \leq N \leq 8$ of $g$ and symplectic automorphisms of any order in that range do actually exist.\footnote{See~\cite{Mukai1988,kondo1998} or~\cite[ch. 15]{MR3586372}.} In this way we only consider CHL models associated to a symplectic automorphism of a K3 surface that has prime order. 
The middle cohomology lattice of the K3
\begin{equation}
\Lambda \coloneqq H^2(\text{K3},\IZ) \cong U^{\oplus 3} \oplus E_8(-1)^{\oplus 2} 
\end{equation}
contains an invariant $\Lambda^g$ and a coinvariant ($ \Lambda_g = \left( \Lambda^g \right)^{\perp} $) lattice with respect to $g$, i.e.,
\begin{equation}\label{eq:cohLatticeCoinv}
\Lambda \supseteq \Lambda^g \oplus \Lambda_g  \quad \text{and} \quad \Lambda^g = \{ v \in \Lambda\, | \, g v = v \} \ .
\end{equation}
We illustrate the case $N=2$, where $g$ is called a Nikulin involution. The induced action on $\Lambda$ exchanges the $E_8(-1)$ sublattices and fixes $U^{\oplus 3}$ pointwise. Equation \eqref{eq:cohLatticeCoinv} becomes 
\begin{equation}
\Lambda^g = U^{\oplus 3} \oplus E_8(-2) \ , \quad \Lambda_g = E_8(-2) \ , \label{eq:H2invlattice}
\end{equation}
with $E_8(-2) \subset E_8(-1)^{\oplus 2}$ denoting the diagonal or the anti-diagonal, respectively. 

On the heterotic side, the $\IZ_N$ orbifold action is asymmetric, i.e., acts by a $\IZ_N$ cyclic permutation and a shift on the left-moving coordinates while the right-moving coordinates are invariant (up to shifts)~\cite{Chaudhuri:1995dj}. For the $N=2$ case this gives an exchange of the internal $E_8 \times E_8$ factors and an order two shift along a circle of $T^6$. The one-loop partition function of this heterotic orbifold is reproduced in section \ref{sec:DHstatesZ2} as it will be needed later.

Moduli of a CHL model are given by the $g$-invariant moduli of the parent theory and take values in
\begin{equation}\label{eq:modspaceCHL}
G_4(\IZ)\backslash \ \left( \, [\OO(2k-2,6)/(\OO(2k-2)\times \OO(6) )] \times  [\SL_2(\IR) / \U(1) ] \, \right)
\end{equation}
for some discrete U-duality group $G_4(\IZ)$ in four non-compact dimensions, which includes a T-duality group $\mathcal{T}$ acting (only) on the first factor and an S-duality group $\mathcal{S}$ acting on the second factor (via M\"{o}bius transformations on the heterotic axio-dilaton~\cite{FONT199035,Sen:1994fa}
),
\begin{equation}
G_4(\IZ) \supset \mathcal{T} \times \mathcal{S} \ .
\end{equation}
The S-duality group turns out to be~\cite{Vafa_1996} 
\begin{equation}
\mathcal{S}=\Gamma_1(N) \subset \SL_2(\IZ) 
\end{equation} for the $\IZ_N$ CHL models, while the T-duality group $\mathcal{T}$ should at least contain\footnote{In practice, we will take this to be an equality and do not rigorously draw distinctions.} the centralizer $C_{(g,\delta)}$ of the orbifold generator $(g,\delta)$ in $\OO(\Lambda_{22,6})$,
\begin{equation}
\mathcal{T} \supset C_{(g,\delta)}  \coloneqq \lbrace h \in \OO(\Lambda_{22,6})  \, |  \,  h(\delta) = \delta , \ hg=gh \rbrace \ .
\end{equation}
A common way to parametrize the moduli associated with the Grassmanian (at least locally) is by means of a real $r \times r$ matrix $M$ subject to
\begin{equation}
MLM^\tp = L \ , \quad M=M^\tp
\end{equation}
where $L=L^{-1}$ is an $\OO(2k-2,6)$-invariant matrix representing the non-degenerate bilinear form on $\IR^{2k-2,6}$. This means $L$ has $2k-2$ eigenvalues $+1$ and 6 eigenvalues $-1$ counted with multiplicity and satisfies\footnote{Equivalently we can write $O^\tp  L O = L$ for all $O\in \OO(2k-2,6)$.}
\begin{equation}\label{eq:TmatrixCondition}
O L O^\tp  = L\ , \quad \text{for all} \ \ O\in \OO(2k-2,6) \ .
\end{equation}

As it has been argued in~\cite{Persson:2015jka}, there should also be a Fricke involution acting as \linebreak[4] $\Shet \mapsto -1/(N \Shet)$ on the axio-dilaton and by an orthogonal, not necessarily integral, transformation on the other moduli, see for instance~\cite{Bossard_2017,Bossard_2019} for further discussion in that direction. For simplicity we will mostly neglect possible Fricke type dualities. Here we think of elements in the T-duality group $\mathcal{T}$ always as automorphisms of the electric charge lattice defined next, 
\begin{equation}
\mathcal{T}\subset \OO(\Lambda_e) \ .
\end{equation}

\paragraph{Electric-magnetic charges.} Electric charges take values in a lattice of rank $r=2k+4$ signature and signature $({2k-2,6})$
\begin{equation}\label{eqn:chle}
\Lambda_e=\left(H^*(\text{K3},\IZ)^g\right)^*\oplus U\oplus U\left(\tfrac{1}{N}\right)
\end{equation}
while the magnetic charges take values in the dual lattice, which again has the same rank and signature,
\begin{equation}\label{eqn:chlm}
\Lambda_m=\Lambda_e^*=H^*(\text{K3},\IZ)^g\oplus U\oplus U\left(N\right).
\end{equation}
Their direct sum gives the electric-magnetic lattice
\begin{equation}\label{eqn:chlem}
\Lambda_{em} = \Lambda_e \oplus \Lambda_m \ .
\end{equation}
For $N>1$ the lattices are no longer self-dual (unimodular). Rather, they are $N$-modular, meaning that $\Lambda_m^* \cong \Lambda_m(\frac{1}{N})$ or $\Lambda_m^* (N) \cong \Lambda_m$ , i.e., they agree with their dual upon rotation and rescaling (see~\cite[eq. (2.10)]{Bossard_2017} for a concrete example):
\begin{equation}\label{eq:NmodularTrans}
\exists \, \sigma_N \in \OO(2k-2,6; \IR)  \ : \ \Lambda_m^* = \frac{\sigma_N}{\sqrt{N}} \Lambda_m \ .
\end{equation}
 The notation $\Lambda_m(\frac{1}{N})$ means that the bilinear form is rescaled by $1/N$. Multiplying \eqref{eq:NmodularTrans} by $N$ from the left and using the natural inclusion $\Lambda_m \subset \Lambda_m^*$ it follows that\footnote{The inclusion $N\Lambda_m^* \subset \Lambda_m $ is claimed in~\cite{Bossard_2017,Bossard_2019}, equivalent to $\langle N v , w \rangle \in \IZ$ for all $v,w \in \Lambda_e$.} 
 \begin{equation}
 N\Lambda_m \ \ \subset \ \ N \Lambda_m^* \, =\,  \sqrt{N}\sigma_N \ \Lambda_m
  \ \  \subset \Lambda_m \ \subset \ \ \Lambda_m^* \ .
 \end{equation}
For later reference we give the electric and magnetic lattice for the $N=2$ orbifold explicitly,
\begin{align}
\Lambda_e & =E_8 \left( -\tfrac{1}{2} \right) \ \oplus U^{\oplus 5}   \oplus U\left(\tfrac{1}{2}\right) \nonumber \\
\Lambda_m & =E_8 \left(-2\right) \,  \ \oplus U^{\oplus 5}  \oplus U\left( 2 \right)  \ . \label{eq:LambdaEMZ2chl}
\end{align}

\paragraph{Duality actions and charge invariants.} An element $ \begin{pmatrix}
 a & b \\ c & d
 \end{pmatrix}
 \in \Gamma_1(N)$ of the S-duality group acts on dyonic states with charge $(Q,P)^\tp \in \Lambda_{em}$ in the standard way~\cite[eq. (2.8)]{Persson:2015jka}:
 \begin{equation}\label{eq:SactionQP}
 \begin{pmatrix}
 Q\\ P
 \end{pmatrix} \mapsto  \begin{pmatrix}
 a & b \\ c & d
 \end{pmatrix}^{-1} \begin{pmatrix}
 Q\\ P
 \end{pmatrix}  \ , \quad \Shet \mapsto \frac{a \Shet + b}{c \Shet + d} \ .
 \end{equation}
The T-duality group $\mathcal{T} \ni O$ fixes $\Shet$ but acts on the remaining moduli and the charges as\footnote{We use the notation $O^\mtp = (O^\tp)^{-1}$.}
\begin{equation}
 \begin{pmatrix}
 Q\\ P
 \end{pmatrix}  \mapsto  \begin{pmatrix}
O^\mtp Q \\ O^\mtp P
 \end{pmatrix} \ , \quad  M\mapsto OMO^\tp \ .
\end{equation}
We denote the quadratic T-invariants as
\begin{equation}\label{eq:TinvsQuadratic}
 Q^2		  =  Q^\tp L Q \ , \qquad
 P^2		  =  P^\tp L P\,  \qquad \text{and} \qquad
Q\cdot P = Q^\tp L P	\ .
\end{equation}
The S-action of $\Gamma_1(N)$ on these follows from \eqref{eq:SactionQP}. For later convenience let us also introduce the map\footnote{Because of \eqref{eqn:chle}, \eqref{eqn:chlm} and \eqref{eqn:chlem} $P^2/2$ and $Q\cdot P$ are actually integral.}
\begin{equation}\label{eq:deftmap}
t : \Lambda_{em} \rightarrow \IQ^3 \ , \ (Q,P) \mapsto \left(  \frac{P^2}{2}, Q\cdot P , \frac{Q^2}{2} \right) \ .
\end{equation}

There are further, discrete T-duality invariants characterizing the duality orbit of a charge $(Q,P)$. Following~\cite{Dabholkar:2007vk}, take some basis of the lattice $\Lambda_{em}$ and denote the integer coordinates of a charge $(Q,P)$ with respect to this basis by $Q_i$ and $P_i$, the greatest common divisor of the integers $(Q_i P_j - Q_j P_i)$, denoted as
\begin{equation}\label{eq:defTorsionQwP}
I= \gcd (Q \wedge P) \ ,
\end{equation}
will then be a T-duality\footnote{As shown in~\cite[section 2]{Banerjee_2008do} a change of basis given by an $\SL_r(\IZ)$ matrix leaves the gcd invariant (there $r=22+6$ was considered). If $\mathcal{T}\subset \OO(\Lambda_e) \subset \SL_r(\IZ)$ this argument also holds for $\IZ_N$ CHL orbifolds of Het[$T^6$].} and S-duality invariant, sometimes called torsion.%
\footnote{We give some remarks. %
(1.) First note that $(Q,P)$ being primitive in $\Lambda_{em}$ does not imply that $Q\in \Lambda_{e}$ or $P \in \Lambda_m$ is primitive. In turn, if $Q$ or $P$ is primitive, then $(Q,P)$ is primitive as well.
(2.) If $Q$ or $P$ is non-primitive then $I>1$. On the other hand, $I>1$ does not imply that $Q$ or $P$ are non-primitive, as the example in~\cite[subsection 6.3]{Banerjee:2008pv} with $I=2$ shows: there both $Q$ and $P$ are primitive (and $Q \pm P$ are both twice a primitive vector). So $I=1$ is a sufficient, but not necessary condition for having both $Q$ and $P$ primitive. %
} %
 It has been shown that for Het[$T^6$] the quantity $I$ and the above quadratic T-invariants are sufficient to uniquely determine a duality orbit under S- and T-transformations in $G_4(\IZ)$. If S-transformations are left out, apart  from $I$ and the quadratic T-invariants three \textit{further} discrete T-invariants (on which the S-duality group acts non-trivially) are needed to characterize a T-orbit unambigously, see~\cite{Banerjee_2008do,Banerjee_2008sd} and~\cite[section 5.3]{Sen:2007qy} for details. Just in the special case $I=1$, which fixes the remaining three discrete T-invariants to unity, there is a single T-orbit.

\label{cor:typo:appB2}As was also pointed out in~\cite[app. B]{Bryan:2018nlv}, the precise duality group $G_4(\IZ)$ of a four-dimensional $\IZ_N$ CHL model with $N>1$ is not yet determined, nor is a complete set of duality invariants that uniquely specifies the distinct charge orbits in $\Lambda_{em}$ with respect to $G_4(\IZ)$. 
In any case, we expect that again finitely many duality invariants suffice to uniquely determine a duality orbit. Having several distinct duality orbits of charges means we should also expect several a priori distinct degeneracies associated to states with charge in the respective orbits. In this work we elaborate on this idea in the case of counting dyonic quarter-BPS states in the $\IZ_2$ CHL model. For simplicity we will focus on charges satisfying $I=1$. However, in contrast to the unorbifolded theory, this alone is not expected to uniquely specify a duality orbit, as there is at least one more discrete (candidate) charge invariant. 

As in~\cite[app. B]{Bryan:2018nlv} the ``residue'' of a charge $(Q,P)\in\Lambda_{em}$ is defined as the class in the discriminant group\footnote{Recall $\Lambda_{e} / \Lambda_e^* \cong \Lambda_{m}^* / \Lambda_m$ so definition \eqref{eq:defresidueQP} is equivalent to the one given in \cite[app. B]{Bryan:2018nlv}.\label{fn:resremark}}
\begin{equation}\label{eq:defresidueQP}
r(Q,P) = [Q] \, \in \Lambda_{e} / \Lambda_e^* \ .
\end{equation}
This quantity was shown to be invariant under $\Gamma_1(N) \times C_{(g,\delta)}$. 
 For the $\IZ_2$ model the discriminant group explicitly reads
\begin{equation}\label{eq:residueZ2}
\Lambda_{e} / \Lambda_e^* = \IZ_2^2 \times \IZ_2^8 \ , 
\end{equation}
where the first factor comes from $U(\tfrac{1}{2})/U(2)$ and the second factor from $E_8(-\tfrac{1}{2})/E_8(-2)$. In the perturbative heterotic description of section \ref{sec:DHstatesZ2} we will interpret, for purely electric half-BPS states, the respective components of $[Q]$ in terms of momentum-winding numbers along the CHL circle and the internal $E_8$ momentum. Especially, one $\IZ_2$ component here distinguishes whether the state lies in the untwisted (i.e., even CHL winding number) or twisted (odd CHL winding number) orbifold sector. Correspondingly, we will simply call electric charges twisted sector charges or untwisted sector charges.

 As already mentioned in the introduction, the dyon partition function introduced in \cite{Jatkar:2005bh} counts unit-torsion quarter-BPS dyons whose electric charge belongs to the twisted sector (see, for instance, the discussion in~\cite[section 5.3]{Sen:2007qy}). 
 Our goal is to propose partition functions belonging to other (unit-torsion) charge sectors. 
 A first step in this direction was undertaken in~\cite[section 6.5]{Banerjee:2008pv} for the $\IZ_2$ model by analyzing a closed subsector of the untwisted sector of unit-torsion dyons. Although no closed formula for the respective partition function was given, strong constraints on the latter coming from charge quantization, wall-crossing and S-duality invariance were given. We will later verify this subsector result in section \ref{sec:sdwcconstraints}.

\subsection{Structure of quarter-BPS partition functions}
\label{ssc:structureQBPSfn}

In this subsection we briefly review the structure of partition functions of quarter-BPS dyons in four-dimensional $\mathcal{N}=4$ string theories, closely following the discussion in~\cite{Banerjee:2008pv}. Many details will be omitted and can be found in the reference. 

\paragraph{BPS multiplets and indices.} Recall that quarter-BPS states transform in $2^6$-dim\-ensional intermediate multiplets. It also follows from the $\mathcal{N}~=~4$ superalgebra that quarter-BPS states with electric-magnetic charge $(Q,P)\in \Lambda_{em}$ must satisfy $Q \nparallel  P$ (i.e., the charges are not collinear as vectors in $\IR^{r}$). Half-BPS states in turn transform in $2^4$-dimensional short multiplets and obey the opposite charge condition, $Q \Vert P$. Non-BPS states transform in $2^8$-dimensional long multiplets.\footnote{See, for instance, the references~\cite{Cheng:2008gx,Dabholkar:2012zz} for explanation.} 

Since a quarter-BPS dyon breaks 12 out of 16 supercharges, an appropriate, i.e., non-trivial, index to ``count'' such states of a given charge $(Q,P) \in \Lambda_{em}$ is the sixth helicity supertrace\footnote{See, for instance,~\cite{Dabholkar:2012zz,Gomes:2011zzc,Kiritsis:1997hj} for explanation or~\cite{Bachas_1997,Gregori_1998} and references therein.}, denoted by $\Omega_6(Q,P; \cdot \, )$. 
Here the dot represents the moduli of the theory. Locally this index is constant, but it changes discontinously once the asymptotic moduli of the theory are varied across certain real codimension one subspaces, called walls of marginal stability.  Each wall is associated to a specific decay of the quarter-BPS dyon into a pair of half-BPS dyons. This wall-crossing phenomenon~\cite{Denef_2000,Bates_2011,Denef_2001,Denef_2002,denef2000correspondence,Denef_2011} is best understood in the case where the decay products carry primitive charges and for simplicity we restrict us to this case. Considering a quarter-BPS dyon with charge $(Q,P) \in \Lambda_{em}$ that decays at a certain (generically present) wall into two half-BPS states via
\begin{equation}\label{eq:z0halfBPSdecay}
(Q,P) \longrightarrow (Q,0) \ + \ (0,P) \ ,
\end{equation}
it is clear that we should restrict us to dyons where both $Q \in \Lambda_e$ and $P \in \Lambda_m$ are primitive lattice vectors. Furthermore we restrict to the case $I=1$. According to~\cite{Dabholkar:2007vk} this is also a necessary condition for the dyon partition function to be related to a chiral genus two partition function of the heterotic string, as we will discuss later.

 In principle there can also be decays where at least one decay product is quarter-BPS, however~\cite{Sen_2007_rare}, if $Q$ and $P$ are both primitive charges these occur in the moduli space at codimension two or higher. Thus generic points in this space can be connected by paths that do not cross these loci and the BPS index is not affected by such decay channels.

\paragraph{BPS charge sets.} For the purpose of analyzing or constraining a (quarter-BPS) dyon partition function it may be convenient to reduce the problem to analyzing charge subsectors, for which the counting problem simplifies. Let us introduce some notation. For a set of electric-magnetic charges $\mathcal{Q} \subset \Lambda_{em}$ we define the following conditions:
\begin{enumerate}[leftmargin=1.5cm]
\pagebreak[3]
\item[(Q1)] \textit{Quarter-BPS condition}:\\
					For all $(Q,P)\in \mathcal{Q}$ we have $Q \nparallel P$.

\item[(Q2)] \textit{Unit-torsion condition}:\\
					For all $(Q,P)\in \mathcal{Q}$ we have $I=\gcd(Q \wedge P)=1$.

\item[(Q3)] \textit{T-closure condition}:\\
For any given triplet $(q_1, q_2, q_3)$ of the quadratic T-invariants the set%
\begin{equation}\label{eq:chargeSlice}
\Big\lbrace \, (Q,P) \in \mathcal{Q} \, \Big| \,  \left( \frac{P^2}{2}, Q\cdot P,\frac{Q^2}{2}\right) = (q_1, q_2, q_3) \, \Big\rbrace \ ,
\end{equation} 
if not empty, maps to itself under the action of the T-duality group $\mathcal{T}$.

\item[(Q4)]  \textit{T-transitivity condition}:\\
Any two elements of subsets of the form \eqref{eq:chargeSlice} are related via $\mathcal{T}$. 

\item[(Q5)]  \textit{Unboundedness condition}:\\
Any of the quadratic T-invariants takes arbitrarily large absolute values on $\mathcal{Q}$.

\item[(Q6)] \textit{Quantization condition}:\\
There are rational numbers $q_{i}\in \IQ^+$ such that for any $(Q,P)\in \mathcal{Q}$ we can find integers $\nu_i \in \mathbb{Z}$ satisfying 
\begin{equation}\label{eq:chargequant}
\frac{P^2}{2} = \nu_1  q_1 \ \, ,\ \, Q \cdot P = \nu_2 q_2 \ \, , \ \, \frac{Q^2}{2}= \nu_3  q_3  \  \ .
\end{equation}
\end{enumerate}
Some remarks are in order. If $Q\parallel P$, then $Q \wedge P =0$, so (Q2) implies (Q1). %
The T-closure condition (Q3) obviously transfers to the whole set $\mathcal{Q}=\mathcal{T}\mathcal{Q}$. Condition (Q4) especially implies that any (further) T-invariants become constant functions on sets of the form \eqref{eq:chargeSlice}. Under both assumptions (Q3) and (Q4) a unique representative can be chosen for any non-empty set of the form \eqref{eq:chargeSlice} and the remaining elements of that set are precisely all $\mathcal{T}$-images of it. 
Furthermore, condition (Q6) is always satisfied for \textit{some} rational numbers $q_i  \in \IQ^+ $ (c.f. eqs. \eqref{eqn:chle} and \eqref{eqn:chlm}) and from now on we consider the \textit{maximal} numbers $\mathsf{q}_i \in \IQ^+$ for which \eqref{eq:chargequant} is satisied.\footnote{This becomes relevant when the charges in $\mathcal{Q}$ satisfy coarser quantization conditions than $\Lambda_{em}$, as applying to the charge sets considered in \cite[section 6]{Banerjee:2008pv}. In their simplest example one has a charge set $\mathcal{Q} \subset \Lambda_{em}$ for which $Q^2/2$ only takes even values, leading to $\mathsf{q}_3 = 2$ in that case, while $\Lambda_e= U^{\oplus 6}\oplus E_8(-1)^{\oplus 2}$ (considering charges of Het[$T^6$]) also allows for odd values of $Q^2/2$ (corresponding to $q_3 = 1$).\label{fn:Q6remark}}
If (Q3) to (Q6) are satisfied the T-orbits \eqref{eq:chargeSlice} are in one-to-one correspondence with points in $t(\mathcal{Q})$, which form a subset of some affine rank-three lattice\footnote{Recall that affine means that it is given by some lattice (including the zero vector), shifted by a non-zero vector.} $\mathsf{L} \subset \IQ^3$. The charge examples in~\cite{Banerjee:2008pv} are constructed such that already the T-representatives form an affine rank-three lattice $\mathsf{L}_{\mathcal{Q}} \subset \Lambda_{em}$ which then bijects to its T-invariants $t(\mathcal{Q}) = \mathsf{L}$ and $\mathcal{Q}$ is obtained by simply taking all T-images, $\mathcal{Q}=\mathcal{T}\mathsf{L}_{\mathcal{Q}}$. In this way (Q1)-(Q6) are satisfied simultaneously.

\paragraph{BPS partition functions.} We make the standard assumption that the sixth helicity supertrace $\Omega_6(Q, P ; \cdot )$ (or simply BPS index in the following) is invariant under S- and T-transformations, i.e., at a given generic point in the moduli space it only depends on the duality orbit of $(Q,P)\in\Lambda_{em}$.  Given $\mathcal{Q}$ satisfying (Q1), (Q3) and (Q4), because of T-invariance the BPS index of dyons with charge $(Q,P)\in \mathcal{Q}$ will already be uniquely determined by specifying the quadratic T-invariants of the charge and for some appropriate  $f_{\mathcal{Q}}$ we have
\begin{equation}\label{eq:Omega6f}
\Omega_6(Q ,P; \cdot \,) = f_{\mathcal{Q}} ( P^2, Q\cdot P, Q^2\, ; \, \cdot \, ) \ .
\end{equation}
One can also introduce a partition function for these numbers via%
\footnote{Following~\cite{Banerjee:2008pv}, we also introduced $\Phi_{\mathcal{Q}} \coloneqq (\mathsf{Z}_{\mathcal{Q}}(\tau,z,\sigma))^{-1}$. Writing the partition function in the form $\mathsf{Z}_{\mathcal{Q}}(\tau,z,\sigma) = \frac{1}{\cP_{\mathcal{Q}}(\tau,z,\sigma)} $ is alluding to the original DVV result $1/\chi_{10}$ and the CHL orbifold analogs considered by Sen et al.\label{fn:Z1overQ}}
\begin{equation}
\mathsf{Z}_{\mathcal{Q}}(\tau,z,\sigma) = \frac{1}{\cP_{\mathcal{Q}}(\tau,z,\sigma)}  \coloneqq \sum_{P^2, Q\cdot P, Q^2} (-1)^{Q\cdot P +1} f_{\mathcal{Q}}(P^2, Q\cdot P, Q^2 \, ; \cdot \, ) \ e^{2\pi i \left(  \crho \frac{P^2}{2} + \cv \, Q\cdot P +\csig\frac{Q^2}{2}  \right) } \, , \label{eq:14BPS_partition_fun_general}
\end{equation}
where a sign factor has been introduced to follow conventions in~\cite{Banerjee:2008pv} and the sum runs over all quadratic values belonging to charge vectors  $(Q,P)\in \mathcal{Q}$. 

Under the condition (Q5) the partition function is expected to have infinitely many non-zero terms.\footnote{Eventually we want $\ZQ$ to be a Siegel modular form (for some congruence subgroup) and we expect that this requires infinitely many non-zero ``Fourier modes'' $\exp(2\pi i k x)$, for each $x\in \lbrace \tau, \sigma, z \rbrace$.} Typically the generalized chemical potentials $\tau,  z,\sigma$ conjugate to  $P^2 /2$, $Q\cdot P$ and $Q^2/2$, must lie in a suitable domain of the Siegel upper half plane $\IH_2$ for this series to converge (see appendix \ref{app:siegelmodforms} for a definition) and we will assume that this is the case. Different domains of convergence admit different Fourier expansions, which in turn give BPS indices valid for different regions of the moduli space. \label{cor:B5}
As $\mathcal{Q}$ satisfies (Q6), the partition function will be periodic:
\begin{equation}
\forall n_1, n_2, n_3 \in \mathbb{Z}: \ \ \ \mathsf{Z}_{\mathcal{Q}}\left (\tau+\frac{n_1}{\mathsf{q}_1},z +\frac{n_2}{\mathsf{q}_2}, \sigma +\frac{n_3}{\mathsf{q}_3} \right) = \mathsf{Z}_{\mathcal{Q}}(\tau,z,\sigma ) \ .
\end{equation}
BPS indices can be extracted from $\ZQ$ by taking an appropriate contour integral
\begin{equation}\label{eq:contourint}
 f_{\mathcal{Q}}( P^2, Q\cdot P , Q^2 \, ; \cdot \, ) = \frac{(-1)^{Q\cdot P +1}}{(\mathsf{q}_1 \mathsf{q}_2 \mathsf{q}_3)^{-1}}  \oint_{\mathcal{C}}  \frac{e^{-2\pi i \left(  \crho \frac{P^2}{2} + \cv \, Q\cdot P + \csig\frac{Q^2}{2}  \right) }}{\PQ(\tau, \sigma, z)} \, \dd \tau \! \wedge \! \dd \sigma\! \wedge \! \dd z  \ 
\end{equation}
over a (minimal) period in each direction at some fixed, large imaginary part. In this work we will stay schematic with regard to the choice of integration contour, which could in principle be analyzed more carefully as in~\cite{Cheng_2007}, see also~\cite{Sen:2007vb,Banerjee:2008pv}. As mentioned before, we are mainly concerned with quarter-BPS dyons of unit-torsion, and for these dyons we assume the validity of the moduli-dependent contour proposed in~\cite{Cheng_2007}.

For quarter-BPS dyons of unit-torsion and we expect that a \textit{finite} number of discrete T-invariants provides a partition of the set
\begin{equation}\label{eq:14BPScharges}
\Big\lbrace \, (Q,P) \in \Lambda_{em} \, \Big| \, \gcd(Q\wedge P)=1 \, \Big\rbrace  
\end{equation}
into a \textit{finite} number of pairwise disjoint subsets $\mathcal{Q}$, each obeying (Q1) to (Q6). The important point is that this yields a finite set of (a priori different) quarter-BPS partition functions $\mathsf{Z}_{\mathcal{Q}}$. 

We remark that for any two of such disjoint charge sets $\mathcal{Q}, \mathcal{Q}'$ with quarter-BPS partition functions $\mathsf{Z}_{\mathcal{Q}},\mathsf{Z}_{\mathcal{Q'}}$, respectively, one can formally define the sum $\mathsf{Z}_{\mathcal{Q}} + \mathsf{Z}_{\mathcal{Q'}}$. 
If there are no common triplets of quadratic T-invariants, $t(Q)\cap t(Q')=\emptyset$, hence no common triple exponents in the respective expansion of the type \eqref{eq:14BPS_partition_fun_general}, $\mathcal{Q} \cup \mathcal{Q'}$ again satisfies (Q1) to (Q6) and $\mathsf{Z}_{\mathcal{Q}} + \mathsf{Z}_{\mathcal{Q'}}$ can be interpreted as $\mathsf{Z}_{\mathcal{Q} \cup \mathcal{Q'}}$. No information is lost upon addition.
On the other hand, if $t(Q)\cap t(Q')\neq \emptyset$, condition (Q4) is no longer satisfied. Extracting from $\mathsf{Z}_{\mathcal{Q}} + \mathsf{Z}_{\mathcal{Q'}}$ Fourier coefficients analogously to \eqref{eq:contourint} in this case yields numbers for which the interpretation \eqref{eq:Omega6f} does not hold, as there is no unique charge orbit (or orbit representative) given the quadratic invariants. Rather it is a sum of two BPS indices. However, such a ``compound'' BPS index can still be a well-behaved object, inheriting for instance the wall-crossing properties of its components that we discuss below (mostly due to linearity), and $\mathsf{Z}_{\mathcal{Q}} + \mathsf{Z}_{\mathcal{Q'}}$ exhibits modular transformation properties consistent with that. Similar remarks can be made for the half-BPS partition functions in section \ref{sec:DHstatesZ2}.\label{rk:compoundBPS}

\subsubsection{Constraints from S-duality symmetry and charge quantization}

Generically a subset $\mathcal{Q}$ will not be preserved (setwise) under the full S-duality group  $\mathcal{S}$ but only under a subgroup $\mathcal{S}_{\mathcal{Q}}\subset \mathcal{S}$ and transformations in $ \mathcal{S}\backslash \mathcal{S}_{\mathcal{Q}}$ map to other subsets $\mathcal{Q}'$. This is in line with the discussion after \eqref{eq:defTorsionQwP} and further examples can be found in~\cite{Banerjee:2008pv}. 
In any case, the invariance under $\mathcal{S}_{\mathcal{Q}}\subset \mathcal{S}$ has important consequences for $\mathsf{Z}_{\mathcal{Q}}$, as we will now discuss.

Recall that the S-duality group acts on the charges via \eqref{eq:SactionQP}.
Those transformations which map $\mathcal{Q}$ to itself form a subgroup $\mathcal{S}_{\mathcal{Q}}$ and for such transformations $\begin{psmallmatrix}
a & b \\ c & d
\end{psmallmatrix}$ S-duality invariance of the BPS indices can be recasted into the (suggestive) form (see~\cite{Banerjee:2008pv} for a derivation)
\begin{equation}\label{eq:smftrafo}
\PQ ((A\cO +B)(C\cO + D)^{-1}) = \det (C \cO +D )^k \, \PQ (\cO)
\end{equation} 
for some $k$ where
\begin{equation}\label{eq:symplecticStrafo}
\cO \coloneqq \begin{pmatrix}
\crho & \cv\\
\cv & \csig
\end{pmatrix}, \qquad \text{and} \quad \begin{pmatrix}
A & B\\
C & D
\end{pmatrix} = \begin{pmatrix}
d & b & 0 & 0\\
c & a & 0 & 0\\
0 & 0 & a & -c\\
0 & 0 & -b & d
\end{pmatrix} .
\end{equation}
At this point $k$ is undetermined, since the determinant is unity. However, for the known CHL examples the integer $k$ agrees with the previously defined $k$ and this is in fact required by wall-crossing and modular invariance (more on this later). The $4 \times 4$ matrix in \eqref{eq:symplecticStrafo} is symplectic and takes the form given in  \eqref{eq:SympEx2} for $U=\begin{psmallmatrix}
d & c \\ b & a
\end{psmallmatrix}$.

Similarly, we can rewrite the periodicity property of $\PQ$ in the form \eqref{eq:smftrafo}, but now with 
\begin{equation}\label{eq:SpMatrixPeriodicities}
\begin{pmatrix}
A & B\\
C & D
\end{pmatrix} = \begin{pmatrix}
1 & 0 & r_1 & r_2 \\
0 & 1 & r_2 & r_3 \\
0 & 0 & 1 & 0\\
0 & 0 & 0 & 1
\end{pmatrix} 
\end{equation}
and suitable periods $r_1, r_2, r_3$ subject to the choice of $\mathcal{Q}$. This is also a special case of a symplectic matrix, see eq. \eqref{eq:SympEx1} with $S=\begin{psmallmatrix}
r_1 & r_2 \\ r_2 & r_3
\end{psmallmatrix}$.

\subsubsection{Constraints from wall-crossing}

\label{sss:wallcrossing}

Let us now explain how wall-crossing puts additional modular constraints on $\PQ$. A general parametrization for the decay of a quarter-BPS dyon into a pair of half-BPS dyons is given by
\begin{equation}\label{eq:genericdecay}
(Q,P) \to (a_0d_0 Q - a_0b_0 P, c_0d_0 Q - c_0b_0 P) +
(-b_0c_0 Q + a_0b_0 P, - c_0d_0 Q + a_0d_0 P) 
\end{equation}
with $a_0
d_0 - b_0c_0 =1$. The decay products on the right hand side of \eqref{eq:genericdecay}, 
\begin{align}
(Q_1,P_1) &\coloneqq (a_0 Q', c_0 Q') = \left( \begin{pmatrix}
a_0 & b_0\\
c_0 & d_0
\end{pmatrix} \begin{pmatrix}
Q' \\
0
\end{pmatrix}\right)^\tp \\ 
(Q_2,P_2) &\coloneqq (b_0 P', d_0 P') = \left( \begin{pmatrix}
a_0 & b_0\\
c_0 & d_0
\end{pmatrix} \begin{pmatrix}
0 \\
P'
\end{pmatrix}\right)^\tp \ ,
\end{align}
where we have set
\begin{equation}
Q' \coloneqq d_0 Q - b_0 P \quad \text{and}  \quad P'\coloneqq-c_0 Q + a_0 P \ ,
\end{equation}
  again have to belong to the charge lattice $\Lambda_{em}$. Note that a charge set $\mathcal{Q}\subset \Lambda_{em}$ always comes along with its allowed decays \eqref{eq:genericdecay} and thus determines charges $(Q', P')$ and $(Q_i, P_i)$.
  
 Following the ansatz that the jump in the BPS index  due the decay \eqref{eq:genericdecay} is determined by a second order pole of $\PQ^{-1}$ at
\begin{equation}\label{eq:vprime_decay}
\cv' \coloneqq  c_0d_0 \, \crho + a_0b_0 \,  \csig+  (a_0d_0 + b_0c_0)\, \cv  = 0 \  ,
\end{equation}
the contour integral \eqref{eq:contourint} for the Fourier coefficient of \eqref{eq:14BPS_partition_fun_general} needs to pick up a residue\footnote{This wall-crossing formula is only valid for \textit{primitive} charges in the decay products, see~\cite[p. 7]{Banerjee:2008pv} and~\cite{Denef_2011}.}
\begin{equation}\label{eq:wallcrossjump}
(-1)^{Q' \cdot P'+1} 
\ \ Q'\cdot P' \ \
d_h(a_0 Q',c_0 Q') \ d_h(b_0 P', d_0 P')\, 
\end{equation}
up to a sign. In this expression $d_h(\tilde{Q},\tilde{P})=\Omega_4(\tilde{Q},\tilde{P})$ denotes the fourth helicity supertrace, an index only sensitive to half-BPS multiplets of dyonic charge $(\tilde{Q},\tilde{P})$ (often simply called half-BPS index). 
As in~\cite{Banerjee:2008pv} we want to restrict to those cases where the half-BPS indices again can be written as Fourier coefficients of a suitable partition function,
\begin{align}
d_h(a_0 Q',c_0 Q') &= \frac{1}{T} \int_{iM-T/2}^{ i M+T/2}  \ \, \, \frac{ e^{-i\pi Q^{\prime 2} \sigma'}}{\phi_e(\sigma';a_0,c_0)}\ \dd \sigma' \\
d_h(b_0 P', d_0 P') &=\frac{1}{T'} \int_{iM-T'/2}^{i M+T'/2} \, \frac{ e^{-i\pi P^{\prime 2}\tau '}}{\phi_m(\tau';b_0,d_0)}\ \dd \tau' \ .
\end{align}
Here the integration contour lies parallel to the real axis and extends over a unit period $T (T')$ of $\phi_{e} (\phi_{m})$ and $M \gg 0$ is large enough to ensure convergence.
Half-BPS partition functions for purely electrically charged states (in the heterotic frame) can, for instance, be found by counting perturbative, heterotic Dabholkar-Harvey (DH) states of the corresponding charge (as reviewed in section \ref{sec:DHstatesZ2} for the $\IZ_2$ model). Requiring the existence of functions $\phi_{e} (\phi_{m})$ as stated imposes constraints\footnote{These are the subtleties mentioned in~\cite[pp. 19 f. and p. 21 f. n. 8]{Banerjee:2008pv}.} on $\mathcal{Q}$: 
\begin{enumerate}[leftmargin=1.5cm]
\item[(Q7)] For any $(Q',P')$ appearing as above, the values $(Q')^2$ takes for fixed $(P')^2$ are independent of the latter. The same holds for their roles reversed.\footnote{An example of an excluded case: $(Q')^2/2$ is odd iff $(P')^2/2$ is even and vice versa.} 

\item[(Q8)] For fixed ``decay code'' $ \begin{psmallmatrix}
a_0 & b_0\\
c_0 & d_0
\end{psmallmatrix}$, all the decay products $(Q_1, P_1)$ obtained from letting $(Q,P)$ run over $\mathcal{Q}$ need to fall into a \textit{single} T-orbit for each value of  $Q^{\prime 2}$ . The same holds for $(Q_2, P_2)$ and $P^{\prime 2}$.

\end{enumerate}
Without (Q8), i.e., if there were several orbits, the half-BPS indices would not be functions of the mere quadratic T-invariants. 

The property (Q8) is similar to (Q4) above. In accordance with the remarks on page~\pageref{rk:compoundBPS} for compound quarter-BPS indices obtained from unions of charge orbits the half-BPS indices (or partition functions) occuring in the wall-crossing formula are again sums, coming from the decay products of the component orbits.
  
A sufficient condition for the jump is that near $\cv' =0$ the function $\PQ$ behaves as
\begin{equation} \label{eq:decaypolebehaviour}
\PQ^{-1}(\crho, \csig, \cv) \propto \Big( \,  \phi_e(\csig'; a_0,c_0)^{-1} \ {\phi_m(\crho' ; b_0, d_0)^{-1}} \
\, \cv^{\prime -2} + \mathcal{O}(\cv^{\prime 0}) \,  \Big)
\end{equation}
in the transformed variables
\begin{equation}\label{eq:polecoords}
\cO' \coloneqq \begin{pmatrix}
\crho' & \cv'\\
\cv' & \csig'
\end{pmatrix} = (A\cO +B)(C\cO + D)^{-1}, \qquad \begin{pmatrix}
A & B\\
C & D
\end{pmatrix} = \begin{pmatrix}
d_0 & b_0 & 0 & 0\\
c_0 & a_0 & 0 & 0\\
0 & 0 & a_0 & -c_0\\
0 & 0 & -b_0 & d_0
\end{pmatrix} .
\end{equation}
More explicitly, $z'$ is as defined in \eqref{eq:vprime_decay} while
\begin{equation}
\crho' =d_0^2 \, \crho + b_0^2 \, \csig + 2b_0 d_0\, \cv \quad \text{and} \quad \csig' =c_0^2 \, \crho + a_0^2 \,\csig + 2a_0 c_0 \, \cv \ .
\end{equation}
Note that (Q7) is generically required for the factorization in \eqref{eq:decaypolebehaviour}.

Given that the functions
$\phi_m(\tau;b_0, d_0)$ and $\phi_e(\tau; a_0, c_0)$ 
transform as weight $k+2$ modular forms under fractional linear transformations (a.k.a. M\"{o}bius transformations) of $\tau$ encoded by $ \SL_2(\mathbb{Z})$-matrices $\begin{psmallmatrix}\alpha_1 & \beta_1\\ \gamma_1 & \delta_1\end{psmallmatrix}$
and $\begin{psmallmatrix}p_1 & q_1 \\ r_1 & s_1\end{psmallmatrix}$,  respectively, we can map these to symplectic transformations of the form
\begin{equation} \label{egensp}
\begin{pmatrix}
d_0 & b_0 &0&0 \\
 c_0 & a_0 &0&0 \\
0&0& a_0 & -c_0\\
 0&0& -b_0 & d_0
 \end{pmatrix}^{-1}
 \begin{pmatrix}
\alpha_1 &0 & \beta_1 &0 \\
0 & 1 &0&0 \\
\gamma_1 & 0 & \delta_1 & 0\\
 0&0& 0 & 1
 \end{pmatrix}
 \begin{pmatrix}
 d_0 & b_0 &0&0 \\
 c_0 & a_0 &0&0 \\
0&0& a_0 & -c_0\\
 0&0& -b_0 & d_0
 \end{pmatrix}
\end{equation}
and
\begin{equation} \label{egensp2}
\begin{pmatrix}
d_0 & b_0 &0&0 \\
 c_0 & a_0 &0&0 \\
0&0& a_0 & -c_0\\
 0&0& -b_0 & d_0
 \end{pmatrix}^{-1}
 \begin{pmatrix}
1 &0 & 0 &0 \\
0 & p_1 &0& q_1 \\
0 & 0 &1& 0\\
 0&r_1& 0 & s_1
 \end{pmatrix}
 \begin{pmatrix}
 d_0 & b_0 &0&0 \\
 c_0 & a_0 &0&0 \\
0&0& a_0 & -c_0\\
 0&0& -b_0 & d_0
 \end{pmatrix} \ ,
\end{equation}
respectively. These in turn act as $\cO\mapsto (A\cO +B)(C\cO + D)^{-1} $ when written in the usual block form. Typically such a half-BPS partition function is a modular form for some congruence subgroup of $\SL_2(\IZ)$. In some cases \eqref{egensp} and \eqref{egensp2} lift to modular symmetries of $\PQ$ in the sense of \eqref{eq:smftrafo}. Also notice the simple relation between the modular weights $k+2$ of the functions $\phi_{e,m}$ and the weight $k$ of the function $\Phi_{\mathcal{Q}}$. Hence, wall-crossing determines the \textit{location} and \textit{coefficients} of quadratic poles in our quarter-BPS partition function together with candidate Siegel \textit{modular symmetries} and the \textit{modular weight}.\footnote{There might be additional (``accidental'') modular symmetries as in~\cite[subsection 6.4]{Banerjee:2008pv} or some of the (genus one) modular symmetries do not lift to the full quarter-BPS partition function, see, for instance, the example in~\cite[subsection 6.2]{Banerjee:2008pv}.} 

We remark that the middle matrix in each \eqref{egensp}, \eqref{egensp2} preserves the locus $z=0$, while the conjugated matrix preserves the locus $z' = 0$.

Formally, \eqref{eq:polecoords} resembles an embedded S-duality transformation (c. f. \eqref{eq:symplecticStrafo}), but the matrix $\begin{psmallmatrix}
a_0 & b_0\\
c_0 & d_0
\end{psmallmatrix}$ does not need to lie in $\SQ \subset \SL_2(\mathbb{Z})$. Indeed, S-duality can be shown to act on a decay code $\begin{psmallmatrix}
a_0 & b_0\\
c_0 & d_0
\end{psmallmatrix}$ from the left. In this way S-duality symmetry of the theory and the behaviour at $\cv=0$, which is related to the decay $(Q,P)\rightarrow (Q,0) + (0,P)$ with the identity matrix as decay code, already imply the location and coefficients of an infinite set of quadratic poles. Furthermore, multiplying a decay code by $\begin{psmallmatrix}
\lambda & 0\\
0 & \lambda^{-1}
\end{psmallmatrix}$ from the right for any real $\lambda\neq 0$ leads to an equivalent decay. The same holds for $\begin{psmallmatrix}
0 & 1\\
-1 & 0
\end{psmallmatrix}$. This makes it clear that for heterotic strings on $T^6$ with the weight 10 Igusa cusp form taking the role of $\PQ$ all decays are related to the one at $\cv =0$ by an $\SL_2(\mathbb{Z})$ transformation, which is known to be the S-duality group of that theory. However, in CHL orbifolds we may find inequivalent walls after modding out the mentioned redundancies. 

As was multiply exemplified in~\cite{Banerjee:2008pv}, the expected properties of $\PQ$ just described lead to a heuristics for finding quarter-BPS counting functions subject to a charge set $\mathcal{Q}$. By the same token, they provide a set of highly non-trivial tests for any given candidate counting function. Since the half-BPS partition functions form a key ingredient of this approach, we will now recall some facts about the latter in case of the heterotic $\IZ_2$ CHL model.


\section{Half-BPS spectra from Dabholkar-Harvey states in the $\mathbb{Z}_2$ model}
\label{sec:DHstatesZ2}

In this section we reproduce from~\cite{Dabholkar:2005dt} the computation of electric half-BPS partition functions in the heterotic $\IZ_2$ CHL orbifold\footnote{See also~\cite{Mikhailov_1998} and~\cite[app. A.1]{Bossard_2017} for closely related results. For the prime order CHL models these half-BPS partition functions, or rather those of the singly twisted sector, have recently been revisited in~\cite{Nally:2018raf} from a macroscopic point of view.} that appear in wall-crossing relations for quarter-BPS partition functions. Doing so we set the notation and collect relevant wall-crossing data for section \ref{sec:sdwcconstraints}. The genus two analysis of section \ref{sec:genustwo} will eventually go along similar lines, so this review section also serves as a warm-up exercise.

By electric half-BPS partition functions we mean generating functions for fourth helicity supertraces that count perturbative heterotic Dabholkar-Harvey (DH) states~\cite{Dabholkar:1989jt,Dabholkar:1990yf} of a given, purely electric charge. These are half-BPS states.\footnote{When we speak of DH states in the following, we will always mean the perturbative heterotic half-BPS states. Otherwise, DH states are not always half-BPS~\cite[f.n. 6]{dabholkar2012quantum}.} For DH states the superconformal side\footnote{Here the convention is made to call the superconformal side of the heterotic string right-moving.} of the heterotic string is restricted to the oscillator ground state and degeneracies can be computed both by direct enumeration of the relevant orbifold-invariant bosonic oscillator configurations or by making use of the helicity supertrace method.\footnote{Helicity supertraces are reviewed in appendix E and G of~\cite{Kiritsis:1997hj}, see also~\cite{Kiritsis:2007zza} for a short textbook chapter.} We make use of the latter.

Consider the generating function~\cite{Kiritsis:1997hj}
\begin{equation}\label{eq:helSTgenFunc}
\mathsf Z(q, \bar{q}; v , \bar{v} ) = \operatorname{Tr}_{\mathcal{H}} \left[ (-1)^F e^{2\pi i v J_3^L} e^{2\pi i \bar{v} J_3^R} q^{L_0} \bar{q}^{\bar{L_0}} \right] \ ,
\end{equation}
where the trace is taken over the Hilbert space $\mathcal{H}$ of  the perturbative heterotic $E_8 \times E_8$ string compactified on $T^6$ or an orbifold thereof. The spacetime fermion number is denoted by $F$ and the physical helicity in the four non-compact spacetime dimensions 
 $J_3 = J_3^R + J_3^L$ 
 is a sum of the left-helicity $J_3^L$ coming from left-movers and the right-helicity $J_3^R$ coming from right-movers. More precisely, the oscillators that contribute to the right-helicity $J_3^R$ come from the right-moving light-cone bosons $\partial \bar{X}^{\pm}=\partial \bar{X}^3 \pm i \partial \bar{X}^4$, contributing helicity $\pm 1$, respectively, and the light-cone fermions $\psi^\pm$, again contributing $\pm 1$ to the right helicity. On the other hand, only $\partial X^\pm$ contribute to the left-helicity $J_3^L$. For instance, the 2+2 chiral light-cone bosons contribute a factor of 
 \begin{equation}
\frac{\xi(v)}{\eta^2} \frac{\bar{\xi}(\bar{v})}{\bar{\eta}^2} = q^{-2/24}\bar{q}^{-2/24} \prod^\infty_{n=1} \frac{1}{(1-q^n e^{2\pi i v})(1-q^n e^{-2\pi i v})} \frac{1}{(1-\bar{q}^n e^{2\pi i \bar{v}})(1-\bar{q}^n e^{-2\pi i \bar{v}})} \ .
 \end{equation}
Note that $ \mathsf Z(q, \bar{q};0,0)= \operatorname{Tr}_{\mathcal{H}} \left[ (-1)^F q^{L_0} \bar{q}^{\bar{L_0}} \right]$ is just the ordinary one-loop partition function of the heterotic string (or its orbifold) including the GSO projection.

 Taking all together the generating function of helicity supertraces for the $\mathbb{Z}_2$ CHL orbifold is\footnote{We keep the $q, \bar{q}$ dependence\label{cor:B8} implicit, where $q=\exp(2 \pi i \tau)$.}
 \begin{align}
\mathsf Z(q, \bar{q};v , \bar{v} )  & = \frac{1}{\tau_2}\frac{\xi(v)\bar{\xi}(\bar{v})}{\eta^2 \bar{\eta}^2}
 \left( 
 \frac{1}{2} \sum_{\alpha, \beta=0}^1 (-1)^{\alpha+\beta+\alpha \beta} \frac{\bar{\theta}^{ }\begin{bsmallmatrix}
 \alpha /2 \\
 \beta /2
\end{bsmallmatrix} (\bar{v}) \ }{\bar{\eta}^{ } }
\frac{\bar{\theta}^{3}\begin{bsmallmatrix}
 \alpha /2  \\
 \beta /2
\end{bsmallmatrix} (0)}{ \bar{\eta}^{3}}
 \right) \nonumber \\
  &  \ \ \times
 \left( \frac{1}{2} \sum_{g,h=0}^1 \frac{
\mathcal{Z}_{6,6}\begin{bsmallmatrix}
 h \\
g 
\end{bsmallmatrix}}
{\eta^6 \bar{\eta}^6}
\mathcal{Z}_8\begin{bsmallmatrix}
 h  \\
g 
\end{bsmallmatrix}
 \right) \label{eq:Zvv_2ndline}  \ .
 \end{align}
Here $\alpha, \beta = 0, 1$ run over the four spin structures, $h=0,1$ indicates the untwisted or twisted sector and $g=0,1$ indicates an insertion of the orbifold involution into the trace. In the above expression we have the partition function of the (shifted) Narain-lattice 
\begin{equation}\label{eq:torusblockdefg1}
 \mathcal{Z}_{6,6}\begin{bsmallmatrix}
h  \\ g
\end{bsmallmatrix} =\sum_{Q \in \Lambda_{6,6}^{[h]}} (-1)^{g \, \delta \cdot  Q} \ e^{i\pi Q_{L} \tau Q_{L} - i \pi Q_{R} \bar{\tau} Q_{R}} \ ,
\end{equation}
where the subscript $L/R$ denotes the left- and right-part of the lattice vectors\footnote{That is, $L,R$ denotes a projection onto the subspace of $\IR^{6,6}$ of negative or positive signature (using the moduli dependent embedding of the charge lattice into $\IR^{6,6}$) and similarly for higher rank lattices (e.g., $\Lambda_{6,6} \oplus  E_8(2)$) appearing below. See, e.g.,~\cite[section 2]{dabholkar2012quantum} for the projection operators.}
and
\begin{equation}\label{eq:toruslattice}
\Lambda^{[h]}_{6,6} = \left( \Lambda_{6,6} + \frac{h}{2} \delta \right)
\end{equation}
is the Narain-lattice associated with $T^6$, possibly shifted by the null vector  $\delta = (0^6 \, ; \, 0^{6-1} , 1)$ such that the CHL action on $T^6$ is given by a half-translation along the last circle in $T^6$ (the CHL circle).
Thus $h=0$ means summation over untwisted sector charges, $Q \in \Lambda_{6,6} \cong U^{\oplus 6} $, the winding number along the CHL circle taking integral values. On the other hand, $h=1$ gives twisted sector charges $Q \in \Lambda_{6,6} + \frac{\delta
}{2}$ with the winding number along the CHL circle taking values in $\IZ + \frac{1}{2}$. The factor $(-1)^{\delta \cdot Q}$, for $g=1$, then becomes $(-1)$ for an odd number of momentum quanta along the CHL circle and $(+1)$ for an even number.
Furthermore, in \eqref{eq:Zvv_2ndline} we introduced the orbifold blocks  $\mathcal{Z}_8\begin{bsmallmatrix}
 h  \\
g 
\end{bsmallmatrix}$ for the 16 chiral bosons compactified on the $E_8 \times E_8$ root lattice, where the orbifold involution exchanges the two $E_8$ factors\footnote{Upon diagonalization, this gives eight invariant chiral bosons and eight chiral bosons that pick up a minus sign under the $\IZ_2$ action.} and one finds
\begin{align}
\mathcal{Z}_8\begin{bsmallmatrix}
\, 0 \, \\
\, 0 \,
\end{bsmallmatrix} = \frac{\left[ \theta_{E_8(1)}(\tau)\right]^2}{\eta^{16}(\tau)} \ , \, \qquad 
&      \ \qquad 
\mathcal{Z}_8\begin{bsmallmatrix}
0 \\
1 
\end{bsmallmatrix} =  \frac{\theta_{E_8(1)}(2\tau)}{\eta^8(2\tau)}
\label{eq:Zcur_untw} \ , \\
\mathcal{Z}_8\begin{bsmallmatrix}
1  \\
0
\end{bsmallmatrix} = \frac{\theta_{E_8(1)}(\frac{\tau}{2})}{\eta^8(\frac{\tau}{2})} \ \qquad \text{and}
&  \ \qquad 
\mathcal{Z}_8\begin{bsmallmatrix}
1  \\
1 
\end{bsmallmatrix} = e^{-2\pi i/3} \frac{\theta_{E_8(1)}(\frac{\tau +1 }{2})}{\eta^8(\frac{\tau +1}{2})} \label{eq:Zcur_tw} \ . 
\end{align}
Especially, $\theta_{E_8(1)}(\tau)=\sum_{v \in E_8(1)}e^{i\pi\tau v^2}=E_4(\tau)$ is the weight four Eisenstein series. 

As a remark, the terms in the first line of \eqref{eq:Zvv_2ndline} should arise for all heterotic $\IZ_N$ CHL orbifolds, as the superconformal sector of the heterotic string is unaffected by the orbifold action. 
On the other hand, the terms in the second line of \eqref{eq:Zvv_2ndline} are the orbifold blocks specific to the order $N=2$ shift along one of the circles of $T^6$ and the order $N$ permutation on the left-moving chiral bosons.

Helicity supertraces can be obtained from the generating function \eqref{eq:helSTgenFunc} by taking appropriate derivatives with respect to the generalized chemical potentials $v$ and $\bar{v}$ coupling to the left and right helicity, respectively:
\begin{equation}
B_n(q,\bar{q}) =\left(\frac{1}{2\pi i} \frac{\partial}{\partial v} +\frac{1}{2\pi i} \frac{\partial}{\partial \bar v} \right)^n \mathsf Z(q, \bar{q};v,\bar{v})\Big|_{v=0=\bar{v}} \ .
\end{equation}
We now want to obtain the fourth helicity supertrace $B_4$.\footnote{Strictly speaking, this is rather another a generating function, not yet a helicity supertrace $\Omega_4$ in the sense of section \ref{sec:countingbpsN4}.}  The fermion terms in the first line of \eqref{eq:Zvv_2ndline} can be rewritten using the Riemann identity to give $\bar{\theta}_1^4(\bar{v}/2)$. This implies that the only combination of $v$- and $\bar{v}$-derivatives that does not vanish when evaluated at $v=\bar{v}=0$ is taking four $\bar{v}$-derivatives, since $\theta_1(0|\tau )=0$. Using further \begin{equation}\label{eq:theta1derivative}
 \partial_{\tilde{v}} \bar{\theta} \begin{bsmallmatrix}
 1 /2 \\
 1 / 2
\end{bsmallmatrix} ( \tilde{v} | \tau ) \Big|_{ \tilde{v} =0} = 2\pi  \bar{\eta} (\tau )^3
\end{equation}
and $\xi(0)=\bar{\xi}(0)=1$ we obtain\footnote{The factor $3/2$ arises as $24\times (1/2)^4$ coming from the $4!=24$ permutations of $ \bar{v}$-derivatives and the inner derivative, c.f. the argument $\tilde{v}=\bar{v}/2$. }
\begin{equation}
B_4(q, \bar{q}) = \frac{3}{2} \frac{1}{\tau_2} \frac{1}{\eta^{2+6}} \times \left( \frac{1}{2}  \sum_{g,h=0}^1 
\mathcal{Z}_{6,6}\begin{bsmallmatrix}
 h \\
g 
\end{bsmallmatrix}
\mathcal{Z}_8\begin{bsmallmatrix}
 h  \\
g 
\end{bsmallmatrix}
 \right) \ . 
\end{equation}
Inserting the identities \eqref{eq:Zcur_untw} and \eqref{eq:Zcur_tw} we can also write
\begin{align}
B_4(q, \bar{q})  & = \frac{3}{2\tau_2}\ \frac{1}{2} \Bigg[ \frac{\theta^2_{E_8(1)}(\tau)}{\eta^{24}(\tau)}  \mathcal{Z}_{6,6}\begin{bsmallmatrix}
 0 \\
0 
\end{bsmallmatrix} + \frac{\theta_{E_8(1)}(2\tau)}{\eta^{8}(\tau)\eta^{8}(2\tau)}  \mathcal{Z}_{6,6}\begin{bsmallmatrix}
 0 \\
1 
\end{bsmallmatrix}  +  \frac{\theta_{E_8(1)}(\frac{\tau }{2} )}{\eta^{8}(\tau)\eta^{8}(\frac{\tau }{2})}  \mathcal{Z}_{6,6}\begin{bsmallmatrix}
 1 \\
0
\end{bsmallmatrix}  \nonumber \\
 & \qquad \qquad  \qquad \qquad  + e^{-2\pi i/3}  \frac{\theta_{E_8(1)}( \frac{\tau + 1 }{2} )}{\eta^{8}(\tau) \eta^{8}(\frac{\tau+1 }{2})}  \mathcal{Z}_{6,6}\begin{bsmallmatrix}
 1 \\
1
\end{bsmallmatrix} \Bigg] \ .\label{eq:B4_Z2chl_etas}
\end{align}

 Before interpreting the result \eqref{eq:B4_Z2chl_etas}, we interlude with a reminder of the unorbifolded case. The contribution with $\mathcal{Z}_{6,6}\begin{bsmallmatrix}
 0 \\
0 
\end{bsmallmatrix}$ corresponds (up to the factor $1/2$) to helicity supertraces of perturbative states in the unorbifolded theory Het$[T^6]$:
\begin{equation}\label{eq:B4unorb}
B_4^{\mathrm{unorb}}(q,\bar{q}) = \frac{1}{\tau_2}\times  \mathcal{Z}_{6,6}\begin{bsmallmatrix}
 0 \\
0 
\end{bsmallmatrix}(q,\bar{q}) \ \theta^2_{E_8(1)}(\tau) \times \frac{3}{2} \frac{1}{\eta^{24} (\tau)} \ .
\end{equation}
Let us pause to recall the semantics of this factorization. We have a continuous degeneracy due to the momenta $p_3, p_4$ in the non-compact directions transverse to the light-cone, leading to a factor of $1/ \tau_2$. Also we have the Narain-lattice sum of vectors $Q \in \Lambda_{22,6} \cong E_8(-1)^{\oplus 2} \oplus  U^{\oplus 6}$ and a factor $\eta^{-24}(\tau)$ corresponding to oscillator modes of 24 chiral bosons (transverse to the light-cone). As seen from the four-dimensional spacetime perspective for each momentum $(p_3, p_4)$ and electric charge vector $Q\in \Lambda_{22,6}$ (momentum and winding) we have the full tower of DH states generated by allowing arbitrary left-moving oscillators while keeping the superconformal sector in the ground state. The latter is, due to the GSO projection, a Weyl spinor with $2^{8/2}=16$ components. Hence for fixed $(p_3,p_4)$ we can relate the fourth helicity supertrace of states with charge $Q$ to the absolute degeneracy of states with charge $Q$ as\footnote{Recall that $p_{24}(N)$ is the number of ways of writing the non-negative integer $N $ as a sum of 24 non-negative integers. This is also the Fourier coefficient of $q^{N-1}$ in $\eta^{-24}(\tau)$. For any $\tau \in \mathbb{H}$ the Fourier series of the latter converges, so there is no ambiguity, i.e., no wall-crossing for these half-BPS states and no moduli dependence in $\Omega_4$.}
\begin{equation}\label{eq:DHexample}
d_h(Q,0) = \Omega_4(Q) = \frac{3}{32} \Omega_{\mathrm{abs}}(Q)  = \frac{3}{2} p_{24}(N)  \ . 
\end{equation}
The level number $N$ (not to be confused with the order of the CHL orbifold group) is related to the charge $Q \in \Lambda_{22,6}$ via the level matching condition
\begin{equation}\label{eq:levelmatching}
N -1 = \frac{1}{2}\left( Q_R^2 - Q_L^2 \right) = \frac{1}{2}Q^2 \ .
\end{equation}
Also recall that in the unorbifolded case the discriminant function $\Delta(\sigma)=\eta^{24}(\sigma)$ appears on the diagonal divisor limit of $\chi_{10}^{-1}(Z)$, which is the (complete $I=1$) quarter-BPS partition function of heterotic strings on $T^6$ (c. f. the discussion in \ref{sss:wallcrossing}). Historically the appearance of this perturbative half-BPS partition function and its magnetic counterpart, together with manifest electric-magnetic (S-)duality between them, was a crucial point in the proposal of~\cite{Dijkgraaf1997d}.

We return to the CHL orbifold and apply a similar logic to $B_4(q,\bar{q})$ in eq. \eqref{eq:B4_Z2chl_etas}, which we split into the untwisted and twisted sector contribution,
\begin{equation}
B_4 = B_4^{\mathrm{untw}} + B_4^{\mathrm{tw}} \ .
\end{equation}

\paragraph{Untwisted sector.} To read off the degeneracies of DH states with fixed electric charge, the Narain-lattice vectors  $(P_1, P_2)\in E_8(-1)^{\oplus 2}$ are decomposed\footnote{Also see~\cite{FORGACS1988477} for a relation to numerators of affine characters of $\hat{E_8}$ at level two.} with respect to their sum  --- which is invariant under $\IZ_2$ and hence a physical charge --- and their difference. That is,\label{cor:P1P2pm} 
\begin{equation}
P_1 \pm P_2 = 2 P_\pm  \pm \mathcal{P}
\end{equation}
for some root lattice vectors $P_+, P_- \in E_8(-1)$ and a shift vector $\mathcal{P} \in E_8(-1)/(2 E_8(-1))$. The latter represents an element of a finite group of rank $2^8$, which is by a simple rescaling by $1/\sqrt{2}$ isomorphic to the residue component from $E_8(-1/2)/E_8(-2)$ in eq. \eqref{eq:residueZ2}. In terms of $E_8(2)$ theta functions with characteristics $\mathcal{P}$, defined as
\begin{equation}\label{eq:defthetaE8P}
\theta_{E_8(2),\mathcal{P}}(\tau) \coloneqq  \sum_{\Delta \in E_8(1)} \exp\left[ \pi i \tau \left( \sqrt{2}\Delta - \frac{\mathcal{P}}{\sqrt{2}}\right)^2 \right] \ ,
\end{equation}
the theta function for $E_8(1)^{\oplus 2}$ may be expressed as
\begin{equation}\label{eq:thetaE8sq}
\theta_{E_8(1)}^2= \theta_{E_8(2),1}^2+ 120 \ \theta_{E_8(2),248}^2+ 135 \ \theta_{E_8(2),3875}^2 \ .
\end{equation}
Here it has been used that $\theta_{E_8(2),\mathcal{P}}$ only depends on the orbit $\mathcal{O}_*$ of $\mathcal{P}$ under the Weyl group of $E_8$. There are three such orbits, namely the orbit of the fundamental weight of the trivial, of the adjoint and of the 3875 representation of respective lengths $1+120+135=2^8$, i.e.,
\begin{equation}
\frac{E_8(-1)}{2E_8(-1)} = \mathcal{O}_1 \cup \mathcal{O}_{248} \cup \mathcal{O}_{3875} \ ,
\end{equation}
where the subscript labels the dimension of the respective representation. 
In general, any vector $Q'$ in $E_8(-1/2)=\tfrac{1}{\sqrt{2}}E_8(-1)$ decomposes as
\begin{equation}
Q'= \frac{1}{\sqrt{2}}(2Q'' + \mathcal{P})
\end{equation}
for appropriate elements $Q''\in E_8(-1)$ and $\mathcal{P} \in E_8(-1)/(2 E_8(-1))$, and therefore one also has
\begin{align}
E_8(-1/2)  \ &= \ E_8(-2) \cup \left( E_8(-2) + \mathcal{O}_{248} \right)\cup \left( E_8(-2) + \mathcal{O}_{3875} \right) \\
\theta_{E_8\left(\frac{1}{2}\right)}(\tau) \ &= \ \theta_{E_8(2),1} \, +\ \ \, 120\, \theta_{E_8(2),248}\   \  + \ \ 135 \, \theta_{E_8(2),3875} \ . \label{eq:thetaE8single}
\end{align} 
Both \eqref{eq:thetaE8sq} and \eqref{eq:thetaE8single} are easily checked by writing $\theta_{E_8(2),\mathcal{P}}(\tau)$ in terms of theta constants (see appendix \ref{app:siegelmodforms}). Note that under $\tau \mapsto \tau + 1$ only the sign of the term corresponding to the 248-orbit  in \eqref{eq:thetaE8single} flips, since $\mathcal{P}^2 \equiv 2 \, (\operatorname{mod} 4)$ for this orbit, while $\mathcal{P}^2 \equiv 0 \, (\operatorname{mod} 4)$ for the other two orbits. 

  The untwisted sector contribution reads in terms of the $\theta_{E_8(2),\mathcal{P}}(\tau)$ functions\footnote{This corrects a typo in~\cite[eq. (3.42)]{Dabholkar:2005dt}, where $\frac{2^4}{\theta_2^4 \eta^{12}} =\frac{\theta_3^4 \theta_4^4}{\eta^{24}}= \frac{1}{\eta^8(\tau)\eta^8(2\tau)}$ .}
\begin{align}\label{eq:B4_Z2chl_untwEps}
B_4^{\mathrm{untw}}(q, \bar{q})   = & \frac{3}{2\tau_2} \times \sum_{\epsilon\in \lbrace+1, -1\rbrace} \frac{\mathcal{Z}_{6,6}\begin{bsmallmatrix}
 0 \\
0 
\end{bsmallmatrix}+\epsilon\mathcal{Z}_{6,6}\begin{bsmallmatrix}
 0 \\
1
\end{bsmallmatrix}}{2}
\Bigg[\theta_{E_8(2),1}\times \frac{1}{2} \left( \frac{\theta_{E_8(2),1}}{ \eta^{24}} + \epsilon  \frac{1}{\eta^8(\tau)\eta^8(2\tau)} \right) \nonumber \\ &+ 120 \ \theta_{E_8(2),248} \times \left( \frac{\theta_{E_8(2),248}}{2 \eta^{24}}\right) + 135 \  \theta_{E_8(2),3875} \times \left( \frac{\theta_{E_8(2),3875}}{2 \eta^{24}}\right)\Bigg] \ .
\end{align} 
In this form $B_4^{\mathrm{untw}}$ corresponds to the non-orbifold counterpart \eqref{eq:B4unorb}, with the modular form on the right-hand side of each ``$\times$''-sign playing the role of $\eta^{-24}$. The $E_8$ theta series inside the parentheses sums only over the unphysical charge $(\tfrac{P_1-P_2}{2})^2$.
The sign $\epsilon$ corresponds to two kinds of DH states in the untwisted sector. It specifies the sign picked up by the oscillator monomial under $\IZ_2$ (c. f.~\cite[eq. (3.14)]{Dabholkar:2005dt}). This goes along with an even ($+1$) or odd ($-1)$ number of momentum quanta along the CHL circle, such that the two phases coming from the (left-moving) oscillators and the (left-moving) zero-mode cancel out to give an invariant state.
As can be seen, e.g., from the explicit form of the T-transformations on charges in~\cite{Sen:2007qy}, this ``momentum parity'' along the CHL circle is also invariant under T-transformations, so we have a splitting into two disjoint T-orbits.
Correspondingly, we find that untwisted sector $\mathcal{P}=0$ DH states with odd (even) momentum parity possess a separate half-BPS partition function
\begin{equation}\label{eq:electricHBPSeps}
\frac{1}{2} \left( \frac{\theta_{E_8(2)}}{ \eta^{24}} + \epsilon \frac{1}{ \eta^{8}(\tau)\eta^{8}(2\tau)} \right)
\end{equation}
with $\epsilon=-1$ ($\epsilon=+1$), as was implicitly used in writing down~\cite[eq. (6.5.12)]{Banerjee:2008pv}. For untwisted sector states with $\mathcal{P}\neq 0$ the parity of the CHL momentum does not play a role in the counting, as seen from eq. \eqref{eq:B4_Z2chl_untwEps}. 

To get the half-BPS index for states with fixed electric charge from these partition functions we reformulate the level matching condition \eqref{eq:levelmatching}, as the quantity $Q\in \Lambda_{22,6}$ in \eqref{eq:levelmatching} is no longer the physical electric charge in the orbifold theory. In the untwisted sector we introduce the modified level number 
\begin{equation}
N'  \coloneqq \ \ N - \frac{(P_1 - P_2)^2}{4}  \ \ = \ N - \left(P_- - \frac{\mathcal{P}}{2} \right)^2
\end{equation} 
and with a physical electric charge $Q \in E_8(-\tfrac{1}{2})\oplus U \oplus U^{\oplus 5}$ we find again $N'-1 = \tfrac{1}{2}Q^2$. Thus, when expanding eq. \eqref{eq:electricHBPSeps} in terms of $q=e^{2\pi i\tau}$, the exponent of $q$ in each term gives $\tfrac{Q^2}{2}$, while the coefficient gives the desired index $\Omega_4(Q,0)$ for $Q$ in the respective charge sector (ignoring the universal factor $3/2$), i.e., $Q\in E_8(-2)\oplus U(2) \oplus U^{\oplus 5}$ for $\epsilon=+1$ and $Q\in E_8(-2) \oplus (U\backslash U(2)) \oplus U^{\oplus 5}$ for $\epsilon=-1$ in the example of \eqref{eq:electricHBPSeps}. Here we have identified $U(2)\subset U$ as the (non-shifted) momentum-winding vectors with an even number of momentum quanta along the CHL circle.

\paragraph{Twisted sector.} The twisted sector part of $B_4$ is
\begin{align}\label{eq:B4_Z2chl_twEps}
B_4^{\mathrm{tw}}(q, \bar{q})   = & \, \frac{3}{2\tau_2} \times \! \sum_{\epsilon\in \lbrace+1, -1\rbrace} \frac{\mathcal{Z}_{6,6}\begin{bsmallmatrix}
 1 \\
0 
\end{bsmallmatrix}+\epsilon\mathcal{Z}_{6,6}\begin{bsmallmatrix}
 1 \\
1
\end{bsmallmatrix}}{2}
\Bigg[
\theta_{E_8(2),1}\times \frac{1}{2} \left( \frac{1}{ \eta^{8}(\tau)\eta^{8}(\tfrac{\tau}{2})} + \epsilon  \frac{e^{-2\pi i/3}}{\eta^{8}(\tau)\eta^{8}(\tfrac{\tau+1}{2})} \right) \nonumber \\ 
&\qquad + 120 \ \theta_{E_8(2),248} \times \left( \frac{1}{ \eta^{8}(\tau)\eta^{8}(\tfrac{\tau}{2})} - \epsilon  \frac{e^{-2\pi i/3}}{\eta^{8}(\tau)\eta^{8}(\tfrac{\tau+1}{2})} \right)\nonumber\\
&\qquad + 135 \  \theta_{E_8(2),3875} \times \left( \frac{1}{ \eta^{8}(\tau)\eta^{8}(\tfrac{\tau}{2})} + \epsilon  \frac{e^{-2\pi i/3}}{\eta^{8}(\tau)\eta^{8}(\tfrac{\tau+1}{2})} \right)
\Bigg] \ .
\end{align} 
Note that the relative sign between the two terms in each pair of parentheses is that of $(-1)^{Q^2}=(-1)^{\mathcal{P}^2/2}(-1)^{Q\cdot \delta}$ with $(-1)^{Q\cdot \delta}=\epsilon$. The twisted sector level-matching\footnote{See \cite[eq. (2.14)]{Mikhailov_1998} or \cite[section 3.3.]{Dabholkar:2005dt}.} equates the exponents in the $q$-expansion of the functions in parentheses in \eqref{eq:B4_Z2chl_twEps} to the value of $\frac{1}{2}(Q_8^2+ Q_8^2 + Q_1^2 + Q_5^2 )\in \tfrac{1}{2}\IZ$, where $(Q_8,Q_1,Q_5) \in E_8(-\tfrac{1}{2}) \oplus (U+\tfrac{\delta}{2}) \oplus U^{\oplus 5}$ is a physical electric charge vector in the twisted sector. The term $Q_8^2$ occurs twice, as the internal $E_8$ momenta in the twisted sector automatically satisfy $P_1=P_2$. With slight abuse of notation we write $Q=(Q_8,Q_8,Q_1,Q_5) \in E_8(-\tfrac{1}{2}) \oplus (U+\tfrac{\delta}{2}) \oplus U^{\oplus 5}$ such that $\frac{1}{2}(Q_8^2+ Q_8^2 + Q_1^2 + Q_5^2 )=\frac{Q^2}{2}$. This allows to treat the untwisted sector and twisted sector on an equal footing.

\paragraph{Comparing the sectors.} In accordance with the analysis of the perturbative spectrum in \cite{Mikhailov_1998}, the degeneracies for certain subsectors of the untwisted sector agree with twisted sector degeneracies. This is due to the modular identities
\begin{align}
\frac{1}{2}\left( \frac{\theta_{E_8(2),1}}{\eta^{24}} + \frac{1}{\eta^8(\tau)\eta^8(2\tau)} \right) &= \frac{1}{2}\left(\frac{1}{\eta^{8}(\tau)\eta^{8}(\frac{\tau}{2})}+  \frac{e^{-2\pi i/3}}{\eta^8(\tau)\eta^8(\frac{\tau + 1}{2})} \right) + \frac{1}{\eta^8(\tau)\eta^8(2\tau)}\label{eq:ModId0}\\
\frac{1}{2}\left(\frac{\theta_{E_8(2),1}}{\eta^{24}} - \frac{1}{\eta^8(\tau)\eta^8(2\tau)}\right) &= \frac{1}{2}\left(\frac{1}{\eta^{8}(\tau)\eta^{8}(\frac{\tau}{2})}+  \frac{e^{-2\pi i/3}}{\eta^8(\tau)\eta^8(\frac{\tau + 1}{2})}\right)\label{eq:ModId1}\\
\frac{\theta_{E_8(2),248}}{2\, \eta^{24}} &= \frac{1}{2}\left(\frac{1}{\eta^{8}(\tau)\eta^{8}(\frac{\tau}{2})}- \frac{e^{-2\pi i/3}}{\eta^8(\tau)\eta^8(\frac{\tau + 1}{2})}\right)\label{eq:ModId2}\\
\frac{\theta_{E_8(2),3875}}{2\, \eta^{24}}  &= \frac{1}{2}\left(\frac{1}{\eta^{8}(\tau)\eta^{8}(\frac{\tau}{2})}+  \frac{e^{-2\pi i/3}}{\eta^8(\tau)\eta^8(\frac{\tau + 1}{2})}\right) \ .\label{eq:ModId3}
\end{align}
Note that on the right-hand-side of eqs. \eqref{eq:ModId1} to \eqref{eq:ModId3} the second term is, up to sign, the first term shifted by $\tau \mapsto \tau +1$, which is
\begin{equation}\label{eq:etaetahalf}
\frac{1}{\eta^{8}(\tau)\eta^{8}(\frac{\tau}{2})}= \frac{1}{\sqrt{q}}+8+52 \sqrt{q}+256 q+1122 q^{3/2}+4352 q^2+15640 q^{5/2}+O\left(q^{3}\right) \ .
\end{equation}
Hence, adding the second term to the first projects to terms with even (eqs. \eqref{eq:ModId1} and \eqref{eq:ModId3}) or odd (eq. \eqref{eq:ModId2}) exponents of $\sqrt{q}=e^{2\pi i\tfrac{\tau}{2}}$. The parity of this exponent modulo two matches the parity of $Q^2/2$ and due to this one might simply regard  $\eta^{-8}(\tau)\eta^{-8}(\tfrac{\tau}{2})$ as the half-BPS partition function for twisted sector DH states --- and in fact as the half-BPS partition function for DH states with charge in any of the sectors listed in eqs. \eqref{eq:ModId1} to \eqref{eq:ModId3}. The only charge sector that is not covered by this is that of even momentum $\mathcal{P}=0$ untwisted states with $\epsilon=+1$ in \eqref{eq:electricHBPSeps}, i.e., electric charges $Q\in E_8(-2)\oplus U(2)\oplus U^{\oplus 5}=\Lambda_m \subset \Lambda_e$. Their degeneracy is not just given by the coefficient of $q^{Q^2/2}$ in $\eta^{-8}(\tau)\eta^{-8}(\tfrac{\tau}{2})$ but gets an extra contribution from the coefficient of $q^{Q^2/2}$ in $\eta^{-8}(\tau)\eta^{-8}(2\tau)$, as also observed in \cite{Bossard_2017}. Another way to arrive at the same conclusion is via the following identity.
Since the exchange of the two $E_8$ factors alone without the shift along a circle of the torus gives back an equivalent theory, there is an equality between the partition functions of the two theories~\cite[app. B]{Dabholkar_2007}:
\begin{equation}\label{eq:E8equivalence}
\frac{E_4(\tau)^2}{\eta^{16}(\tau)} = 
\frac{1}{2} \frac{E_4(\tau)^2}{\eta^{16}(\tau)} +
\frac{1}{2} \frac{E_4(2 \tau )}{\eta^{8}(2\tau)} +  
\frac{1}{2} \frac{E_4(\frac{\tau}{2})}{\eta^{8}(\frac{\tau}{2})} +
\frac{ e^{-2\pi i /3}}{2} \frac{E_4(\frac{\tau + 1}{2})}{\eta^{8}(\frac{\tau +1}{2})}  \ .
\end{equation}
Using this  \eqref{eq:B4_Z2chl_etas} can be re-expressed as
\begin{equation} \label{eq:DHsublatticesEM}
B_4(q,\bar{q}) = \frac{3}{2 \tau_2} \left[ \frac{\Gamma_{\Lambda_e^*}}{\eta^8(\tau)\eta^8(2\tau)} +  \frac{1}{2}\frac{\Gamma_{\Lambda_e}}{\eta^8(\tau)\eta^8(\tfrac{\tau}{2})} + \frac{1}{2} \frac{\Gamma_{\Lambda_e}[(-1)^{Q^2}]}{ e^{2\pi i /3}\eta^8(\tau)\eta^8(\tfrac{\tau+1}{2})}\right] \ ,
\end{equation}
where the notation~\cite{Bossard_2017}
\begin{equation}
\Gamma_{\Lambda_0}[\mathcal{X}] = \sum_{Q\in \Lambda_0} \mathcal{X}\  q^{\frac{1}{2}Q_L^2}\, \bar{q}^{\frac{1}{2}Q_R^2}
\end{equation}
was introduced. Pairs of (Narain) theta functions multiplying the same eta-quotient have been recasted into a single lattice sum for the electric lattice $\Lambda_e$ or magnetic lattice $\Lambda_e^* \subset \Lambda_e$, as defined in  \eqref{eq:LambdaEMZ2chl}. An equivalent representation is
\begin{equation} \label{eq:DHsublatticesEM2}
B_4(q,\bar{q}) = \frac{3}{2 \tau_2} \sum_{Q \in\Lambda_e} q^{\frac{Q^2}{2}} \left[ \frac{\delta_{Q\in \Lambda_e^*}}{\eta^8(\tau)\eta^8(2\tau)} +  \frac{1}{2}\frac{1}{\eta^8(\tau)\eta^8(\tfrac{\tau}{2})} + \frac{1}{2} \frac{(-1)^{Q^2}}{ e^{2\pi i /3}\eta^8(\tau)\eta^8(\tfrac{\tau+1}{2})} \right] \ ,
\end{equation}
where $(-1)^{Q^2}=(-1)^{\frac{\mathcal{P}^2}{2}}(-1)^{h\,Q\cdot \delta}$ with $h$ as in \eqref{eq:toruslattice}. This also nicely demonstrates the assertion that the DH states are electrically charged with respect to $\Lambda_e$ as given in \eqref{eq:LambdaEMZ2chl}. 
In section \ref{sec:genustwo} a genus two analog of \eqref{eq:E8equivalence} will become important.


\section{Quarter-BPS spectra from genus two partition function in the $\IZ_2$ model}
\label{sec:genustwo}

Our analysis in section \ref{sec:countingbpsN4} mostly concerned generic quarter-BPS partition functions. We now turn specifically to unit-torsion quarter-BPS dyons in the $\IZ_2$ CHL model, the prime interest being dyons whose electric charge in the heterotic frame belongs to the untwisted sector.
The goal of this section is to obtain closed expressions for the relevant partition functions by relating them to a genus two chiral partition function for the four-dimensional heterotic $\IZ_2$ CHL model.\footnote{Left- and right-moving partition function should be understood as in~\cite[f. n. 2]{Dabholkar_2007}.} Properties of the candidate dyon partition functions thus obtained will be addressed in section \ref{sec:sdwcconstraints}.

According to~\cite{gaiotto2005rerecounting,Dabholkar_2007,Dabholkar07Borcherds,Banerjee_2009g2} quarter-BPS dyons can be represented as string webs~\cite{Sen:1997xi,Aharony:1997bh}, which via an M-theory lift are related to a chiral genus two partition function of the heterotic string. As was argued in~\cite{Dabholkar:2007vk}, the genus $g$ of the M-theory lift of the string web is actually given by $g=I+1$, so the genus two partition function is expected to only capture unit-torsion dyons ($I=1$). Indeed, in~\cite{Dabholkar_2007} the \textit{twisted} sector dyon partition function of~\cite{Jatkar:2005bh,David:2006ji} was re-derived by identifying appropriate contributions to the genus two orbifold partition function that can be interpreted as arising from states of the relevant charge type.\footnote{The contour prescription and wall-crossing phenomenon can also be studied in the genus two picture~\cite{Banerjee_2009g2,Cheng:2009hm}, though the analysis was mostly spelled out for the maximal rank theory.} Our \textit{untwisted} sector quarter-BPS partition functions should in a similar fashion be found in this heterotic genus two partition function. The latter was recently revisited in~\cite[section B.2]{Bossard_2019}, expanding the results of~\cite{Dabholkar_2007} by, for instance, also writing down the remaining orbifold blocks. For the sake of a clear and coherent presentation, we will reproduce parts of~\cite{Bossard_2019} and collect the relevant formulae that are needed in the subsequent analysis.

\paragraph{Genus two orbifold blocks.} As in the one-loop case, the chiral partition function is given by a sum of orbifold blocks, each associated to a choice of periodicity conditions $[h_1, h_2]$ and $[g_1, g_2]$ along the A- and B-cycles of a genus two surface with period matrix $\Omega = \begin{psmallmatrix}
\tau & z \\ z & \sigma
\end{psmallmatrix} = \begin{psmallmatrix}
\Omega_{11} & \Omega_{12} \\ \Omega_{21} & \Omega_{22}
\end{psmallmatrix} = \Omega_1 + i \Omega_2 \in \mathbb{H}_2$, i.e.,
\begin{equation}\label{eq:ZgenustwoALL}
\mathcal{Z}(\Omega) = \frac{1}{2^2} \sum_{\substack{h_1, h_2 \in \lbrace 0,1 \rbrace \\ g_1, g_2 \in \lbrace 0,1 \rbrace}} \mathcal{Z}\begin{bsmallmatrix}
h_1 & h_2 \\ g_1 & g_2
\end{bsmallmatrix} \ .
\end{equation}
At least on the locus of the moduli space where the Narain-lattice splits as $E_8 \oplus E_8 \oplus \Lambda_{6,6}$ we may factorize the orbifold blocks into a contribution of the ten-dimensional $E_8 \times E_8$ string and the contribution of the bosonic zero-modes of the chiral bosons on $T^6$,
\begin{equation}
\mathcal{Z}\begin{bsmallmatrix}
h_1 & h_2 \\ g_1 & g_2
\end{bsmallmatrix} = \mathcal{Z}_8 \begin{bsmallmatrix}
h_1 & h_2 \\ g_1 & g_2
\end{bsmallmatrix} \ \mathcal{Z}_{6,6}\begin{bsmallmatrix}
h_1 & h_2 \\ g_1 & g_2
\end{bsmallmatrix}  \ .
\end{equation}
Here we have 
\begin{equation}\label{eq:torusblockdef}
 \mathcal{Z}_{6,6}\begin{bsmallmatrix}
h_1 & h_2 \\ g_1 & g_2
\end{bsmallmatrix} =\sum_{(Q_1,Q_2)\in \Lambda^{[h_1,h_2]}_{6,6}} (-1)^{\delta \cdot (g_1 Q_1 +g_2 Q_2)} \ e^{i\pi Q_{L}^r \Omega_{rs} Q_{L}^s - i \pi Q_{R}^r \bar{\Omega}_{rs} Q_{R}^s}
\end{equation}
with summation over  $r,s \in \lbrace 0 , 1 \rbrace$ here and in the following (no distinction between upper and lower indices made). Let us abbreviate the exponential by $e_{Q_1,Q_2}(\Omega)$. Also we have
\begin{equation}\label{eq:toruslatticeh1h2}
\Lambda^{[h_1,h_2]}_{6,6} = \left( \Lambda_{6,6} + \frac{h_1}{2} \delta \right)\oplus \left( \Lambda_{6,6} +  \frac{h_2}{2} \delta \right) \ ,
\end{equation}
the genus two analog of the Narain-lattice associated with $T^6$, shifted by half of the null vector $\delta = (0^6 \, ; \, 0^{6-1} , 1)$.\footnote{\label{fn:CHLcircleboson}We interpret $\delta \cdot  Q_i$ as the momentum of the ``CHL circle boson'' flowing along the $i$-th B-cycle of the genus two worldsheet. This should correct a typo below~\cite[eq. (B.52)]{Bossard_2019} (there: ``winding'' instead of ``momentum'') and restore consistency with~\cite[section A.1]{Bossard_2017}. Also note that we have dropped a factor of $(\det \Omega_2 )^{6/2}$ in $ \mathcal{Z}_{6,6}\begin{bsmallmatrix}
h_1 & h_2 \\ g_1 & g_2
\end{bsmallmatrix}$, which will not be relevant in our discussion.}

In view of the twisted sector dyon states, the authors of~\cite{Dabholkar_2007} computed the orbifold block
\begin{equation}\label{eq:Z80001}
\mathcal{Z}_8\begin{bsmallmatrix}
0 & 0 \\ 0 & 1
\end{bsmallmatrix}\left( \begin{psmallmatrix}
\tau & z \\ z & \sigma
\end{psmallmatrix} \right) = 
\frac{\Theta_{E_8}^{(2)}(2\tau, 2z, 2\sigma)}{\Phi_{6,0}} + \frac{\Theta_{E_8}^{(2)}(\frac{\tau}{2}, z, 2\sigma)}{16\, \Phi_{6,1}} +
\frac{\Theta_{E_8}^{(2)}(\frac{\tau+1}{2}, z, 2\sigma)}{16\, \Phi_{6,2}} 
\end{equation}
building on the results of~\cite{dijkgraaf1988}.
Here we have the genus two theta series for the $E_8$ root lattice,
\begin{equation}\label{eq:ThetaE8gen2def}
\Theta_{E_8}^{(2)}(\Omega) = \sum_{(Q_1,Q_2)\in {E_8}\oplus {E_8}} \ e^{i \pi \, Q^r \Omega_{rs} Q^s} = E_4^{(2)}(\Omega) \, ,
\end{equation}
agreeing with the Siegel-Eisenstein series $ E_4^{(2)}(\Omega)$, as well as the weight six Siegel modular forms $\Phi_{6,k}$ defined in appendix \ref{app:siegelmodforms} (one of which is given by a multiplicative lift of the K3 twining genera of class $2A$).
Rescalings and shifts in the arguments of the $E_8$ theta series can be rewritten in terms of the theta series for 2-modular lattices and insertions of sign factors, for instance:\footnote{Later we need theta series related to the ones in \eqref{eq:Z80001} by an exchange in the roles of $(\tau, Q_1), (\sigma, Q_2)$, see eqs. \eqref{eq:ThetaE8halfdouble} and \eqref{eq:ThetaE8halfdoubleB}.\label{fn:corB9}}
\begin{align}
\Theta_{E_8}^{(2)}(2\tau,2z,2\sigma)=& \ \quad  \ \ \sum_{\substack{(Q_1,Q_2)\in\\ E_8(2)\oplus E_8(2)}}e^{i\pi Q^r \Omega_{rs} Q^s} = E_4^{(2)}(2\Omega) 
\label{eq:ThetaE82Omega} \\
\Theta_{E_8}^{(2)}(\tfrac{\tau}{2},z, 2\sigma)=& \ 2^{-4}\sum_{\substack{(Q_1,Q_2)\in\\ E_8(2)^* \oplus E_8(2)}}e^{i\pi Q^r \Omega_{rs} Q^s}\label{eq:ThetaE8doublenoshift}
\\
\Theta_{E_8}^{(2)}(\tfrac{\tau +1}{2},z,2\sigma)=&\ 2^{-4}\sum_{\substack{(Q_1,Q_2)\in\\ E_8(2)^*\oplus E_8(2)}}(-1)^{Q_1^2}\,e^{i\pi Q^r \Omega_{rs} Q^s} \ . \label{eq:ThetaE8doubleshift}
\end{align}
Further expressions of this kind we will encounter below are moved to appendix \ref{app:siegelmodforms}.
The third and second term in \eqref{eq:Z80001} turn out to be modular images of the first under the Petersson slash operator,
\begin{equation}\label{eq:slashorbiblocks}
\mathcal{Z}_8\begin{bsmallmatrix}
0 & 0 \\ 0 & 1
\end{bsmallmatrix} = \sum_{\gamma \in \Gamma^{(2)}_{e_1}(2) / \Gamma^{(2)}_{0,e_1}(2)}
\left( \frac{\Theta_{E_8}^{(2)}(2\tau, 2z, 2\sigma)}{\Phi_{6,0}} \right) \Bigg|_{\gamma} \ ,
\end{equation}
see \eqref{eq:slashmatrices1} and \eqref{eq:slashmatrices2} for explicit $\gamma$. Per definition $\Gamma^{(2)}_{e_1}(2) \subset \Sp_4(\IZ)$ is the index 15 subgroup that preserves the periodicity conditions (characteristics) $\begin{bsmallmatrix}
0 & 0 \\ 0 & 1
\end{bsmallmatrix}$ modulo 2, while the group $\Gamma^{(2)}_{0,e_1}(2)$ is its intersection with the level two congruence subgroup $\Gamma_{0}^{(2)}(2) \subset \Sp_4(\IZ)$. This intersection has index 3 in $\Gamma^{(2)}_{e_1}(2)$, the three cosets correspond to three terms in eq. \eqref{eq:slashorbiblocks} or eq. \eqref{eq:Z80001}. Since $E_4^{(2)}(2\Omega)$ and $ \Phi_{6,0}(\Omega)$ are Siegel modular forms for $\Gamma_0^{(2)}(2)$ (they are invariant under $(\cdot)|_\gamma$ for $\gamma \in \Gamma_0^{(2)}(2)$), the other summands in \eqref{eq:slashorbiblocks} are Siegel modular forms with respect to subgroups conjugate to $\Gamma_0^{(2)}(2)$. From \eqref{eq:slashorbiblocks} it is clear that $\mathcal{Z}_8\begin{bsmallmatrix}
0 & 0 \\ 0 & 1
\end{bsmallmatrix}$ is indeed invariant under the group $\Gamma^{(2)}_{e_1}(2) \subset \Sp_4(\IZ)$.
An analogous formula also holds when the torus contribution is taken into account,
\begin{equation}\label{eq:Z8Z66block}
\mathcal{Z}_8\begin{bsmallmatrix}
0 & 0 \\ 0 & 1
\end{bsmallmatrix}\mathcal{Z}_{6,6}\begin{bsmallmatrix}
0 & 0 \\ 0 & 1
\end{bsmallmatrix} = \sum_{\gamma \in \Gamma^{(2)}_{e_1}(2) / \Gamma^{(2)}_{0,e_1}(2)}
\left( \frac{\Gamma^{(2)}_{U^{\oplus 6} \oplus E_{8}(2)}[(-1)^{\delta\cdot Q_2}]}{\Phi_{6,0}} \right) \Bigg|_{\gamma} \ ,
\end{equation}
where we adopted the notation
\begin{equation}
\Gamma^{(2)}_{\Lambda_0}[\mathcal{X}] = \sum_{(Q_1, Q_2)\in(\Lambda_0)^{\oplus 2}} \mathcal{X} \, e^{i\pi Q_{r,L} \Omega_{rs} Q_{s,L} - i \pi Q_{r,R} \bar{\Omega}_{rs} Q_{s,R}} \ 
\end{equation}
for the case  $\Lambda_0 = U^{\oplus 6} \oplus E_{8}(2), \ \mathcal{X}=(-1)^{\delta \cdot Q_2}$ (the $E_8$ charges being only ``left-moving'', as consistent with \eqref{eq:ThetaE8gen2def}). 

Further modular transformations on the above block \eqref{eq:Z8Z66block} with $\tilde{\gamma} \in \Sp_4(\IZ)/ \Gamma^{(2)}_{e_1}(2) $ generate the remaining 14 of the $2^4-1 =15$ orbifold blocks with non-trivial boundary conditions.
The respective part from $\mathcal{Z}_8\begin{bsmallmatrix}
h_1 & h_2 \\ g_1 & g_2
\end{bsmallmatrix}$ is displayed in Table \ref{tab:g2orbblocks} for convenience. The orbifold block $\mathcal{Z}\begin{bsmallmatrix}
0 & 0 \\ 0 & 0
\end{bsmallmatrix}$ forms a separate orbit, which is the genus two chiral partition function of the parent model, the (left-moving) heterotic string on $T^6$ and the same holds for $\mathcal{Z}\begin{bsmallmatrix}
0 & 0 \\ 0 & 0
\end{bsmallmatrix}$ in eq. \eqref{eq:E8equivgenus2} below.\label{fn:singletorbit}

\begin{table}
\vspace*{-0.3cm}
$$
\begin{array}{|c|c|c|}
\hline
\ar{h_1 h_2}{g_1 g_2} & \mathcal{Z}_8 \ar{h_1 h_2}{g_1 g_2}  & \tilde{\gamma}\in \Sp_4(\IZ)/\Gamma^{(2)}_{e_1}(2) \\
\hline
\ar{00}{01} & \frac{\Theta^{(2)}_{E_8}(2\tau,2\sigma,2z)}{\Phi_{6,0}}
+\frac{\Theta^{(2)}_{E_8}(\frac{\tau}{2},\sigma,z)}{2^4\Phi_{6,1}}
+\frac{\Theta^{(2)}_{E_8}(\frac{\tau+1}{2},\sigma,z)}{2^4\Phi_{6,2}}
&{\scriptsize
\left(
\begin{array}{cccc}
 1 & 0 & 0 & 0 \\
 0 & 1 & 0 & 0 \\
 0 & 0 & 1 & 0 \\
 0 & 0 & 0 & 1 \\
\end{array}
\right)}
\\
\hline
\ar{00}{10} & \frac{\Theta^{(2)}_{E_8}(2\tau,2\sigma,2z)}{\Phi_{6,0}}
+\frac{\Theta^{(2)}_{E_8}(2\tau,\frac{\sigma}{2},z)}{2^4\Phi_{6,3}}
+\frac{\Theta^{(2)}_{E_8}(2\tau,\frac{\sigma+1}{2},z)}{2^4\Phi_{6,4}}
&{\scriptsize
\left(
\begin{array}{cccc}
 0 & 1 & 0 & 0 \\
 1 & 0 & 0 & 0 \\
 0 & 0 & 0 & 1 \\
 0 & 0 & 1 & 0 \\
\end{array}
\right)}
\\
\hline
\ar{01}{00} &\frac{\Theta^{(2)}_{E_8}(2\tau,\frac{\sigma}{2},z)}{2^4\Phi_{6,3}}
+\frac{\Theta^{(2)}_{E_8}(\frac{\tau}{2},\frac{\sigma}{2},\frac{z}{2})}{2^8\Phi_{6,5}}
+\frac{\Theta^{(2)}_{E_8}(\frac{\tau+1}{2},\frac{\sigma}{2},\frac{z}{2})}{2^8\Phi_{6,6}}
&{\scriptsize
\left(
\begin{array}{cccc}
 1 & 0 & 0 & 0 \\
 0 & 0 & 0 & -1 \\
 0 & 0 & 1 & 0 \\
 0 & 1 & 0 & 0 \\
\end{array}
\right)}
\\
\hline
\ar{10}{00}& \frac{\Theta^{(2)}_{E_8}(\frac{\tau}{2},2\sigma,z)}{2^4\Phi_{6,1}}
+\frac{\Theta^{(2)}_{E_8}(\frac{\tau}{2},\frac{\sigma}{2},\frac{z}{2})}{2^8\Phi_{6,5}}
+\frac{\Theta^{(2)}_{E_8}(\frac{\tau}{2},\frac{\sigma+1}{2},\frac{z}{2})}{2^8\Phi_{6,7}}
&{\scriptsize
\left(
\begin{array}{cccc}
 0 & 0 & 0 & -1 \\
 1 & 0 & 0 & 0 \\
 0 & 1 & 0 & 0 \\
 0 & 0 & 1 & 0 \\
\end{array}
\right)}
\\
\hline
\ar{11}{00}& 
\frac{\Theta^{(2)}_{E_8}(\frac{\tau}{2},\frac{\sigma}{2},\frac{z}{2})}{2^8\Phi_{6,5}}\
+\frac{\Theta^{(2)}_{E_8}(\frac{\tau+1}{2},\frac{\sigma+1}{2},\frac{z+1}{2})}{2^8\Phi_{6,9}}
+\frac{\Theta^{(2)}_{E_8}(2\tau,\frac{\sigma-2z+\tau}{2},z-\tau)}{2^4\Phi_{6,13}}
&{\scriptsize
\left(
\begin{array}{cccc}
 1 & 0 & 0 & 0 \\
 -1 & 0 & 0 & -1 \\
 0 & 1 & 1 & 0 \\
 0 & 1 & 0 & 0 \\
\end{array}
\right)}
\\
\hline
\ar{01}{01}& 
\frac{\Theta^{(2)}_{E_8}(2\tau,\frac{\sigma+1}{2},z)}{2^4\Phi_{6,4}}
+\frac{\Theta^{(2)}_{E_8}(\frac{\tau}{2},\frac{\sigma+1}{2},\frac{z}{2})}{2^8\Phi_{6,7}}\
+\frac{\Theta^{(2)}_{E_8}(\frac{\tau+1}{2},\frac{\sigma+1}{2},\frac{z}{2})}{2^8\Phi_{6,8}}
&{\scriptsize
\left(
\begin{array}{cccc}
 1 & 0 & 0 & 0 \\
 0 & 1 & 0 & -1 \\
 0 & 0 & 1 & 0 \\
 0 & 1 & 0 & 0 \\
\end{array}
\right)}
\\
\hline
\ar{10}{10}& 
\frac{\Theta^{(2)}_{E_8}(\frac{\tau+1}{2},2\sigma,z)}{2^4\Phi_{6,2}}
+\frac{\Theta^{(2)}_{E_8}(\frac{\tau+1}{2},\frac{\sigma}{2},\frac{z}{2})}{2^8\Phi_{6,6}}\
+\frac{\Theta^{(2)}_{E_8}(\frac{\tau+1}{2},\frac{\sigma+1}{2},\frac{z}{2})}{2^8\Phi_{6,8}}
&{\scriptsize
\left(
\begin{array}{cccc}
 0 & 1 & 0 & -1 \\
 1 & 0 & 0 & 0 \\
 0 & 1 & 0 & 0 \\
 0 & 0 & 1 & 0 \\
\end{array}
\right)}
\\
\hline
\ar{01}{1 0}&
\frac{\Theta^{(2)}_{E_8}(2\tau,\frac{\sigma}{2},z)}{2^4\Phi_{6,3}}
+\frac{\Theta^{(2)}_{E_8}(\frac{\tau}{2},\frac{\sigma}{2},\frac{z+1}{2})}{2^8\Phi_{6,10}}
+\frac{\Theta^{(2)}_{E_8}(\frac{\tau+1}{2},\frac{\sigma}{2},\frac{z+1}{2})}{2^8\Phi_{6,11}}
&{\scriptsize
\left(
\begin{array}{cccc}
 0 & 1 & 0 & 0 \\
 0 & 0 & -1 & 0 \\
 0 & 0 & 1 & 1 \\
 1 & -1 & 0 & 0 \\
\end{array}
\right)}
 \\
 \hline
\ar{10}{11}& 
\frac{\Theta^{(2)}_{E_8}(\frac{\tau+1}{2},2\sigma,z)}{2^4\Phi_{6,2}}
+ \frac{\Theta^{(2)}_{E_8}(\frac{\tau+1}{2},\frac{\sigma+1}{2},\frac{z+1}{2})}{2^8\Phi_{6,9}}
+\frac{\Theta^{(2)}_{E_8}(\frac{\tau+1}{2},\frac{\sigma}{2},\frac{z+1}{2})}{2^8\Phi_{6,11}}
&{\scriptsize
\left(
\begin{array}{cccc}
 0 & 1 & -1 & -1 \\
 1 & -1 & 0 & 0 \\
 0 & 1 & 0 & 0 \\
 0 & 0 & 1 & 0 \\
\end{array}
\right)}
 \\
 \hline
\ar{10}{01}& 
\frac{\Theta^{(2)}_{E_8}(\frac{\tau}{2},2\sigma,z)}{2^4\Phi_{6,1}}
+\frac{\Theta^{(2)}_{E_8}(\frac{\tau}{2},\frac{\sigma}{2},\frac{z+1}{2})}{2^8\Phi_{6,10}}
+\frac{\Theta^{(2)}_{E_8}(\frac{\tau}{2},\frac{\sigma+1}{2},\frac{z+1}{2})}{2^8\Phi_{6,12}}
&{\scriptsize\left(
\begin{array}{cccc}
 0 & 0 & -1 & -1 \\
 1 & -1 & 0 & 0 \\
 0 & 1 & 0 & 0 \\
 0 & 0 & 1 & 0 \\
\end{array}
\right)}
\\
\hline
\ar{01}{11}&
\frac{\Theta^{(2)}_{E_8}(2\tau,\frac{\sigma+1}{2},z)}{2^4\Phi_{6,4}}
+\frac{\Theta^{(2)}_{E_8}(\frac{\tau+1}{2},\frac{\sigma+1}{2},\frac{z+1}{2})}{2^8\Phi_{6,9}}
+\frac{\Theta^{(2)}_{E_8}(\frac{\tau}{2},\frac{\sigma+1}{2},\frac{z+1}{2})}{2^8\Phi_{6,12}}
&{\scriptsize\left(
\begin{array}{cccc}
 0 & 1 & 0 & 0 \\
 1 & -1 & -1 & 0 \\
 0 & 0 & 1 & 1 \\
 1 & -1 & 0 & 0 \\
\end{array}
\right)}
\\
\hline
\ar{00}{11}& 
\frac{\Theta^{(2)}_{E_8}(2\tau,2\sigma,2z)}{\Phi_{6,0}}
+\frac{\Theta^{(2)}_{E_8}(2\tau,\frac{\tau-2z+\sigma}{2},z-\tau)}{2^4\Phi_{6,13}}
+\frac{\Theta^{(2)}_{E_8}(2\tau,\frac{\tau-2z+\sigma+1}{2},z-\tau)}{2^4\Phi_{6,14}}
&{\scriptsize\left(
\begin{array}{cccc}
 0 & 1 & 0 & 0 \\
 1 & -1 & 0 & 0 \\
 0 & 0 & 1 & 1 \\
 0 & 0 & 1 & 0 \\
\end{array}
\right)}
\\
\hline
\ar{11}{01}&
\frac{\Theta^{(2)}_{E_8}(\frac{\tau}{2},\frac{\sigma+1}{2},\frac{z}{2})}{2^8\Phi_{6,7}}
+\frac{\Theta^{(2)}_{E_8}(\frac{\tau+1}{2},\frac{\sigma}{2},\frac{z+1}{2})}{2^8\Phi_{6,11}}
+\frac{\Theta^{(2)}_{E_8}(2\tau,\frac{\tau-2z+\sigma+1}{2},z-\tau)}{2^4\Phi_{6,14}}
&{\scriptsize\left(
\begin{array}{cccc}
 1 & 0 & 0 & 0 \\
 -1 & 1 & 0 & -1 \\
 0 & 1 & 1 & 0 \\
 0 & 1 & 0 & 0 \\
\end{array}
\right)}
\\
\hline
\ar{11}{10}&
\frac{\Theta^{(2)}_{E_8}(\frac{\tau+1}{2},\frac{\sigma}{2},\frac{z}{2})}{2^8\Phi_{6,6}}
+\frac{\Theta^{(2)}_{E_8}(\frac{\tau}{2},\frac{\sigma+1}{2},\frac{z+1}{2})}{2^8\Phi_{6,12}}
+\frac{\Theta^{(2)}_{E_8}(2\tau,\frac{\tau-2z+\sigma+1}{2},z-\tau)}{2^4\Phi_{6,14}}
&{\scriptsize\left(
\begin{array}{cccc}
 1 & 1 & 1 & 0 \\
 -1 & 0 & 0 & -1 \\
 0 & 1 & 1 & 0 \\
 0 & 1 & 0 & 0 \\
\end{array}
\right)}
\\
\hline
\ar{11}{11}&
\frac{\Theta^{(2)}_{E_8}(\frac{\tau+1}{2},\frac{\sigma+1}{2},\frac{z}{2})}{2^8\Phi_{6,8}}
+\frac{\Theta^{(2)}_{E_8}(\frac{\tau}{2},\frac{\sigma}{2},\frac{z+1}{2})}{2^8\Phi_{6,10}}
+\frac{\Theta^{(2)}_{E_8}(2\tau,\frac{\sigma-2z+\tau}{2},z-\tau)}{2^4\Phi_{6,13}}
&{\scriptsize\left(
\begin{array}{cccc}
 1 & 1 & 1 & 0 \\
 -1 & 1 & 0 & -1 \\
 0 & 1 & 1 & 0 \\
 0 & 1 & 0 & 0 \\
\end{array}
\right)}
\\
\hline
\end{array}
$$
\caption{Chiral genus two orbifold blocks for the heterotic $\IZ_2$ CHL model (taken from \cite{Bossard_2019}).}\label{tab:g2orbblocks}
\end{table}

As in the one-loop partition function there is a modular identity arising from the equivalence of the $E_8 \times E_8$ theory with its orbifold obtained by exchange of the $E_8$ factors (without any shift along $T^6$):
\begin{equation}\label{eq:E8equivgenus2}
\mathcal{Z}_8\begin{bsmallmatrix}
0 & 0 \\ 0 & 0 
\end{bsmallmatrix} = \frac{\left[ \Theta_{E_8}^{(2)}(\Omega)\right]^2}{\chi_{10}}= \sum_{\substack{h_1, h_2 \in \lbrace 0,1 \rbrace \\ g_1, g_2 \in \lbrace 0,1 \rbrace}}' \mathcal{Z}_8 \begin{bsmallmatrix}
h_1 & h_2 \\ g_1 & g_2
\end{bsmallmatrix} \ .
\end{equation}
This is the genus two analog of \eqref{eq:E8equivalence}.

Using the behaviour of $E_4^{(2)}(\Omega)$ and the genus two Thetanullwerte $\theta_{a_1 a_2 b_1 b_2}(\Omega )$ (which appear in $\Phi_{6,k}$) in the dia\-gonal limit $z \rightarrow 0$ together with some simple theta identities (see appendix \ref{app:siegelmodforms}), it is straightforward to verify that each orbifold block factorizes into two genus one orbifold blocks:
\begin{equation}\label{eq:Z8diagonalfactors}
\mathcal{Z}_8\begin{bsmallmatrix}
h_1 & h_2 \\ g_1 & g_2
\end{bsmallmatrix} \stackrel{z \rightarrow 0}{\longrightarrow} -\frac{1}{4 \pi z^2} \, \mathcal{Z}_8\begin{bsmallmatrix}
h_1 \\ g_1 
\end{bsmallmatrix}(\tau) \, \mathcal{Z}_8\begin{bsmallmatrix}
h_2 \\ g_2 
\end{bsmallmatrix}(\sigma) + \mathcal{O}(z^0) \ .
\end{equation}
This limiting behaviour mirrors the wall-crossing constraints of quarter-BPS partition functions.

\paragraph{Identification of quarter-BPS partition functions.}

In the following we will identify the genus two period matrix $\Omega \in \mathbb{H}_2$ with the chemical potentials conjugate to the quadratic T-duality invariants obtained from the electric and magnetic components of a dyonic charge, 
\begin{equation}\label{eq:OmegaId}
 \Omega\stackrel{!}{=} Z  \quad \left( = \begin{psmallmatrix}
\tau & z \\ z & \sigma
\end{psmallmatrix} \right) \ .
\end{equation}
This means $\tau$ is conjugate to the magnetic charge $\tfrac{1}{2}P^2$, whereas $\sigma$ is conjugate to the electric charge $\tfrac{1}{2}Q^2$.%
%
\footnote{The roles of the chemical potentials $\tau, \sigma$ on the diagonal of the $2\times 2$ period matrix are switched with respect to~\cite[p. 8]{Dabholkar_2007}.
As a remark, switching the diagonal entries of a period matrix corresponds to the action of the symplectic matrix \eqref{eq:SympEx2}, $U=\begin{psmallmatrix}
0 & 1 \\ 1 & 0
\end{psmallmatrix}$, switching the periodicity conditions along the pairs of cycles $(A_1, B_1)$, $(A_2, B_2)$, i.e., $\begin{psmallmatrix}
h_1 & h_2 \\ g_1 & g_2
\end{psmallmatrix} \mapsto \begin{psmallmatrix}
h_2 & h_1 \\ g_2 & g_1
\end{psmallmatrix}$.
However, in the sequel paper~\cite{Dabholkar:2007vk} the authors also use the convention employed here.}
%
 It has important consequences for finding the contributions in $\mathcal{Z}$ that can be interpreted as arising from appropriate dyonic charges $(Q,P)=(Q_2,Q_1)$ in the lattice sums. The most convenient way to write $\mathcal{Z}$ for the following discussion is
 \begin{equation}
 \mathcal{Z} = \frac{1}{2^2} \sum_{\substack{h_1, h_2 \in \lbrace 0,1 \rbrace \\ g_1, g_2 \in \lbrace 0,1 \rbrace}}' \mathcal{Z}_8 \begin{bsmallmatrix}
h_1 & h_2 \\ g_1 & g_2
\end{bsmallmatrix} \ \left( \mathcal{Z}_{6,6}\begin{bsmallmatrix}
0 & 0\\ 0 & 0
\end{bsmallmatrix} + \mathcal{Z}_{6,6}\begin{bsmallmatrix}
h_1 & h_2\\ g_1 & g_2
\end{bsmallmatrix} \right) \ .
 \end{equation}
We first address the toroidal part $\mathcal{Z}_{6,6}\begin{bsmallmatrix}
h_1 & h_2\\ g_1 & g_2
\end{bsmallmatrix} $ and recall \eqref{eq:toruslatticeh1h2} and  \eqref{eq:LambdaEMZ2chl}, finding that the summation over $Q_1 = P$ in the lattice sums must go over the non-shifted lattice for the interpretation as a magnetic charge being possible, i.e., we must consider terms with $h_1=0$. Also $P \in \Lambda_m$ has components only along the sublattice $U(2)\subset U$. This further restriction will naturally be satisfied for terms in the $\mathcal{Z}_8 \begin{bsmallmatrix}
h_1 & h_2 \\ g_1 & g_2
\end{bsmallmatrix}$ blocks that appear both for $g_1=0$ and $g_1=1$. The reason is that this effectively means the presence of the desired projector $\tfrac{1}{2}(1+(-1)^{Q_1\cdot \delta})$ to $U(2)\subset U$ in the toroidal lattice sum. For untwisted sector charges $Q\in E_8(-1) \oplus U \oplus U^{\oplus 5}$, i.e. $h_2=0$, such terms can only arise for
\begin{equation}\label{eq:charUntwisted}
\begin{bsmallmatrix}
h_1 & h_2 \\ g_1 & g_2
\end{bsmallmatrix} \in \Big\{ \,
\begin{bsmallmatrix}
0 & 0\\ 1 & 0
\end{bsmallmatrix} , 
\begin{bsmallmatrix}
0 & 0\\ 1 & 1
\end{bsmallmatrix},
 \begin{bsmallmatrix}
0 & 0\\ 0 & 1
\end{bsmallmatrix} \, \Big\} \ ,
\end{equation}
while for twisted sector charges $Q\in E_8(-1) \oplus (U+\frac{\delta}{2}) \oplus U^{\oplus 5}$, i.e., $h_2=1$, the analogous statement is
\begin{equation}\label{eq:chartwisted}
\begin{bsmallmatrix}
h_1 & h_2 \\ g_1 & g_2
\end{bsmallmatrix} \in \Big\{ \,
\begin{bsmallmatrix}
0 & 1\\ 1 & 0
\end{bsmallmatrix} , 
\begin{bsmallmatrix}
0 & 1\\ 1 & 1
\end{bsmallmatrix},
 \begin{bsmallmatrix}
0 & 1\\ 0 & 1
\end{bsmallmatrix},
 \begin{bsmallmatrix}
0 & 1\\ 0 & 0
\end{bsmallmatrix} \, \Big\} \ .
\end{equation}
Note that due to the replacement \eqref{eq:E8equivgenus2} the characteristic $ \begin{bsmallmatrix}
0 & 0\\ 0 & 0
\end{bsmallmatrix}$ is not listed in \eqref{eq:charUntwisted}. Inspecting Table \ref{tab:g2orbblocks}, terms in the above blocks that appear for both $g_1$ cases are the ones with denominators
\begin{equation}
\Phi_{6,0}, \, \Phi_{6,3}, \, \Phi_{6,4} \quad \text{and} \quad  \Phi_{6,3}, \, \Phi_{6,4} \ ,
\end{equation}
respectively. Collecting these and writing out the sum over the torus lattice gives for the untwisted case ($h_2 = 0$)
\begin{align}\label{eq:g2untwTerms}
 \frac{1}{2^2}  \sum_{\substack{Q_1 \in  U^{\oplus 6} \\ Q_2  \in U^{\oplus 6} }}
 \quad \    & e_{Q_1,Q_2}(\Omega ) \, \Bigg[  \frac{\Theta_{E_8}^{(2)}(2\tau, 2z, 2\sigma)}{\Phi_{6,0}}  (1+(-1)^{\delta\cdot Q_1}+(-1)^{\delta \cdot Q_2}+(-1)^{\delta \cdot Q_1 + \delta\cdot Q_2}) \nonumber \\
   + &
 \frac{\Theta_{E_8}^{(2)}(2\tau, 2z, \frac{\sigma}{2})}{16\, \Phi_{6,3}}  (1+(-1)^{\delta\cdot Q_1})  +  
 \frac{\Theta_{E_8}^{(2)}(2\tau, 2z, \frac{\sigma + 1}{2})}{16\, \Phi_{6,4}}  (1+(-1)^{\delta\cdot Q_1})     
 \Bigg] 
\end{align}
and
\begin{align}\label{eq:g2twTerms}
 \frac{1}{2^2} \sum_{\substack{Q_1 \in  U^{\oplus 6} \\ Q_2  \in (U+\frac{\delta}{2}) \oplus U^{\oplus 5} }}
 \    & \!\! e_{Q_1,Q_2}(\Omega ) \, \Bigg[  
 \frac{\Theta_{E_8}^{(2)}(2\tau, 2z, \frac{\sigma}{2})}{16\, \Phi_{6,3}}  (1+(-1)^{\delta\cdot Q_1})  +  
 \frac{\Theta_{E_8}^{(2)}(2\tau, 2z, \frac{\sigma + 1}{2})}{16\, \Phi_{6,4}}  \\ &((-1)^{\delta\cdot Q_2}+(-1)^{\delta\cdot Q_1+ \delta\cdot Q_2})     
 \Bigg] 
\end{align}
for the twisted case ($h_2 = 1$).
As announced, we may factor a projector $\tfrac{1}{2}(1+(-1)^{Q_1\cdot \delta})$ and henceforth restrict to summation over $U(2)\subset U$. 

Next we address the $E_8$ part. Recall that the charge components $Q' = \sqrt{2}Q''+ \tfrac{\mathcal{P}}{\sqrt{2}}$ along $E_8(-\tfrac{1}{2})\subset \Lambda_e$ come in three classes , where $Q'' \in E_8(-1)$ and $\mathcal{P} \in E_8(-1)/(2E_8(-1))$, labelled by the orbit $\mathcal{O}_1,\mathcal{O}_{248},\mathcal{O}_{3875}$ under the Weyl group of $E_8$ that $\mathcal{P}$ belongs to. For these orbits $\mathcal{O}_x$ define
\begin{equation}
\Theta_x \coloneqq \sum_{\mathcal{P}\in \mathcal{O}_x}  \sum_{\substack{(Q_1,Q_2)\in\\ E_8(2) \oplus \left(E_8(2)+ \frac{\mathcal{P}}{\sqrt{2}}\right)}} \, e^{i\pi Q^r \Omega_{rs} Q^s} \ .
\end{equation}
The Siegel theta functions in the numerators of \eqref{eq:g2untwTerms} may be re-expressed as
\begin{align}
 \Theta_{E_8}^{(2)}( 2\tau , 2z ,  2)= \, \qquad &\sum_{\substack{(Q_1,Q_2)\in\\ E_8(2) \oplus E_8(2)}}\,e^{i\pi Q^r \Omega_{rs} Q^s} & = \Theta_1 \qquad \qquad \qquad \ \ \ \, \label{eq:Theta1} 
 \\
 \Theta_{E_8}^{(2)}( 2\tau , z ,  \tfrac{\sigma}{2})=\ 2^{-4}& \sum_{\substack{(Q_1,Q_2)\in\\ E_8(2) \oplus E_8(2)^*}}\,e^{i\pi Q^r \Omega_{rs} Q^s} & = \Theta_1 + \Theta_{248} + \Theta_{3875} \ \, \label{eq:Theta2} 
 \\
  \Theta_{E_8}^{(2)}( 2\tau , z ,  \tfrac{\sigma + 1}{2})=\ 2^{-4}&\sum_{\substack{(Q_1,Q_2)\in\\ E_8(2) \oplus E_8(2)^*}} \,e^{i\pi Q^r \Omega_{rs} Q^s} \, (-1)^{Q_2^2} & = \Theta_1 - \Theta_{248} + \Theta_{3875}  \label{eq:Theta3} \ .
\end{align}
The second of these relations is a genus two analog of \eqref{eq:thetaE8single}. Collecting $\Theta_x$ gives
\begin{align}\label{eq:g2untwTerms2}
 \sum_{\substack{Q_1 \in  U(2)\oplus U^{\oplus 5} \\ Q_2  \in U^{\oplus 6} }}
 \quad \    & e_{Q_1,Q_2}(\Omega ) \, \Bigg[  \Theta_1 \times \left( \frac{\frac{1}{2}(1+(-1)^{\delta \cdot Q_2})}{\Phi_{6,0}} + \frac{1}{2} \left(\frac{1}{16\, \Phi_{6,3}} +  \frac{1}{16\, \Phi_{6,4}} \right)\right)  \nonumber \\
   + &
\Theta_{248} \times \frac{1}{2} \left(\frac{1}{16\, \Phi_{6,3}} -  \frac{1}{16\, \Phi_{6,4}} \right)
  \ \  + 
\Theta_{3875} \times \frac{1}{2} \left(\frac{1}{16\, \Phi_{6,3}} +  \frac{1}{16\, \Phi_{6,4}} \right) 
 \Bigg] .
\end{align}
This in turn is the genus two analog of \eqref{eq:B4_Z2chl_untwEps}, the terms on the right-hand-side of each ``$\times$''-symbol give the quarter-BPS partition function in the respective subsector of the untwisted charge sector, depending on $\mathcal{P}$ and $(-1)^{\delta
\cdot Q_2}$ of the electric charge of the dyon. The sign between $\Phi_{6,3}^{-1}$ and $\Phi_{6,4}^{-1}$ matches $(-1)^{\frac{\mathcal{P}^2}{2}}$. Note also the presence of a projector in the term with $\Phi_{6,0}$. It is zero unless the winding along the CHL circle is even, and since this term only occurs for $\mathcal{P}=0=h_2$, we can equivalently say that it only arises for $Q\in \Lambda_m \subset \Lambda_e$ (or $r(Q,P)=[Q]=[0]$). With the identities \eqref{eq:ModId0} to \eqref{eq:ModId3} we recognize pairs of corresponding modular forms $(\Phi_{6,0},\eta^{8}(\sigma)\eta^{8}(2\sigma))$, $(\Phi_{6,3},\eta^{8}(\sigma)\eta^{8}(\frac{\sigma}{2}))$ and $(\Phi_{6,4},\eta^{8}(\sigma)\eta^{8}(\frac{\sigma+1}{2}))$. The first pair contains the cusp form for the level two congruence subgroup  $\Gamma_0(2)$ of the (Siegel) modular group, the second is obtained from it via an (embedded) S-duality transformation $\begin{psmallmatrix}
0 & -1\\ 1 & 0
\end{psmallmatrix}$ on $\sigma$ and the third pair is the $\sigma \mapsto \sigma +1 $ translate of the latter. Besides this we have
\begin{align}
\frac{1}{ \Phi_{6,0}} &= \frac{1}{(2\pi i z)^2} \ \frac{1}{\eta^{8}(\tau)\eta^8(2\tau)} \ \frac{1}{\eta^{8}(\sigma)\eta^8(2\sigma)}  \ \, \, \left( 1 + O(z^0)\right)\label{eq:divisorPhi60}\\
\frac{1}{16 \,  \Phi_{6,3}} &= \frac{1}{(2\pi i z)^2} \ \frac{1}{\eta^{8}(\tau)\eta^8(2\tau)}\  \frac{1}{\eta^{8}(\sigma)\eta^8(\frac{\sigma}{2})}\ \,\,\,\, \left( 1 + O(z^0)\right)\label{eq:divisorPhi63}\\
\frac{1}{16 \,  \Phi_{6,4}} &= \frac{1}{(2\pi i z)^2} \ \frac{1}{\eta^{8}(\tau)\eta^8(2\tau)} \ \frac{e^{-2\pi i/3}}{\eta^{8}(\sigma)\eta^8(\frac{\sigma+1}{2})}  \left( 1 + O(z^0)\right) \ ,\label{eq:divisorPhi64}
\end{align}
 by the help of which we immediately see that \eqref{eq:ModId0} to \eqref{eq:ModId3} re-appear in the linear combinations of \eqref{eq:g2untwTerms2} near the diagonal locus $z=0$. The same holds for twisted sector electric charges and
\begin{align}\label{eq:g2twTerms2}
 \sum_{\substack{Q_1 \in  U(2)\oplus U^{\oplus 5} \\ Q_2  \in (U+\frac{\delta}{2}) \oplus U^{\oplus 5} }}
 \quad \    & e_{Q_1,Q_2}(\Omega ) \, \Bigg[  \Theta_1 \times  \frac{1}{2} \left(\frac{1}{16\, \Phi_{6,3}} +  \frac{(-1)^{\delta \cdot Q_2}}{16\, \Phi_{6,4}}\right)  \nonumber \\
   + &
\Theta_{248} \times \frac{1}{2} \left(\frac{1}{16\, \Phi_{6,3}} -  \frac{(-1)^{\delta \cdot Q_2}}{16\, \Phi_{6,4}} \right)
   \ \ + 
\Theta_{3875} \times \frac{1}{2} \left(\frac{1}{16\, \Phi_{6,3}} +  \frac{(-1)^{\delta \cdot Q_2}}{16\, \Phi_{6,4}} \right) 
 \Bigg] ,
\end{align}
 corresponding in turn to \eqref{eq:B4_Z2chl_twEps}. As in the genus one case, these linear combinations of $\Phi_{6,3}^{-1}$ and $\Phi_{6,4}^{-1}$ basically imply the projection to Fourier modes with even or odd exponents of $e^{2\pi i \sigma}$, depending on the parity of the momentum along the CHL circle encoded in $(-1)^{\delta \cdot Q_2}$. One may thus argue that the quarter-BPS partition function of unit-torsion dyons with twisted sector electric charge $Q\in E_8(- \frac{1}{2}) \oplus (U+\frac{\delta}{2}) \oplus U^{\oplus 5}$ is simply $2^{-4} \Phi_{6,3}^{-1}$, in agreement with the result of \cite{Jatkar:2005bh,David_2006,Dabholkar_2007}.\footnote{Note that $\Phi_{6,3}(\tau,\sigma,z)=\Phi_{6,1}(\sigma,\tau,z)$ upon swapping the diagonal elements, so this is the same Siegel modular form as in~\cite{Dabholkar_2007} once the meaning of the chemical potentials is properly matched. 
 }
 
Comparing the untwisted and twisted sector results (eqs. \eqref{eq:g2untwTerms2} and \eqref{eq:g2twTerms2}) and following the logic of section \ref{sec:DHstatesZ2}, we similarly find that the quarter-BPS index of unit-torsion dyons is given by the Fourier coefficient of $\Phi_{6,3}^{-1}$ (understood with the moduli-dependent contour prescription in \eqref{eq:contourint}) plus an extra contribution in case that $Q\in \Lambda_m\subset \Lambda_e$, coming from the Fourier coefficient of $\Phi_{6,0}^{-1}$. By analogy with eq. \eqref{eq:DHsublatticesEM2} we can write
\begin{align}\label{eq:g2allTerms}
 \sum_{\substack{Q_1 \in  \Lambda_m \\ Q_2  \in \Lambda_e }}
 \quad \    & e_{Q_1,Q_2}(\Omega ) \, \Bigg[\frac{\delta_{Q_2\in \Lambda_e^*}}{ \Phi_{6,0}}  +  \frac{1}{2} \left( \frac{1}{16\, \Phi_{6,3}} +  \frac{(-1)^{Q_2^2}}{16\, \Phi_{6,4}} \right) 
 \Bigg]  \ .
\end{align}
For later convenience let us introduce some notation for the basic partition functions that are encountered here
\begin{align}
\Zo & \coloneqq  \frac{1}{2} \left(\frac{1}{16\, \Phi_{6,3}} +  \frac{1}{16\, \Phi_{6,4}} \right)+\frac{1}{\Phi_{6,0}}\\
\mathsf{Z}^{(\pm)} & \coloneqq \frac{1}{2} \left(\frac{1}{16\, \Phi_{6,3}} \pm  \frac{1}{16\, \Phi_{6,4}} \right) \ .
\end{align}
The forms $\Zo$ and $\Zp$ may also be rewritten in terms of modular forms $W,Y,T$ for the Iwahori subgroup $B(2)\subset \Sp_4(\IZ)$ (see appendix \ref{app:siegelmodforms}) via
\begin{align}\label{eq:Phi6ktoB2}
\frac{1}{W}=\frac{1}{\Phi_{6,0}} \ , \qquad \frac{1}{16\, \Phi_{6,3}} +  \frac{1}{16\, \Phi_{6,4}} = \frac{16 T}{Y W} \ , 
\end{align}
where $YW=\chi_{10}$ is the Igusa cusp form.

Our findings are compatible with the findings of \cite[eq. (2.14)]{Bossard_2019}, if our partition functions are subject to the condition $P \in \Lambda_m \backslash 2\Lambda_e$.\label{pg:Pnotin2Le} In section \ref{sec:sdwcconstraints} we will support this statement by considering wall-crossing.
To conclude, the M-theory lift of string webs argument of \cite{gaiotto2005rerecounting,Dabholkar_2007,Dabholkar:2007vk}, which lead us to analyzing the chiral fluctuations of the genus two heterotic CHL string, provides quarter-BPS indices for a large class of unit-torsion dyons that are compatible with indices obtained from suitable six-derivative couplings in the 3D CHL vacuum in the circle decompactification limit \cite{Bossard_2019}. As we have just shown, by analyzing the genus two orbifold partition function in greater detail, one can make the previous point not just for states with twisted sector electric charge (to which \cite{Dabholkar_2007} was limited), but also for untwisted electric charge sectors.
 

\section{Modular and polar constraints in the $\IZ_2$ model}
\label{sec:sdwcconstraints}

In the previous section we have proposed quarter-BPS partition functions for unit-torsion dyons in various subsectors of the untwisted and twisted charge sector.
In light of subsection \ref{ssc:structureQBPSfn} there are non-trivial constraints, especially from S-duality symmetry and wall-crossing, such a partition function must satisfy. These constraints will be addressed in the following.
In fact, this analysis already highly constrains these partition functions. With only few assumptions one might already ``guess'' the form of the latter.


\subsection{Quantization of the charge invariants}

First recall that for $Q\in \Lambda_e$ the parity of
\begin{equation}\label{eq:Q2parity}
Q^2 \equiv \frac{\mathcal{P}^2}{2} + h\, Q\cdot \delta \quad \operatorname{mod} \ 2
\end{equation}
depends on the Weyl orbit of the shift vector $\frac{\mathcal{P}}{\sqrt{2}}\in E_8(-\frac{1}{2})/E_8(-2)$, the twistedness $h\in\lbrace 0 ,1 \rbrace$ and the CHL circle momentum $Q\cdot \delta \in \IZ$. In fact $Q^2 \equiv [Q]^2$ modulo two.
 This parity is fixed within each of the charge subsectors considered in section \ref{sec:genustwo} and determines the periodicity the respective partition function must obey in the variable $\sigma$. For even $Q^2$ the period is 1, for odd $Q^2$ the period is 2. For each charge sector of eqs. \eqref{eq:g2untwTerms2} and \eqref{eq:g2twTerms2} (also see \eqref{eq:g2allTerms}) this expected periodity is indeed satisfied by the respective partition function, as $\Zo$ and $\Zp$ have period 1, while $\Zm$ is only periodic under $\sigma \mapsto \sigma + 2$. 
 Thus, according to \eqref{eq:SpMatrixPeriodicities}, all symplectic matrices of the form
\begin{equation}
\label{eq:SpMatrixPeriodicities2}
 \begin{pmatrix}
1 & 0 & r_1 & r_2 \\
0 & 1 & r_2 & r_3 \\
0 & 0 & 1 & 0\\
0 & 0 & 0 & 1
\end{pmatrix} \in  \begin{pmatrix}
1 & 0 & \IZ & \IZ \\
0 & 1 & \IZ & \IZ\\
0 & 0 & 1 & 0\\
0 & 0 & 0 & 1
\end{pmatrix}
\end{equation}
acting on $Z\in \mathbb{H}_2$ in the usual way leave the former two partition functions invariant, while for $\Zm$ this is only the case if also $r_3$ is even. For $r_3=1$ the form $\Zm$ picks up a minus sign. In all cases $\frac{P^2}{2}, Q\cdot P \in \IZ$, so the period in both the $\tau$ and $z$ direction is unity.

\subsection{S-duality symmetry}

As a second constraint, the S-duality group for the $\IZ_2$ CHL model is $\Gamma_1(2)=\Gamma_0(2)$ 
and leaves unchanged the residue $r(Q,P)=[Q]$ of a dyon charge, so for $\left(\begin{smallmatrix}
 a&b \\
c&d
\end{smallmatrix}\right)\in \Gamma_1(2) $ the embedded S-transformation \eqref{eq:symplecticStrafo}, i.e., $\begin{psmallmatrix}
A & B \\ C & D
\end{psmallmatrix}$ in the form
\begin{equation}\label{eq:symplecticStrafo2}
\begin{pmatrix}
d & b & 0 & 0\\
c & a & 0 & 0\\
0 & 0 & a & -c\\
0 & 0 & -b & d
\end{pmatrix}  \in \begin{pmatrix}
2\IZ+1 &\IZ & 0 & 0\\
2 \IZ & 2\IZ+1 & 0 & 0\\
0 & 0 & 2\IZ+1& 2\IZ\\
0 & 0 & \IZ & 2\IZ+1
\end{pmatrix} \cap \Sp_4(\IZ) \ ,
\end{equation}
 should describe the symmetry \eqref{eq:smftrafo} of each quarter-BPS partition function, here simply
 \begin{equation}\label{eq:zuntwSinv}
\mathsf{Z}(Z) = \mathsf{Z}(Z'') \ , \quad Z'' = (AZ+B)(CZ+D)^{-1} \ .
\end{equation} 
 
 For $\Zo$ and $\Zp$ eq. \eqref{eq:Phi6ktoB2} shows that \eqref{eq:symplecticStrafo2} indeed is a valid modular symmetry as the matrix lies in $B(2)$, and for Sen's partition function $2^{-4}\Phi_{6,3}^{-1}$ this symmetry is also known. The combination of these facts then demonstrates that $\Zm$ is also $\Gamma_1(2)$ S-duality invariant.

\subsection{Wall-crossing relations}

We now apply the general lessons from \ref{sss:wallcrossing} and study the implications of wall-crossing. 

\paragraph{First wall.} Regarding unit-torsion dyon charges $(Q,P) \in \Lambda_e \oplus \Lambda_m$ for any of the subsectors of section \ref{sec:genustwo} we first consider the decay into half-BPS states
\begin{equation}\label{eq:firstdecay}
(Q,P) \rightarrow (Q,0) + (0,P) \ .
\end{equation}
This decay is encoded by the matrix $\begin{psmallmatrix}
a_0 & b_0  \\
 c_0  & d_0
\end{psmallmatrix}=\begin{psmallmatrix}
1 & 0  \\
 0  & 1
\end{psmallmatrix}$ and demands that the respective partition function $\mathsf{Z}$ exhibits a quadratic pole at $z=0$, with coefficient given by
\begin{equation}\label{eq:ZuntwDivisor}
\mathsf{Z} \left( \begin{psmallmatrix}
\tau & z \\
z & \sigma
\end{psmallmatrix}\right) \propto \frac{1}{z^2} \ \phi_e^{-1}(\sigma)  \  \phi_m^{-1}(\tau)    + \mathcal{O}(z^0)\ .
\end{equation}
The functions $\phi_e^{-1}(\sigma)$ and $\phi_m^{-1}(\tau) $ are the half-BPS counting functions of the decay products $(Q,0)$ and $(0,P)$, respectively.

We start with the magnetic part. On page \pageref{pg:Pnotin2Le} we have already made the assertion that our dyon partition functions are subject to the restriction $P \in \Lambda_m \backslash 2 \Lambda_e$ on the magnetic charges. This was required by matching the results of \cite{Bossard_2019}. It is also consistent with wall-crossing. 
To give some background, we first remark that in~\cite{Bossard_2017}, in accordance with $\Gamma_1(2)$ S-duality and Fricke symmetry, it was shown that the half-BPS index (fourth helicity supertrace) for primitive charges $(Q,P) \in (\Lambda_{e}\oplus \Lambda_m)\backslash (\Lambda_m \oplus 2\Lambda_e)$ is given by
\begin{equation}\label{eq:Omega4one}
\Omega_4(Q,P) = c_8\left(- \frac{ \gcd(2Q^2, P^2, Q\cdot P)}{2} \right) \ ,
\end{equation}
while for charges in the complement  $(Q,P) \in (\Lambda_m \oplus 2\Lambda_e)$ it is given by
\begin{equation}\label{eq:Omega4two}
\Omega_4(Q,P) = c_8\left(- \frac{ \gcd(2Q^2, P^2, Q\cdot P)}{2} \right) + c_8\left(- \frac{ \gcd(2Q^2, P^2, Q\cdot P)}{2\cdot 2} \right) \ .
\end{equation}
The numbers $c_8(...)$ are the (always positive) Fourier coefficients of the $\Gamma_0(2)$ modular form
\begin{equation}\label{eq:etaprodFourier}
\frac{1}{\eta^{8}(\tau)\eta^8(2\tau)} = \sum_{m=-1}^{\infty} c_8(m) \, q^m = \frac{1}{q} + 8 + 52q + 256q^2 + O(q^3) \ .
\end{equation}
Since $\frac{P^2}{2}\in \IZ$ in general, for purely magnetic charges $(0,P)$ the first term in \eqref{eq:Omega4two} always contributes $c_8(\frac{P^2}{2})$ while the second term $c_8(\frac{P^2}{4})$ in \eqref{eq:Omega4two} vanishes unless $P^2\in 4\IZ$.
Furthermore, considering $Q=0$ states where trivially $Q\in\Lambda_m$, the condition $P\in \Lambda_m \backslash 2 \Lambda_e$
holds if and only if $(Q,P) \in (\Lambda_{e}\oplus \Lambda_m)\backslash (\Lambda_m \oplus 2\Lambda_e)$, which in turn is equivalent to \eqref{eq:Omega4one}. 
  Otherwise \eqref{eq:Omega4one} holds. 
Hence $P\in \Lambda_m \backslash 2 \Lambda_e$ is a sufficient condition for \eqref{eq:Omega4one}, and half-BPS states of charge $(0,P)$ being counted by
\begin{equation}\label{eq:halfBPSmagn}
\phi_m^{-1}(\tau) = \frac{1}{\eta^8(\tau)\eta^8(2 \tau)} \ .
\end{equation}
Indeed, the latter occurs on the diagonal divisor of all our quarter-BPS partition functions by eqs. \eqref{eq:divisorPhi60} to \eqref{eq:divisorPhi64}, suggesting that our counting formula should be understood as holding for states with $P \in \Lambda_m \backslash 2\Lambda_e$.

The magnetic charge assumption for our unit-torsion quarter-BPS partition functions in section \ref{sec:genustwo} is also consistent with results in the literature that rely on charge configurations for which this magnetic condition is explicitly satisfied.
Regarding twisted sector unit-torsion dyons, the derivation in \cite{David_2006,Sen:2007qy}, which is independent from the ansatz pursued here and in \cite{Dabholkar_2007}, starts indeed from charge representatives $(Q,P)$ satisfying $P\in \Lambda_m \backslash 2 \Lambda_e$ and arrives at the quarter-BPS counting function $2^{-4}\Phi_{6,3}^{-1}$. This clearly exhibits \eqref{eq:halfBPSmagn} at $z=0$, counting half-BPS states with charge $(0,P)$.
Regarding untwisted sector unit-torsion dyons in a certain (sub-)subsector, an analysis starting from explicit charge representatives $(Q,P)$ satisfying $P\in \Lambda_m \backslash 2 \Lambda_e$ was presented in \cite[section 6.5]{Banerjee:2008pv} and again leads to constraints consistent with our untwisted sector quarter-BPS partition functions (discussed further below).

Now we turn to the electric part. Here we can refer to eq. \eqref{eq:g2allTerms}. Via the identities \eqref{eq:divisorPhi60} to \eqref{eq:divisorPhi64}, the half-BPS partition function of states with charge $(Q,0)$ always reduces to the respective one in \eqref{eq:DHsublatticesEM2}. This consistently works out for all types $[Q]$ of electric charge.

Since $\eta^{-8}(\tau)\eta^{-8}(2\tau) $ and $\theta_{E_8(2),1}(\sigma)/\eta^{24}(\sigma)$ transform as weight $-8$ modular forms for $\Gamma_0(2)=\Gamma_1(2)$ (recall \eqref{eq:ModId0}, \eqref{eq:ModId1} and \eqref{eq:ModId3}), the weight of $\Zo$ and $\Zp$ must be $-6$, which is indeed the case. They should also transform as Siegel modular forms under modular transformations given by
\begin{equation}\label{eq:SpMatrixWall1}
\begin{pmatrix}
2\IZ +1 &0 & \IZ &0 \\
0 & 1 &0&0 \\
2\IZ & 0 &2\IZ +1 & 0\\
 0&0& 0 & 1
 \end{pmatrix}\cap\Sp_4(\IZ) \ , \qquad  
 \begin{pmatrix}
1 &0 & 0 &0 \\
0 &2\IZ +1 &0&\IZ\\
0 & 0 &1& 0\\
 0&2 \IZ& 0 & 2\IZ +1
 \end{pmatrix} \cap \Sp_4(\IZ) \ ,
\end{equation}
the first coming from the magnetic and the second coming from the electric part. Indeed, as these matrices belong to $B(2)$, the correct transformation of $\Zp$ and $\Zo$ is guaranteed. 

The form $\Zm$, where $\phi_e^{-1}(\sigma)=\frac{1}{2}(\eta^{8}(\sigma)\eta^{8}(\frac{\sigma}{2})-\eta^{8}(\sigma)\eta^{8}(\frac{\sigma+1}{2}))$ is a modular form for $\Gamma(2)$ (or for $\Gamma_0(2)$ with multiplier $(-1)^{q_1}$), transforms correctly under
 \begin{equation}\label{eq:SpMatrixWall1B}
\begin{pmatrix}
2\IZ +1 &0 & \IZ &0 \\
0 & 1 &0&0 \\
2\IZ & 0 &2\IZ +1 & 0\\
 0&0& 0 & 1
 \end{pmatrix}\cap\Sp_4(\IZ) \ , 
 \qquad  
 \begin{pmatrix}
1 &0 & 0 &0 \\
0 &2\IZ +1 &0&2\IZ\\
0 & 0 &1& 0\\
 0&2\IZ& 0 & 2\IZ +1
 \end{pmatrix} \cap \Sp_4(\IZ) \ ,
\end{equation}
which of course must be the case as this is formally just a projection of the partition function $2^4 \Phi_{6,3}^{-1}$ to odd $Q^2/2$. 
The latter is known to satisfy the modular and polar constraints mentioned in subsection \ref{ssc:structureQBPSfn} (see \cite[section 6.4]{Banerjee:2008pv} or \cite{Sen:2007qy}), and $\Zm$ will inherit this property.

\paragraph{Second wall.} Next we want to investigate the decay into half-BPS states
\begin{equation}
(Q,P) \rightarrow (Q-P,0) + (P,P) \ .
\end{equation}
This decay is now encoded by the matrix $\begin{psmallmatrix}
1 & 1  \\
 0  & 1
\end{psmallmatrix}$ and demands that $\mathsf{Z}$ exhibits a quadratic pole at $z' =0$ (recall eq. \eqref{eq:vprime_decay}), with coefficient given by
\begin{equation}\label{eq:ZuntwDivisor2}
\mathsf{Z} \left( \begin{psmallmatrix}
\tau & z \\
z & \sigma
\end{psmallmatrix}\right) \ \propto \  \frac{1}{{z'} {}^2} \ \phi_e^{-1}(\sigma ' ;1,0 )  \  \phi_m^{-1}(\tau ' ; 1, 1)   \, + \, \mathcal{O}( {z'} {}^0 )\ .
\end{equation}
The variables for this decay are related via \eqref{eq:polecoords}, explicitly
\begin{equation}\label{eq:2ndwallpolecoords}
Z' = \begin{psmallmatrix}
\tau ' & z' \\
z' & \sigma'
\end{psmallmatrix} = \begin{psmallmatrix}
\tau+\sigma+2z & z+\sigma \\
z+\sigma  & \sigma
\end{psmallmatrix} \ .
\end{equation}
Even though this decay is related to the previous one by an S-duality transformation in $\Gamma_1(2)$, we shall briefly analyze it to further illustrate the appearance of the Iwahori subgroup $B(2)$ for $\Zo$ and $\Zp$ on physical grounds. 
Furthermore, it allows to better test the untwisted sector partition functions against the analysis presented in~\cite[section 6.5]{Banerjee:2008pv}.

Now note that adding any vector from $\Lambda_m$ to $Q\in \Lambda_e$ cannot change the residue $[Q]\in \Lambda_e \backslash \Lambda_e^*$. 
As we have seen in section \ref{sec:DHstatesZ2}, the residue selects the half-BPS partition function of purely electric states, so the partition function for decay products $(Q-P,0) \in \Lambda_e$ will be the same as the one for decay products $(Q,0)$, i.e.,
\begin{equation}\label{eq:phiewall2}
\phi_e^{-1}(\sigma ' ;1,0 ) = \phi_e^{-1}(\sigma')  \ .
\end{equation}
For unit-torsion dyons the electric component $(Q,0)$ must be primitive and consistency requires $(Q-P,0)$ to be primitive as well.
Namely, if $Q-P=n Q'$ for some integer $n$ and primitive $Q' \in\Lambda_e$, then $Q \wedge P= n (Q' \wedge P)$, but we know that $I=\gcd(Q\wedge P)=1$ for unit-torsion dyons, so $n=1$.
Similarly, in the new duality frame obtained by the S-duality transformation $\begin{psmallmatrix}
1 & 1  \\
 0  & 1
\end{psmallmatrix}$ we still count dyons of unit-torsion, so again $Q-P$ must be primitive in $\Lambda_e$.

Since S-duality also relates $(P,P)^\tp =(\begin{psmallmatrix}
1 & 1 \\ 0 & 1
\end{psmallmatrix}(0,P))^\tp$ to $(0,P)$, we know that
\begin{equation}\label{eq:phimwall2}
\phi_m^{-1}(\tau ' ; 1, 1) = \frac{1}{\eta^{8}(\tau')\eta^8(2\tau')} \ .
\end{equation}
Both \eqref{eq:phiewall2} and \eqref{eq:phimwall2} also follow from S-duality invariance for elements \eqref{eq:symplecticStrafo2} by combining \eqref{eq:zuntwSinv}, \eqref{eq:ZuntwDivisor} and \eqref{eq:2ndwallpolecoords}.

We have already mentioned that the functions appearing here are, in the case of $\Zo$ and $\Zp$, modular forms for the congruence subgroup $\Gamma_0(2)=\Gamma_1(2) \ni \begin{psmallmatrix}\alpha_1 & \beta_1\\ \gamma_1 & \delta_1\end{psmallmatrix}, \ \begin{psmallmatrix}p_1 & q_1 \\ r_1 & s_1\end{psmallmatrix}$. Employing \eqref{egensp} and \eqref{egensp2}, each $\mathsf{Z}$ is required to transform as a Siegel modular form with respect to 
\begin{equation}\label{eq:SpMatrixWall2}
\begin{pmatrix}
\alpha_1 & \alpha_1-1 & \beta_1 &0 \\
0 & 1 &0&0 \\
\gamma_1 & \gamma_1 & \delta_1 & 0\\
 \gamma_1 & \gamma_1& \delta_1-1 & 1
 \end{pmatrix}\cap\Sp_4(\IZ) \ , \qquad  
 \begin{pmatrix}
1 & 1-p_1 & q_1 & -q_1  \\
0 & p_1 & -q_1 & q_1 \\
0 & 0 &1& 0\\
 0& r_1 & 1-s_1 & s_1
 \end{pmatrix} \cap \Sp_4(\IZ) \ , 
\end{equation}
where $\gamma_1$ and $r_1 $ are even, while $\alpha_1, \delta_1, p_1$ and $s_1$ are odd. For $\Zm$ the integer $q_1$ must be even.

Again we compare these constraints to the explicit form of $\Zo$, $\Zp$ and $\Zm$ proposed before. Since \eqref{eq:polecoords} describes an S-duality transformation for this decay code, $\mathsf{Z}(Z) = \mathsf{Z}(Z')$ holds via \eqref{eq:zuntwSinv}. This immediately translates \eqref{eq:ZuntwDivisor} into
\begin{equation}\label{eq:Zuntw2ndWallEM}
\mathsf{Z}(Z)\propto   \frac{1}{{z'}^{2}}\  \phi_{e}^{-1}(\sigma') \ \frac{1}{\eta^8(\tau')\eta^8(2\tau')} + \mathcal{O}({z'}^{0}) \ ,
\end{equation}
and therefore matches \eqref{eq:ZuntwDivisor2} with \eqref{eq:phiewall2} and \eqref{eq:phimwall2}.

\paragraph{Third wall.} There is one decay channel only possible for dyons with untwisted sector charge $(Q,P)$ subject to the extra condition $Q\in \Lambda_m \subset \Lambda_e$, namely 
\begin{equation}
(Q,P) \rightarrow (Q,Q) + (0, -Q + P) \ .
\end{equation}
The decay code  $\begin{psmallmatrix} a_0&b_0 \\ c_0 &d_0 \end{psmallmatrix}=\begin{psmallmatrix} 1&0 \\ 1 &1 \end{psmallmatrix}$ does not lie in $\Gamma_0(2)$
and coordinates appropriate for this pole are now by \eqref{eq:polecoords}
\begin{equation}
\cO' = \begin{pmatrix}
\crho ' & \cv ' \\ \cv' & \csig'
\end{pmatrix} = \begin{pmatrix}
\crho							 & \cv +  \crho  \\ 
\cv +  \crho & \crho + \csig + 2 \cv
\end{pmatrix} \quad =M_{3w} \cO \ 
\end{equation}
with 
\begin{equation}\label{eq:decay3sympl}
M_{3w} \coloneqq \begin{pmatrix}
1 & 0 & 0 &0 \\ 
1 & 1 & 0 & 0 \\
0 & 0 & 1 & -1 \\
0 & 0 & 0 & 1
\end{pmatrix} \ \in \Gamma_0^{(2)}(2)\,  \backslash \, B(2) \ .
\end{equation}

Recall that $Q$ is primitive in $\Lambda_e$ since we consider unit-torsion, so the first decay product $(Q,Q) \in (\Lambda_m \backslash 2 \Lambda_e)^{\oplus 2}$ is again counted by
\begin{equation}\label{eq:phiewall3}
\phi_e^{-1}(\sigma ' ; 1, 1) = \frac{1}{\eta^{8}(\sigma')\eta^8(2\sigma')} \ ,
\end{equation}
in accordance with~\eqref{eq:Omega4one}. Note that $(Q,Q)^\tp =\begin{psmallmatrix} 1 & 1 \\ 0 & 1 \end{psmallmatrix}\begin{psmallmatrix}0 \\ Q \end{psmallmatrix} $ is related via an S-transformation in $\Gamma_0(2)$ to a purely magnetic charge of the form $(0,\tilde{P})\in \Lambda_m\backslash 2\Lambda_e$.\footnote{So the subscript ``$e$'' in $\phi_e^{-1}(\sigma ' ; 1, 1)$ is a notational artifact inherited from~\cite{Banerjee:2008pv}.}

It was mentioned before that $Q-P$ is primitive for unit-torsion dyons, so the second decay product, a purely magnetic half-BPS state of charge $(0,-Q+P)\in \Lambda_m \backslash 2 \Lambda_e$, also corresponds to the eta-quotient
\begin{equation}\label{eq:phimwall3}
\phi_m^{-1}(\tau ' ; 0, 1) = \frac{1}{\eta^{8}(\tau')\eta^8(2\tau')} \ .
\end{equation}

Combining the ingredients we infer that wall-crossing demands that for $z' \rightarrow 0$
the quadratic pole in $\Zo$ becomes
\begin{equation}\label{eq:ZuntwDivisor3}
\Zo \left( \begin{psmallmatrix}
\tau & z \\
z & \sigma
\end{psmallmatrix}\right) \ \propto \  \frac{1}{{z'} {}^2} \ \phi_e^{-1}(\sigma ' ;1,1)  \  \phi_m^{-1}(\tau ' ; 0, 1)   \, + \, \mathcal{O}( {z'} {}^0 )
\end{equation}
with the given eta-quotients. As a consequence, by \eqref{egensp} and \eqref{egensp2} the partition function $\Zo$ should also transform as a Siegel modular form with respect to the embedded transformations
\begin{equation} \label{eq:SpMatrixWall3}
\begin{pmatrix}
\alpha_1 & 0 & \beta_1 & \beta_1 \\
- \alpha_1 -1 & 1 & -\beta_1 & - \beta_1 \\
\gamma_1 & 0 & \delta_1 & \delta_1 +1 \\
0 & 0 & 0 & 1
\end{pmatrix} \, , \quad  \begin{pmatrix}
1 & 0 & 0 & 0\\
 p_1-1 & p_1 & 0 & q_1\\
r_1 &r_1 & 1 & s_1 -1 \\
r_1 & r_1 & 0 & s_1
\end{pmatrix}
\end{equation}
where $\begin{psmallmatrix}
\alpha_1 & \beta_1 \\ \gamma_1 & \delta_1
\end{psmallmatrix}, \begin{psmallmatrix}
p_1 & q_1 \\ r_1 & s_1
\end{psmallmatrix} \in \Gamma_1(2)$.

Let us see whether \eqref{eq:ZuntwDivisor3} is also satisfied for the concrete $\Zo$ proposed before. Starting from \eqref{eq:Phi6ktoB2}, we consider the tautology $\Zo(Z)=\Zo(M_{3w}^{-1}Z')$. Since $M_{3w}^{-1} \in \Gamma_0^{(2)}(2)\,  \backslash \, B(2)$, only the $B(2)$ modular form $T$ transforms non-trivially, so for $T(M_{3w}^{-1}Z')$ we may use the transformation formula for theta characteristics (see appendix \ref{app:siegelmodforms}) to find a new characteristic
\begin{equation}
M_{3w}^{-1} \left\{ \begin{psmallmatrix}
a_1 \\ a_2 \\ b_1 \\ b_2 
\end{psmallmatrix} \right\} = \begin{psmallmatrix}
a_1 + a_2 \\ a_2 \\ b_1 \\ b_1 + b_2 
\end{psmallmatrix} \ .
\end{equation}
This means that
\begin{equation}
16^2 \, T(M_3^{-1}\cO') = \theta_{1100}^4 (\cO ') \, \theta_{1111}^4 (\cO ') \ 
\end{equation}
and thus
\begin{equation}\label{eq:Zuntw3rdwall}
\Zo(\cO) = - \frac{1}{2} \frac{1}{W(\cO')} \ - \ \frac{1}{32} \frac{ \theta_{1100}^4 (\cO ') \, \theta_{1111}^4 (\cO ')}{YW(\cO ')} \ .
\end{equation}
Now use the behaviour of the theta constants $\theta_{a_1 a_2 b_1 b_2}(\cO')$ under $\cv' \rightarrow 0$  (again see appendix \ref{app:siegelmodforms}) to find that the second term in \eqref{eq:Zuntw3rdwall}, being proportional to $\theta_{1111}^2 (\cO ')$, vanishes quadratically in $\cv'$ for $\cv' \rightarrow 0$. Only the first term contributes to the quadratic pole in $\cv'$ which is relevant for the BPS indices. More precisely, for $\cv' \rightarrow 0$ we have
\begin{equation}\label{eq:Zuntw3rdHBPS}
\Zo(\cO) = - \frac{1}{2} \, \frac{1}{(2\pi i)^2} \, \frac{1}{\cv '{}^2} \ \frac{1}{\eta^8(\crho') \eta^8(2\crho')} \ \frac{1}{\eta^8(\csig')\eta^8(2 \csig') } \ + \ \mathcal{O}(\cv'\, {}^{0}) \ ,
\end{equation}
nicely matching our wall-crossing expectations. The calculation also shows that only $\Phi_{6,0}^{-1}$ contributes to the pole, while $\Zp$ does not.

\paragraph{Modular symmetry group.} In close parallel to~\cite[section 6.5]{Banerjee:2008pv} the question emerges, whether the symplectic matrices \eqref{eq:SpMatrixPeriodicities2}, \eqref{eq:symplecticStrafo2}, \eqref{eq:SpMatrixWall1}, \eqref{eq:SpMatrixWall2} and \eqref{eq:SpMatrixWall3} fit into a subgroup of $\Sp_4(\IZ)$ defined by some congruence relation. Affirmative answer can be given for the group 
\begin{equation}\label{eq:congrCandGroup}
\begin{pmatrix}
2\IZ+1 & \IZ & \IZ & \IZ \\
2\IZ & 2\IZ+1 & \IZ & \IZ \\
2\IZ & 2\IZ & 2\IZ+1 & 2 \IZ \\
2\IZ & 2\IZ & \IZ  & 2\IZ+1  
\end{pmatrix}\cap \Sp_4(\IZ) = \begin{pmatrix}
\IZ &  \IZ & \IZ & \IZ \\
2\IZ & \IZ & \IZ & \IZ \\
2\IZ & 2\IZ &  \IZ  & 2 \IZ \\
2\IZ & 2\IZ & \IZ  & \IZ 
\end{pmatrix}\cap \Sp_4(\IZ) \ ,
\end{equation}
where the group on the right hand side is in fact the Iwahori subgroup $B(2)$.
 To see why \eqref{eq:congrCandGroup} is an equality, we use eq. \eqref{eq:SpInv}.
Then for $M, M^{-1} \in B(2)\subset \Sp_4(\IZ)$ we inspect entries in $M M^{-1} - 1_4 \equiv 0 \!\! \mod 2$ to find what is claimed.

Thus, by analyzing the polar and modular constraints one is led naturally to $B(2)$ as the symmetry group for the charge sectors with even $Q^2=Q^2_2$. As we know by section \ref{sec:genustwo} (recall eqs. \eqref{eq:g2untwTerms2} and \eqref{eq:g2twTerms2} or \eqref{eq:g2allTerms}), for these sectors the partition function is either given by $\Zo$ (if $[Q]=0$) or $\Zp$ (if $[Q]\neq 0$) and both of them indeed are modular forms for $B(2)$.

\paragraph{Remark on a subsector.}
Let us comment on the relation between our findings and that of~\cite[section 6.5]{Banerjee:2008pv}.

The unit-torsion quarter-BPS partition function considered there concerns dyons with untwisted sector electric charge subject to the constraints
\begin{align}
\frac{1}{2}Q^2 &\in 2\IZ+1 , & \frac{1}{2}P^2 &\in 2\IZ+1 , & Q\cdot P &\in 2\IZ , \nonumber \\
h&=0 , & \mathcal{P}&=0, & Q\cdot \delta & \in 2\IZ+1 \ . \label{eq:BanerjeeSubsector}
\end{align}
As these restrictions are only preserved for S-transformations in $\Gamma(2) \subset \Gamma_1(2)$, the partition function for this subsector does not need to be invariant under \textit{all} elements in \eqref{eq:symplecticStrafo2}, but only under those where $b$ is even.

The partition function for unit-torsion dyons that have odd $Q^2/2$ and satisfy all constraints in the second line of \eqref{eq:BanerjeeSubsector}, but have generic values of $\frac{1}{2}P^2 \in \IZ$ and $ Q\cdot P \in \IZ$, are counted by  $\Zm$.
The additional parity restrictions on the latter quadratic T-invariants can be implemented in $\Zm$ by applying suitable projections. For odd $P^2/2$, for instance, one has the lower sign in
 \begin{equation}
\Zm(\tau, \sigma, z)  \rightarrow \frac{1}{2} \left( \Zm(\tau, \sigma, z) \pm  \Zm(\tau + \tfrac{1}{2}, \sigma , z)   \right) \ .
\end{equation}
With this and the properties of $\Zm$ it is straightforward to check that also on the subset \eqref{eq:BanerjeeSubsector} the modular and polar constraints discussed in \cite[section 6.5]{Banerjee:2008pv} are met. 

As mentioned already, for magnetic half-BPS states $(0,P)$ counted at $z=0$, our assumption \eqref{eq:halfBPSmagn} is compatible with the explicit representatives $P$ chosen in~\cite[section 6.5]{Banerjee:2008pv}. These are primitive vectors $P \in \Lambda_m \backslash 2\Lambda_e$. Indeed, these are also the same magnetic charges as occuring in the twisted sector quarter-BPS states~\cite[section 6.4]{Banerjee:2008pv} (up to restriction to odd ``$K$'' quantum number there, causing $P^2/2$ to be odd for the untwisted case).

Befor proceeding, we remark that affirmative consistency checks starting from charge representatives in other subsectors can be performed in complete analogy to~\cite[section 6.5]{Banerjee:2008pv}, however, these are mostly straightforward and in light of our preceeding analysis rather redundant so we will not display them here.

\paragraph{Stringency of the constraints.}\label{par:ansatz} 
In summary, the constraints from quantization laws, S-duality and wall-crossing suggest that $\Zp$ and $\Zo$ transform as Siegel modular forms for the Iwahori subgroup $B(2)$ with weight $-6$. As announced earlier, we may now conclude that the modular and polar constraints alone are (almost) restrictive enough to guess the respective $\mathsf{Z}$ in closed form. 
Of course, the analysis of section \ref{sec:genustwo} already provides explicit expressions, which we have shown to satisfy all constraints, nevertheless, it is instructive to have an alternative approach that gives consistent results.

Since explicit generators for the ring of even (positive) weight Siegel modular forms for $\Gamma_0^{(2)}(2), K(2)$ and $B(2)=\Gamma_0^{(2)}(2) \cap K(2)$ are known in the mathematics literature (references are given in appendix \ref{app:siegelmodforms}), a suitable ansatz might reduce the problem of fixing $\mathsf{Z}$ to a determination of a finite number of coefficients.
This Siegel modular form must exhibit the quadratic poles in \eqref{eq:ZuntwDivisor}, \eqref{eq:ZuntwDivisor2} and \eqref{eq:ZuntwDivisor3}. Indeed, any decay code $\begin{psmallmatrix}
a_0 & b_0 \\ c_0 & d_0
\end{psmallmatrix}\in \SL_2(\IZ)$ is related to either $\begin{psmallmatrix}
1 & 0\\ 0 & 1
\end{psmallmatrix}$ (first wall) or $\begin{psmallmatrix}
1 & 0\\ 1 & 1
\end{psmallmatrix}$ (third wall) by an S-duality transformation in $\Gamma_1(2)$, which has index two in $ \SL_2(\IZ)$. We can therefore demand that $\mathsf{Z}(Z)$ must exhibit a quadratic pole at all images of the diagonal locus $\begin{psmallmatrix}
\tau &0 \\ 0 & \sigma 
\end{psmallmatrix}$ under the group generated by $\SL_2(\IZ)$-transformations \eqref{eq:symplecticStrafo} and integer translations \eqref{eq:SpMatrixPeriodicities2}.\footnote{As an aside, motivated by CHL dyon counting functions Cl\'{e}ry and Gritsenko~\cite{MR2806099} classified and constructed all so-called \textit{dd-modular forms}, i.e., Siegel modular forms for the Hecke congruence subgroups $\Gamma_0^{(2)}(N)$ which vanish precisely along the $\Gamma_0^{(2)}(N)$-translates of the diagonal divisor $z=0$ (with vanishing order one; possibly with a multiplier system). Especially, this includes the square roots of the Igusa cusp form and the Siegel modular form $\Phi_{6,0}$ appearing in the $N=1,2$ CHL models. However, this does not characterize the partition functions $\Zo$, $\Zp$ or $\Zm$.}
The arguably simplest compatible ansatz one might choose for $\mathsf{Z}(Z)$ is $F(Z)/\chi_{10}(Z)$, where the Igusa cusp form $\chi_{10}$, i.e., the product of the square of the ten even genus two Thetanullwerte, vanishes quadratically at all $\Sp_4(\IZ)$-images of the diagonal. The latter is also the partition function for unit-torsion quarter-BPS dyons in the parent theory and at least the untwisted sector dyons of interest might be regarded as an invariant subset thereof.  In this ansatz $F(Z)$ is a weight four Siegel modular form for $B(2)$, which is expected to be holomorpic in $\IH_2$ such that there are no additional, spurious poles. Zeroes in $F(Z)$ however might cancel any additional, spurious poles in $\chi_{10}^{-1}$ (if there are such). Working, for instance, with the ring generators given in \eqref{eq:B2ring} and the properties of the theta constants, the behaviour of $\mathsf{Z}$ at the wall-crossing divisors fixes $F(Z)$ eventually to $F^{(+)}=8T$ or $F^{(0)}=Y+8T$. This gives precisely back $\Zp$ and $\Zo$ found via the chiral genus two partition function in section \ref{sec:genustwo}.


\section{Black hole entropy}
\label{sec:bhentropy}

There is one more physical constraint on a quarter-BPS partition function in four-dimensional $\mathcal{N}=4$ theories. For large dyon charges $(Q,P)$ the microscopic BPS index should yield the macroscopic entropy of an extremal black hole carrying these charges, for instance, as computed in the supergravity approximation. The following analysis will focus on the least intricate features, namely the Bekenstein-Hawking term and the first correction in inverse powers of the charges. We simply quote the mathematical consequences for our $\Zo$, $\Zp$ and $\Zm$ functions (which may be regarded as untwisted sector partition functions), paralleling the discussion for the twisted sector~\cite{Dijkgraaf1997d,Cardoso_2005,Jatkar:2005bh,David_2006,Banerjee:2008pv,Sen:2007qy}.

Generic for all CHL models, the leading term in the entropy of an extremal black hole carrying large charges $(Q,P)$ is the just mentioned Bekenstein-Hawking area term. Together with the leading correction in inverse powers of the charge this gives an entropy
\begin{equation}\label{eq:SBHentropy}
S_{BH}= \pi \sqrt{Q^2 P^2 - (Q \cdot P)^2} + 64 \pi^2 \, \phi\left( \frac{Q \cdot P }{P^2 } , \frac{\sqrt{Q^2 P^2 - (Q \cdot P)^2}}{P^2}\right) + \cdots \ .
\end{equation}
 This correction has been determined in~\cite{Sen_2006entrophet} from the entropy function~\cite{Sen:2005higher} by including the Gauss-Bonnet term in the effective supergravity action~\cite{Sen:2005BHelemStrings,Gregori_1998},
\begin{equation}
\int \operatorname{d}^4x \ \sqrt{-\det g} \ \phi(a, S) \, \left(  R_{\mu\nu\rho\sigma} R^{\mu\nu\rho\sigma} - 4 R_{\mu\nu} R^{\mu\nu}+ R^2 \right)  \ ,
\end{equation}
where $\tau = a + iS$ denotes the axio-dilaton modulus. We also have a model dependent function $\phi$ which for our $\IZ_2$ model becomes
\begin{equation}
\phi(a,S) = - \frac{1}{64 \pi^2} \left[ 8 \log S + \log g(a+iS) + \log g(-a+iS) \right]+ \mathrm{const.}\ 
\end{equation}
with
\begin{equation}
g(\tau) = \eta^8(\tau) \eta^8(2\tau) \ .
\end{equation}

In order to formulate the resulting constraint for our microscopic $\mathsf{Z}$ we follow~\cite{Dijkgraaf1997d,Cardoso_2005,Jatkar:2005bh,David_2006,Sen:2007qy}. In the known examples the asymptotic growth of the Fourier coefficients of the respective Siegel modular form is estimated by a saddle-point approximation in $(\tau, \sigma)$ after picking up the dominant pole in the $z$-plane. For unit-torsion dyons in heterotic string theory on $T^6$ and the Igusa cusp form $\chi_{10}$, as well as generic twisted sector unit-torsion dyons in the $\IZ_2$ CHL model and the Siegel modular form $\Phi_{6,3}$, the dominant contribution comes from the divisor
\begin{equation}
	\mathcal D	\coloneqq	\tau \sigma - z^2 + z	=	0 \ .
\label{divisorbh}
\end{equation}
At $\mathcal{D}=0$ the partition function exhibits a quadratic pole in both of these cases. Since our untwisted sector partition functions $\mathsf{Z}$ can be written as $F/\chi_{10}$ with $F$ holomorphic, its poles are controlled by the zeroes of $\chi_{10}$. An ad hoc adaption of the analysis in~\cite[section 4]{David_2006} (i.e., ignoring the presence of $F$ while extremizing the exponential in \eqref{eq:contourint} at a generic divisor of $\chi_{10}$) suggests that $\mathcal{D}=0$ again gives the dominant contribution to the Fourier integral for large charges, with other divisors corresponding to exponentially suppressed contributions to the entropy.\footnote{This corresponds to the analyis of~\cite[section 4]{David_2006} up to eq. (4.14) there. } 

Let us thus address what behaviour of $\mathsf{Z}$ is expected near $\mathcal{D}=0$. Using a symplectic transformation
\begin{equation}
	M_A	=	\begin{pmatrix} 1& 0 & 0 & 0 \\	1 & 0 & 0 & -1 \\	0 & -1 & 1 & 0 \\	0 & 1 & 0 & 0 	\end{pmatrix}
\end{equation}
to introduce new coordinates around the divisor,
\begin{equation}
	\tau^\prime = \frac{\tau\sigma - z^2}{\sigma}~,	\quad	\sigma^\prime = \frac{\tau\sigma-(z-1)^2}{\sigma}	\quad\text{and}\quad		z^\prime = \frac{\tau\sigma-z^2+z}{\sigma} \ ,
\end{equation}
the partition function should behave near $\mathcal{D}=0$ (now $z'=0$) like
\begin{equation}
	\mathsf{Z}	\propto	\frac{1}{(2z^\prime-\tau^\prime-\sigma^\prime)^6} \left(	\frac{1}{{z^\prime}^2} \frac{1}{g(\tau^\prime)} \frac{1}{g(\sigma^\prime)}	+ \mathcal O({z^\prime}{}^4)	\right)~.
\label{limitbhpartition}
\end{equation}
As noted in~\cite[section 6.5]{Banerjee:2008pv}, if this is the case the macroscopic entropy \eqref{eq:SBHentropy} will be reproduced (upon standard procedure executed already for the twisted sector case).

We shall now check whether the partition function $\mathsf{Z} \in \lbrace \Zo , \Zp , \Zm \rbrace$ satisfies \eqref{limitbhpartition}. Since the symplectic transformation $M_A$ is not an element of $\Gamma_0^{(2)}(2)$ or $B(2)$, we have to express $\mathsf{Z}$ in terms of the genus two theta functions using \eqref{eq:SMFY1}-\eqref{eq: defTSMF}. By the transformation law \eqref{genus2trafo} their characteristics transform as
\begin{equation}
	M_A^{-1}	\{ (a_1,a_2,b_1,b_2)\tp \}	=	(a_1+a_2,b_1+b_2,b_1,a_2)\tp~.
\end{equation}
For instance, we have
\begin{equation}
	\theta_{0010}(Z)	=	\theta_{0010}(M_A^{-1}Z^\prime)	\ \propto	\ (2z^\prime-\tau^\prime-\sigma^\prime)^{1/2} \theta_{1111}(Z^\prime)~.
\end{equation}
Making use of \eqref{limit1} and \eqref{limit2} we find for the limit $z^\prime\rightarrow 0$ that only the term with $Y'$ in the numerator, formally the same as the twisted sector partition function $2^{-4}\Phi_{6,3}$, contributes to the double pole. The calculation gives
\begin{equation}
	\mathsf{Z}	\propto	\frac{1}{{z^\prime}^2} \frac{1}{(2z^\prime-\tau^\prime-\sigma^\prime)^6} \frac{1}{\eta^{12}(\rho^\prime)\theta_2^4(\tau^\prime)} \frac{1}{\eta^{12}(\sigma^\prime)\theta_2^4(\sigma^\prime)} + \mathcal O(z^\prime{}^{4})~,
\end{equation}
which indeed reproduces the expectation \eqref{limitbhpartition} by virtue of the first eta-product identity in \eqref{etaproducts}. In other words, any untwisted sector partition function $\mathsf{Z}$ gives rise not just to the leading Bekenstein-Hawking term, but also to the correct subleading correction in inverses powers of the charges \eqref{eq:SBHentropy}, very similar to the twisted sector quarter-BPS partition function $2^{-4}\Phi_{6,3}^{-1}$. We leave it as an open problem to perform more careful, extensive analyses as, e.g., in~\cite{Banerjee:2008ky,Murthy:2009dq} and to check whether a difference in the entropy of twisted sector and untwisted sector (quarter-BPS unit-torsion) dyons can be found in further subleading terms (say at exponentially suppressed orders). If so, one might ask for a macroscopic explanation in the quantum entropy function (say as certain sub-leading saddles to the supergravity path integral), see the references in the introduction for similar research. 

Having successfully passed the test of black hole entropy, we finally compare the dyon partition functions to the Donaldson-Thomas partition functions.

\section{Comparison to results from Donaldson-Thomas theory}
\label{sec:DTtheory}

The spectrum of quarter-BPS states in four-dimensional $\mathcal{N}=4$ string theories has been linked to the enumerative geometry of algebraic curves in Calabi-Yau threefolds. Predictions from string duality have thus led to precise mathematical conjectures~\cite{Oberdieck2016,Bryan:2018nlv}, some of which have been proven in recent years ~\cite{MR3827207,Oberdieck_2019}. 
Here we explore the connection between quarter-BPS indices and reduced Donaldson-Thomas (DT) invariants by comparing the BPS partition functions in the $\IZ_2$ CHL model with recently conjectured formulas for (tentative) DT counterparts~\cite{Bryan:2018nlv}.

\paragraph{Summary of DT result.} For this let us briefly collect some definitions and (conjectural) formulas for DT invariants of the $\mathbb{Z}_2$ CHL model from~\cite{Bryan:2018nlv}.
The geometric $N=2$ CHL model is given by the Calabi-Yau threefold $X=(S\times E) / \IZ_N$, where $S$ is a non-singular projective K3 surface and $E$ is a non-singular elliptic curve. In accordance with subsection~\ref{ssc:ConstructionCHLmodels} the orbifold group $\IZ_2$ acts by a symplectic involution $g : S \to S$ on $S$ and a translation in $E$ by some two-torsion point $e_0$. Correspondingly, there is a projection operator
\begin{equation}\label{eq:cohomProjector}
\Pi = \frac{1}{2} (1 + g_{\ast}) : H^{*}(S,\IQ) \to  H^*(S,\IQ)  
\end{equation}
and an isomorphism~\cite[app. B]{Bryan:2018nlv}\label{cor:typo:appB3}
\begin{equation}\label{eq:projectorIsom}
\Pi( H^{\ast}(S,\IZ) ) \ \cong \ \left( H^{\ast}(S,\IZ)^g \right)^{\ast} \ \cong \ E_8(-\tfrac{1}{2}) \oplus U^{\oplus 4}.
\end{equation}
By the \emph{divisibility} of a curve class $\gamma \in \mathrm{Image}(\Pi |_{N_1(S)})$ one means the biggest integer $m \in \mathbb{N}_{>0}$ for which
\begin{equation}
\frac{\gamma}{m} \in \mathrm{Image}(\Pi|_{N_1(S)}) \subset \frac{1}{2} H_2(S,\IZ) 
\end{equation} 
is satisfied, where $N_1(\cdot)$ denotes the group of algebraic one-cycles. 
If its divisibility is $1$, $\gamma$ is called a \textit{primitive} class, which is further called \emph{untwisted} if $\gamma \in H_2(S,\IZ)$, or \emph{twisted} if $\gamma \in \frac{1}{2} H_2(S,\IZ) \setminus H_2(S,\IZ)$.
 
We consider the curve class\footnote{By \cite[eq. (9), Lemma 1.4]{Bryan:2018nlv} we have $H_2(X,\IZ)=\operatorname{Im}(\Pi)\oplus \IZ[E/\IZ_2]$  and $N_1(X)=\Pi(N_1(S))\oplus \IZ[E/\IZ_2]$, both modulo torsion.}
\begin{equation}
\beta = (\gamma, d) \in N_1(X) \subset H_2(X,\IZ) 
\end{equation}
for some primitive, non-zero $\gamma$ with self-intersection
\begin{equation}
 \langle \gamma, \gamma \rangle 
= 2 s,
\quad
s \in 
\begin{cases}
\IZ & \text{ if } \gamma \text{ untwisted} \\
\frac{1}{2} \IZ & \text{ if } \gamma \text{ twisted}
\end{cases} \ .
\end{equation}
The reduced Donaldson--Thomas invariant $\DT^X_{n, (\gamma,d)}$ only depends on $n,s,d$ and whether $\gamma$ is untwisted or twisted, so one may also write $\DT^{\text{untw}}_{n, s, d}$ and $\DT^{\text{tw}}_{n, s, d}$ for the two cases. Introducing respective partition functions
\begin{align} \label{defn_part_fn}
\mathsf{Z}^{\text{untw}}(q,t,p) & \coloneqq \, \sum_{\substack{s \in \IZ \\ s \geq -1}}  \, \, \sum_{d \geq 0} \sum_{n \in \IZ} \DT^{\text{untw}}_{n, s, d} \ q^{d-1} \, t^{s} \, (-p)^{n} \\
\mathsf{Z}^{\text{tw}}(q,t,p) & \coloneqq \sum_{\substack{s \in \frac{1}{2} \IZ \\ s \geq -1/2}} \sum_{d \geq 0} \sum_{n \in \IZ}  \DT^{\text{tw}}_{n, s, d} \  q^{d-1} \, t^{s} \, (-p)^{n} \ \label{defn_part_fn2}
\end{align}
and writing
\begin{equation}
q=e^{2\pi i \tau}, \quad t=e^{2\pi i \sigma}, \quad   p = e^{2\pi i z}, \ \ \text{and} \ \ Z =\begin{psmallmatrix}
\tau & z \\
z & \sigma
\end{psmallmatrix} \in \IH_2 \label{eq:defSiegelZ}
\end{equation}
one obtains tentative Siegel modular forms.

The partition function for the twisted primitive DT invariants on $X$ is conjecturally given by the negative reciprocal of the Borcherds lift of the corresponding twisted-twined elliptic genera,
\begin{equation}\label{eq:Ztw}
\mathsf{Z}^{\mathrm{tw}}(q,t,p) = - \frac{1}{\tilde{\Phi}_2(Z)} \ ,
\end{equation}
and thus agrees with the quarter-BPS counting function obtained in~\cite{David:2006ji,Jatkar:2005bh}, which is (possibly up to a multiplicative constant) the function $2^{-4}\Phi_{6,3}^{-1}$.

On the other hand, the untwisted primitive DT invariants are determined by
\begin{align}
\mathsf{Z}^{\mathrm{untw}}(q,t,p) &=\frac{-8 F_4(Z) + 8 G_4(Z) - \frac{7}{30} E_4^{(2)}(2Z)}{\chi_{10}(Z)}  \ , \label{eq:Zuntw_quotient}
\end{align}
where $\chi_{10}$ is the weight ten Igusa cusp form appearing already in the partition function of the unorbifolded model, namely DT theory on $ S\times E$, physically IIA[$ S\times E$] or Het[$T^6$]. 
In the numerator we have two Siegel modular forms $G_4(Z)$ and  $E_4^{(2)}(2Z)$, both of weight four for the level two congruence subgroup $\Gamma_{0}^{(2)}(2) \subset \operatorname{Sp}_4(\mathbb{Z})$.  The function $F_4(Z)$  is a weight four Siegel paramodular form of degree two for the paramodular group $K(2)$. All of them can be expressed within the ring of even genus two theta constants, see appendix~\ref{app:siegelmodforms}. Thus, $\mathsf{Z}^{\mathrm{untw}}$ is a weight $-6$ Siegel modular form for the level two Iwahori subgroup $
B(2) = K(2) \cap \Gamma_{0}^{(2)}(2).
$
We remark that with the help of \eqref{eq:SMFY}, \eqref{eq: defTSMF}, \eqref{eq:SMFY2} and \eqref{eq:TYYrelation}, $\mathsf{Z}^{\mathrm{untw}}$ might be recasted into the form
\begin{align}
\mathsf{Z}^{\mathrm{untw}}
& =-\frac{1}{2} \left( \frac{1}{W} \, + \, \frac{16 T}{YW} \right) \nonumber \\ 
& = -\frac{1}{2} \frac{Y + \tfrac{1}{16}Y ' + \tfrac{1}{16}Y ''}{YW} \nonumber \\	
 &=  -\frac{1}{2}\left( \frac{1}{\Phi_{6,0}} + \frac{1}{2^4 \, \Phi_{6,3}} + \frac{1}{2^4\, \Phi_{6,4}} \right)
 \ \label{eq:Zuntw2} \ .
\end{align}

\paragraph{DT invariants as BPS indices.} A connection to physics was already outlined in the appendix of~\cite{Bryan:2018nlv}, which we shall reproduce and build on. 

DT invariants on Calabi-Yau threefolds are believed to give virtual counts of D6-D2-D0 bound states in type IIA theory, which in turn engineer dyonic BPS states. Recall that a BPS D$(2n)$-brane wraps an algebraic $n$-cycle in $X$ and especially has support in $H_{2n}(X,\IZ)$. These D-branes source various components of the dyon charge $(Q,P)$. The translation to the heterotic duality frame 
and others is given in Table \ref{tab:N=4_brane_dualities}, which we have adapted from the K3$\times T^2$ case described in \cite{Cheng:2008gx}. The magnetic charges are sourced by the non-perturbative objects of the parent theory surviving the orbifolding procedure (see~\cite[section 4]{Persson:2015jka}, for instance). Those D4-branes supported on the elliptic curve times a curve in the K3 which survive the orbifold projection are charged in the invariant lattice $H^2(S,\IZ)^g=E_8(-2) \oplus U^{\oplus 3}$. 
Since the sympletic involution on the K3 leaves invariant the $H^0$ and $H^4$ components of the cohomology spanning a $U$ summand, we have simply kept the notation of \cite{Cheng:2008gx} for the D0- and D4(K3)-charges.
The fundamental (heterotic) string winding number F1($3$) along the CHL circle $S^1_{(3)}$ is quantized in units of $\tfrac{1}{2}$ and the momentum p(3) along the CHL circle in integer units, giving rise to $U(\tfrac{1}{2})\subset \Lambda_e$. Moreover, a configuration of two NS5-branes localized in $S^1_{(3)}$, denoted by NS5($\hat{3}$), with a separation of $\delta / 2$ survives the orbifolding, so this charge will be quantized in units of 2 and gives rise to the $U(2)\subset \Lambda_m$ summand. An integer unit of KKM($\hat{3}$) charge belongs to a Kaluza-Klein monopole with the CHL circle $S^1_{(3)}/\IZ_2$ as asymptotic circle.

\begin{table} 
\centering
\begin{tabular}{ccccc}
\toprule
\multicolumn{5}{c}{\bf Electric and magnetic charges $(Q,P) \in \Lambda_{e}\oplus \Lambda_{m}$} \\
\toprule
&Het & IIA &M & IIB \\
$\IZ_2 \backslash$&
\tiny ${S_{(2)}^1\!\!\times\!\! S_{(3)}^1\!\!\times\! \!S_{(4)}^1\!\!\times\! \!T^3}$
&
\tiny $S_{(2)}^1\!\!\times\!\! S_{(3)}^1\!\!\times\!\! K3$
&
\tiny $S_{(1)}^1\!\!\times\!\! S_{(2)}^1\!\!\times\!\! S_{(3)}^1\!\!\times\!\! K3$
& 
\tiny $S_{(1)}^1\!\!\times\!\! S_{(3)}^1\!\!\times\!\! K3$\\
\rowcolor[gray]{0.9}{}
&& &&\\
\multirow{2}*{$U$} 
& p(4) & \cellcolor[gray]{0.8} D0 & p(1) & F1(1) \\
&F1(4) & D4(K3) & M5(1,K3) & NS5(1,K3)\\
\midrule
\multirow{2}*{$U$} 
& p(2) & p(2) & p(2) & 
D1(1) \\
&F1(2) & NS5(2,K3) & M5(2,K3) & 
D5(1,K3)\\
\midrule
\multirow{2}*{$U(\frac{1}{2})$} 
& p(3) & p(3) & p(3) &
p(3) \\
&F1(3) & NS5(3,K3) & M5(3,K3) & KKM($\hat{1}$)\\
\midrule
$E_8(-\frac{1}{2})\! \oplus \! U^{\oplus 3}$& $q_A$ & \cellcolor[gray]{0.8}D2$(\alpha^A)$ & M2$(\alpha^A)$ & D3$(1,\alpha^A)$ \\
\rowcolor[gray]{0.9}{}
&&&&\\
\multirow{2}*{$U$} 
& NS5($\hat{4}$) &\cellcolor[gray]{0.8} D2(2,3) & M2(2,3)& F1(3) \\
&KKM($\hat{4}$) & \cellcolor[gray]{0.8}D6(2,3,K3) & TN(2,3,K3) & NS5(3,K3)\\
\midrule
\multirow{2}*{$U$} 
& NS5($\hat{2}$) & F1(3) &  M2(1,3)& D1(3) \\
&KKM($\hat{2}$)& KKM($\hat{2}$) & KKM($\hat{2}$)& D5(3,K3)\\
\midrule
\multirow{2}*{$U(2)$} 
& NS5($\hat{3}$) & F1(2) & M2(1,2) & 
p(1) \\
&KKM($\hat{3}$) &KKM($\hat{3}$)  & KKM($\hat{3}$)& 
KKM($\hat{3}$)\\
\midrule
$E_8(-2)\! \oplus\!  U^{\oplus 3}$& $p^A$ & D4$(2,3,C_{AB}\alpha^B)$ & M5$(1,2,3,C_{AB}\alpha^B)$ & D3$(3,C_{AB}\alpha^B)$  \\
\bottomrule
\end{tabular}
\caption{
Sources of the dyon charge $(Q,P)$ in different duality frames of the four-dimensional $\mathcal{N}=4$ $\IZ_2$ CHL model. The $\alpha_{A}$'s are a basis of the 14-dimensional lattice $ E_8(-2)\oplus U^{\oplus 3} \cong H^2(S,\IZ)^g$ with bilinear form denoted by $C_{AB}$. (Table adapted from \cite[Table 3.1]{Cheng:2008gx}.)}
\label{tab:N=4_brane_dualities}
\end{table}

Now in the case of primitive DT invariants on $S \times E$ and unit-torsion dyons of IIA[$S\times E$] (or of Het[$T^6$]) an explicit charge assignment $(Q,P)$ subject to the requirement\footnote{We suppress the dependence on the moduli domain
. Also note the relative overall minus sign between eqs. \eqref{defn_part_fn}-\eqref{defn_part_fn2} and \eqref{eq:14BPS_partition_fun_general}.}
\begin{equation}
\DT^{S\times E}_{n,(\gamma ,d)}= f(P^2, Q \cdot P, Q^2 )
\end{equation}
for $(\gamma,d) \in H_2(S\times E,\IZ)$ is given by
\begin{equation}\label{eqn:SEcharges}
Q=(ne_1,0,0,\gamma) \quad \text{and}
\quad P=((d-1)e_1 + e_2,0,0,0) \ .
\end{equation}
Here $e_1$ and $e_2$ denote the generators of the hyperbolic lattice $U$, $n$ is the D0-charge, $\gamma$ the D2-charge. We have a single unit of D6-charge. These charges have been highlighted in Table \ref{tab:N=4_brane_dualities}, where $E_8(-\frac{1}{2})\oplus U^{\oplus 3}$ should be understood as $E_8(-1)^{\oplus 2}\oplus U^{\oplus 3}$ before orbifolding and similar for other sublattices.
Again $f$ expresses the sixth helicity supertrace (the quarter-BPS index) of unit-torsion states in terms of the quadratic T-invariants
\begin{equation}\label{eq:ChargeIdent}
Q^2=\gamma^2 = 2s\, ,\ \ \ P^2=2(d-1)\, , \ \ \ Q\cdot P=n \ .
\end{equation}
Matching notations, we are lead to identify the Siegel coordinate $Z$ in \eqref{eq:defSiegelZ} with the chemical potentials $\cO$ in \eqref{eq:symplecticStrafo} conjugate to the quadratic T-invariants. In the non-orbifold theory on $S\times E$ the quarter-BPS index of the D6-D2-D0 configuration and the DT invariant are both obtained from $1/\chi_{10}$.

Now returning to the CHL model $X$, note that if $\gamma \in \Pi(H_2(S,\IZ))$ then already $\gamma \in \Lambda_e$ since $ \Pi(H_2(S,\IZ)) \subset \left(H^*(S,\IZ)^g\right)^* \subset \Lambda_e$ (c. f. eq. \eqref{eq:projectorIsom} and eqs. \eqref{eqn:chle}, \eqref{eq:LambdaEMZ2chl}).  Thus, the charges assigned in \eqref{eqn:SEcharges} indeed belong to the CHL electric lattice \eqref{eqn:chle} and CHL magnetic lattice \eqref{eqn:chlm}, respectively. In other words, the assignment is still meaningful.

Moreover, for primitive untwisted $\gamma \in H_2(S,\IZ)^g = E_8(-2)\oplus U^{\oplus 3}$, the charge assignment \eqref{eqn:SEcharges} gives electric charge with $\mathcal{P}=0$.
So regarding DT invariants $\DT^X_{n, (\gamma,d)}$, we may expect that the charge formulas \eqref{eqn:SEcharges} are still valid for the orbifold case $X=(S\times E) / \IZ_2$ if $\gamma \in H_2(S,\IZ)^g$. 
However, the function \eqref{eq:Zuntw2} is not found amongst the 
untwisted sector partition functions in \eqref{eq:g2untwTerms2} (nor amongst those of the twisted sectors in \eqref{eq:g2twTerms2}). 
Formally, the function \eqref{eq:Zuntw2} is the \textit{average} of the modular forms $\Zo$ and $\Zp$. 
In \eqref{eq:g2untwTerms2} these two functions belong to orbits $(-1)^{Q\cdot \delta}=+1$ and $-1$, respectively (but both with $\mathcal{P}=0$ and $h=0$). 
Alternatively, for fixed value $(-1)^{Q\cdot \delta}=+1$ the functions $\Zo$ and $\Zp$ distinguish between the $h=0$ and $h=1$ case, respectively (i.e., the $\mathcal{P}=0$ terms of \eqref{eq:g2untwTerms2} and \eqref{eq:g2twTerms2}, respectively).  
Note also that the charge residue component $((-1)^h,(-1)^{Q\cdot \delta}) \in U(\frac{1}{2})/U(2)$ is apparently independent of any D-brane charges in the type IIA theory (c.f. Table \ref{tab:N=4_brane_dualities}) and especially the heterotic CHL winding number is not seen by the type II D-branes (nor in the data specifying the DT invariant). In any case, there does not seem to be a unique charge (orbit) whose partition function reduces to $\Zuntw$, but rather a pair (union) thereof. 

For a primitive twisted class $\gamma \in E_8(-\frac{1}{2})\oplus U^{\oplus 3}$ (with $\mathcal{P}\neq 0$) the DT formula for $\mathsf{Z}^{\text{tw}}$ is not in tension with the results of \eqref{eq:g2untwTerms2} for the respective quarter-BPS generating functions $\mathsf{Z}^{\mp}$, since the two possible cases for $\mathcal{P}\in \mathcal{O}_{248} \cup \mathcal{O}_{3875}$ via \eqref{eq:Q2parity} belong to different modes in the Fourier expansion of $\mathsf{Z}^{\text{tw}}$, collected in $\mathsf{Z}^{\mp}$. 
Formally, this again agrees with the (in this case trivial) average over $(-1)^{Q\cdot \delta}$ ($h=0$ fixed) for each Weyl orbit of $\mathcal{P}$ or, alternatively, the average over $h=0,1$ ($(-1)^{Q\cdot \delta}=+1$ fixed).

Whether the DT invariants computed in \cite{Bryan:2018nlv} really should be interpreted as averages of suitable quarter-BPS indices or whether the relation is more subtle than that remains an interesting open question to be clarified by future research.


\section{Conclusion}
\label{sec:conclusion}

In this paper we have studied the spectrum of quarter-BPS states in the $\IZ_2$ CHL model. Partition functions for unit-torsion dyons in various charge classes, especially such with untwisted sector electric charge, have been (re-)derived from a chiral genus two orbifold partition function in the heterotic frame, paralleling and refining the original derivation~\cite{Dabholkar_2007} for well-known twisted sector dyons. Stringent  constraints coming from charge quantization, wall-crossing and S-duality symmetry are shown to be satisfied. As we have argued, these do not only elucidate the role of the Siegel modular symmetry group underlying the counting function, but also allow for a ``modular bootstrap derivation'' of the latter in cases where the appropriate ring of modular forms is sufficiently well understood. As a partition function of black hole microstates each also reproduces the correct macroscopic entropy of a dyonic extremal black hole in the large charge limit, including the Bekenstein-Hawking term and the first power-suppressed correction that can be accounted for by the inclusion of the Gauss-Bonnet term in the 4D effective action. 
According to $\mathcal{N}=4$ heterotic-type IIA duality, these dyons have type IIA realizations on $X=( \text{K3} \times T^2 )/ \IZ_2$.
In this perspective BPS indices are expected to be closely related to Donaldson-Thomas invariants of the CHL orbifold $X$. For the primitive ``untwisted'' DT partition function conjectured in~\cite{Bryan:2018nlv} an alternative, but equivalent, expression is found, which opens up a tentative interpretation of DT invariants as suitable sums of BPS indices. It is an important open question whether this interpretation is indeed correct or whether there are further subtleties missed out here, which eventually could restore a more conservative interpretation in which a DT invariant really \textit{is} a quarter-BPS index of a suitably chosen charge (orbit), and not a sum of such.

Let us comment on other possible extensions of the $\IZ_2$ model analysis presented here. In many points this means performing checks and derivations that were hitherto only done with the twisted sector counting function (possibly because handy closed formulae were not known otherwise). One can ask whether the untwisted sector partition function(s) can also be derived in the type IIB frame, say from a rotating D1-D5-Kaluza-Klein monopole-system similar to the twisted sector analysis in~\cite{David_2006}, or from an M-theory compactification.
 
A careful analysis of the contour prescription and the asymptotic growth of the BPS indices, which addresses further (say exponentially suppressed) corrections to the black hole entropy, also seems relevant for comparing the microscopic degeneracies to the macroscopic degeneracies in quantum gravity.
Roughly speaking, compatibility with discrete duality groups (as seen in string theory) demands special arithmetic properties of the black hole degeneracies, such as the dependence on discrete duality invariants and the formation of distinct charge orbits. For instance, higher torsion configurations are known to give rise to exponentially suppressed terms in the entropy, which in the quantum gravity description correspond to orbifold geometries in the AdS${}_2$ path integral (see~\cite{Gomes:2017Uduality} for recent progress). It would be interesting to see whether the entropy of twisted sector dyons can be distinguished from the one of untwisted sector dyons in subleading contributions and how to account for this distinction (non-perturbatively) on the quantum gravity side. 

Furthermore, it has been conjectured in~\cite{Sen:2010mz} that for single-centered black holes with regular horizon, i.e., quarter-BPS states subject to a certain restriction on their quadratic charge invariants, the quarter-BPS index should be positive. Numerical evidence was given in~\cite{Sen:2010mz} for low lying (twisted sector) charges in the order $N\in \lbrace 1,2,3,5,7\rbrace$ orbifolds, and for $N$ non-prime in~\cite{Chattopadhyaya:2017ews}.  For the unorbifolded theory (i.e., $N=1$), this conjecture has rigorously been proven for a subset of the corresponding charges~\cite{Bringmann:2012zr}.\footnote{In the case of $\mathcal{N}=4$ type II toroidal orbifolds the conjecture, however, seems to be violated~\cite{Chattopadhyaya:2018xvg}.} We leave it to future work to check the conjecture for the untwisted sector quarter-BPS states considered here. 

Clearly, there is also room for extending the present analysis to untwisted sector dyons in other $\mathcal{N}=4$ CHL models. On one side, this includes models for symplectic K3 automorphisms for higher (prime or composite) order $N$ as well as non-cyclic orbifold groups. On the other side, we may consider CHL models arising from ``non-geometric'' symmetries of a K3 non-linear sigma model~\cite{Gaberdiel:2011fg,Persson:2015jka}. This could lead to interesting enumerative predictions for the symplectic invariants of these theories, as was also remarked in~\cite{Bryan:2018nlv}.\footnote{In addition,~\cite{Persson:2017lkn} proposed a yet broader notion of CHL compactifications. Speculatively, there might be CHL versions of heterotic strings on $T^8$ whose BPS spectrum could have an enumerative geometric meaning on appropriate Calabi-Yau duals, say on K3$\times T^4$ or orbifolds thereof. This was also speculated in~\cite{Harrison:2016pmb}, where for the case of heterotic strings on $T^8$ a BPS counting function was  written down in terms of the Borcherds automorphic form $\Phi_{12}$. It is natural to guess the appearance of similar automorphic forms in the corresponding CHL orbifolds.}

Since one of the biggest challenges in counting stable BPS states in different regions of the moduli space 
and relating them to the black hole entropies is to consider less supersymmetric --- in particular $\mathcal{N}=2$ ---
compactifications, let us conclude here with a comparison of the formalism used in this paper with recent 
BPS counts on related  $\mathcal{N}=2$ Calabi-Yau geometries. The closest  $\mathcal{N}=2$ cousin  
of the CHL $(\text{K3}\times  T^2)/\IZ_N$ compactification is the Ferrara--Harvey--Strominger--Vafa (FHSV) compactification~\cite{Ferrara:1995yx}, which  is 
likewise a  $(\text{K3}\times  T^2)/\IZ_2$,  with the difference that the $\IZ_2$ action is the Enriques involution on the $\text{K3}$ and the 
hyperelliptic involution of $T^2$ leaving the four branch points fixed. The  threefold has a $\text{K3}$ fibration with four 
Enriques fibers, $\text{SU}(2)\times \IZ_2$ holonomy,  Euler number $\chi=0$  and  a considerably milder scale dependence 
than generic  $\mathcal{N}=2$ compactifications. In particular it has no genus zero world-sheet instantons and the 
D2-D0-brane states that are  counted by the topological string theory at higher genus are related  to the direct 
integration of  the  holomorphic anomaly~\cite{Weiss:2007tk}, starting with the automorphic function 
constructed by a Borcherds lift as product formula in~\cite{BorcherdsInfProd} for the K3 fiber and the Dedekind eta function for the 
base~\cite{Harvey:1996ts,Klemm:2005pd}.  In $\mathcal{N}=2$ compactifications the duality symmetries are
realized in a much more complicated  way, which makes it hard to extract information about different brane states 
from the same modular object. However, in~\cite{Weiss:2007tk} it was observed that the automorphic form 
of~\cite{BorcherdsInfProd} has two cusp expansions: one yielding the D2-D0-brane states and one 
involving light D4-brane states  wrapping the K3 fiber. It would be very important to understand the 
non-perturbative  completion of the FHSV model by an extension of the  duality and wall-crossing 
arguments used here for the CHL orbifold model. 

Heterotic strings on CHL $\IZ_N$ orbifolds are dual to type IIA compactifications on elliptically fibered K3 
with $N$ sections, completed by the $T^2$ and modded out by the shift symmetry~\cite{Bryan:2018nlv}. 
As mentioned the primitive DT partition functions  $\mathsf{Z}_N$ of~\cite{Bryan:2018nlv} are related to  the inverse of Siegel modular forms  
of (Iwahori) congruence subgroups of ${\rm Sp}_4(\IZ)$. These forms generalize the weight ten Igusa cusp form of  
${\rm Sp}_4(\IZ)$, whose inverse $\mathsf{Z}_1$ describes the BPS states of $\text{K3} \times T^2$.   
Moreover, the Fourier expansions of  $\mathsf{Z}_N$ specialize in the non-perturbative limit to inverses of the cusp forms $\Delta_N(\tau)$ of $\Gamma(N)$ of weight $\lceil \frac{24}{N+1}\rceil$. 
For $N$ prime these are the same cusp forms that have been obtained in the leading non-perturbative expansion in~\cite{Jatkar:2005bh}, see eq. (2.18) there, and they are associated to the action of $\Gamma_1(N)$ on the perturbative CHL string.  On the ${\cal N}=2$ side it has been realized in~\cite{Huang:2015sta} that on elliptically fibered  Calabi-Yau threefolds with   
one section and no singularities from the fiber, the topological string has an expansion in terms of Jacobi forms 
that bears similarity to the expansion of $\mathsf{Z}_1$ in terms of Jacobi forms  of $\SL_2(\IZ)$  as discussed in 
section 5  of~\cite{dabholkar2012quantum}, with the difference that the index of the Jacobi form grows cubic 
with the exponent of the expansion parameter $Q_\beta$ rather than linear as 
in~\cite{dabholkar2012quantum}. This fact has been used to check the microscopic 
entropies of spinning black holes~\cite{Haghighat:2015ega}.  In~\cite{Cota:2019cjx}  the work 
of~\cite{Huang:2015sta} has been extended to genus one fibrations with $N$-sections as well as
to elliptic fibrations to $N$ sections in the limit of the K\"ahler parameters that corresponds to the 
additional sections. In the former case Jacobi forms of $\Gamma_1(N)$ occur in the BPS expansion 
related  to D2-D0-branes also for non-prime $N$, while in the latter case these BPS  expansions  are expressible in terms of 
Jacobi forms  of $\Gamma(N)$ and in particular the cusp forms $\Delta_N(\tau)$ occur in the 
denominators.  This indicates that some features of the analysis of BPS degeneracies 
related to the more complicated fibration structures carry over from the ${\cal N}=4$ to the ${\cal N}=2$ 
case. \\                                        
                             
\acknowledgments
\label{ack}
We would like to thank Georg Oberdieck for useful explanations concerning his work and Cesar Fierro Cota for discussion on Jacobi forms of congruence subgroups of the modular group.  Furthermore, we thank Boris Pioline and an anonymous referee for helpful comments on the manuscript. F. F. and C. N. also thank the Bonn-Cologne Graduate School of Physics and Astronomy (BCGS) for generous support.



\appendix

\section{Siegel modular forms}
\label{app:siegelmodforms}

In this appendix we collect basic definitions and useful formulae for the Siegel modular forms appearing in the main text. Our main references are ~\cite[chapter VII]{freitag2009funktionentheorie},~\cite[section 2]{Bryan:2018nlv} and~\cite[appendix A]{Bossard_2019}, also see~\cite{dabholkar2012quantum} for a review that emphasizes the relation between the theory of Siegel modular forms, mock modular forms and quantum black holes. 

\paragraph{Preliminaries.} 
By  $\Sp_4(\IZ) $ we denote the symplectic group of integer $4\times 4$ matrices $M=\begin{psmallmatrix}
A & B \\C&D
\end{psmallmatrix}$ that satisfy
\begin{equation}\label{eq:SympProp}
M^\tp J M = J \  \qquad \text{and} \qquad J=\begin{pmatrix}
0 & \mathbb{1}_2 \\ -\mathbb{1}_2 & 0
\end{pmatrix} \ ,
\end{equation}
which is equivalent to 
\begin{equation}
A^\tp C = C^\tp A \ , \quad B^\tp D = D^\tp B \  \quad \text{and} \quad A^\tp D - C^\tp B = \mathbb{1}_2 \ 
\end{equation}
 for the $2\times 2$ block matrices in $M$. The groups $\Sp_4(\IQ)$ and $\Sp_4(\IR)$ are defined analogously.
 If $M \in \Sp_4(\IZ)$ as above then the inverse of $M$ is given by
  \begin{equation}\label{eq:SpInv}
M^{-1}= \begin{pmatrix}
D^\tp & -B^\tp\\ -C^\tp & A^\tp
\end{pmatrix} \ 
 \end{equation} 
and by using  this in (\ref{eq:SympProp}) we see that also $M^\tp\in  \Sp_4(\IZ)$.  Taking the Pfaffian  and using ${\rm Pf}(M^\tp J M)= 
\det(M) {\rm Pf} (J)$ one concludes that $\det(M)=1$, which more conceptually is equivalent to the fact that symplectic 
transformations are orientation preserving.

Special examples of symplectic matrices that also play a role for the quarter-BPS partition functions are (for $\IK=\IZ, \IQ, \IR$, respectively)
\begin{align}\label{eq:SympEx1}
\begin{pmatrix}
\mathbb{1}_2 & S \\ 0 & \mathbb{1}_2
\end{pmatrix} \qquad \quad & \text{with} \quad S^\tp = S \\
\text{and} \qquad \begin{pmatrix}
U^\tp & 0 \\ 0 & U^{-1}
\end{pmatrix} \qquad & \text{with} \quad U\in \mathrm{GL}_2(\IK) \ . \label{eq:SympEx2}
\end{align}
Any symplectic matrix with $C=0$ can be written as a product of the form ``\eqref{eq:SympEx2} times \eqref{eq:SympEx1}''. 
The prinicipal congruence subgroup $\Gamma^{(2)} (N)$ (with $N\geq 1$) is defined by
\begin{equation}
\Gamma^{(2)} (N) = \Bigg \lbrace \begin{pmatrix}
A & B \\C&D
\end{pmatrix} \in  \Sp_4(\IZ) \, \Bigg|  \begin{pmatrix}
A & B \\C&D
\end{pmatrix}  \equiv \begin{pmatrix}
\mathbb{1}_2 & 0 \\ 0 & \mathbb{1}_2 
\end{pmatrix} \mod N \Bigg  \rbrace \ .
\end{equation}
A congruence subgroup $\Gamma \subset \Sp_4(\IZ)$ is a subgroup that contains a principal congruence subgroup, for instance,
\begin{equation}
\Gamma_0^{(2)}(N) =
\Bigg \lbrace
\begin{pmatrix} A&B \\ C&D \end{pmatrix} \in  \Sp_4(\IZ)  \, \Bigg| \, C \equiv 0 \mod  N \Bigg  \rbrace \supset \Gamma^{(2)}(N) \ .
\end{equation}
For a prime number $p \geq 1$ the group $K(p)$ is defined by~\cite{Ibukiyama1997, gritsenko1995minimal}
\begin{equation}
 K(p) = \Sp_4(\IQ) \cap 
\begin{pmatrix}
\IZ & \IZ & p^{-1} \IZ& \IZ \\ 
p \IZ & \IZ & \IZ& \IZ \\ 
p \IZ & p \IZ & \IZ& p \IZ \\ 
p \IZ & \IZ & \IZ& \IZ
\end{pmatrix},
\end{equation}
while the Iwahori subgroup is defined by the intersection
\begin{equation}\label{eq:defBpIwahori}
 B(p) = K(p) \cap \Gamma_0^{(2)}(p)
=
 \Sp_4(\IZ) \cap 
\begin{pmatrix}
\IZ & \IZ & \IZ& \IZ \\ 
p \IZ & \IZ & \IZ& \IZ \\ 
p \IZ & p \IZ & \IZ& p \IZ \\ 
p \IZ & p \IZ & \IZ& \IZ
\end{pmatrix} \ .
\end{equation}
By conjugation in $\mathrm{GL}_4(\IQ)$ (see~\cite{Ibukiyama1997} for references) the group $K(p)$ is related to the Siegel paramodular group $\Gamma^{\mathrm{para}}(p)$, formed by integer $4\times 4$ matrices that obey \eqref{eq:SympProp} with $J$ replaced by $J_2(p) = \begin{psmallmatrix}
0 & P \\ -P & 0
\end{psmallmatrix}$ with $P=\mathrm{diag} (1,p)$.

Let $\IH_2$ be the (genus two) Siegel upper half space, i.e., the set of $2\times 2$ symmetric complex matrices 
\begin{equation}
Z =
\begin{pmatrix} \tau & z \\ z & \sigma \end{pmatrix}  
\end{equation} 
with positive definite imaginary part, explicitly
\begin{equation}
\Im (\tau) > 0, \quad \Im(\sigma) > 0, \qquad \text{and} \quad \Im(\tau)\Im(\sigma) - \Im(z)^2 >0 \ .
\end{equation}
A group action of $\Sp_4(\IR)\ni M, M'$ on $\IH_2 \ni Z$ is defined by
\begin{equation}
 MZ \coloneqq (AZ+B) (CZ + D)^{-1} \  \ ,
\end{equation}
where $M$ and $M'$ define the same action if and only if they differ by their sign. The special examples \eqref{eq:SympEx1} and \eqref{eq:SympEx2} above act via
\begin{align}
Z \mapsto Z+S  \qquad \text{and} \qquad Z \mapsto U^\tp Z U \ ,
\end{align}
respectively. Important for wall-crossing relations are the following embedded, commuting $\SL_2(\IR)$ subgroups of $\Sp_4(\IR)$:
\begin{align}
\SL_2(\IR)_\tau \ : \ \begin{pmatrix}
a & b \\ c & d
\end{pmatrix}_\tau &= \begin{pmatrix}
a & 0 & b & 0 \\ 0 & 1 & 0 & 0 \\ c & 0 & d & 0 \\ 0 & 0& 0 & 1
\end{pmatrix} \label{eq:SL2embedded1}
\\
\SL_2(\IR)_\sigma \ : \ \begin{pmatrix}
a & b \\ c & d
\end{pmatrix}_\sigma &= \begin{pmatrix}
1 & 0 & 0 & 0 \\ 0 & a & 0 & b \\0& 0 & 1 & 0 \\ 0 & c& 0 & d
\end{pmatrix} . \label{eq:SL2embedded2}
\end{align}
Their action on the Siegel coordinate $Z$ is given by
\begin{equation}
\begin{pmatrix}
a & b \\ c & d
\end{pmatrix}_\tau Z = \begin{pmatrix}
\frac{a \tau + b}{c \tau + d} & \frac{z}{c\tau + d} \\\frac{z}{c \tau + d} & \sigma- \frac{c z^2}{c\tau + d}
\end{pmatrix}
\end{equation}
and
\begin{equation}
\begin{pmatrix}
a & b \\ c & d
\end{pmatrix}_\sigma Z = \begin{pmatrix}
\tau- \frac{c z^2}{c \sigma + d} & \frac{z}{c \sigma + d} \\ \frac{z}{c \sigma+ d} &  \frac{a \sigma + b}{c \sigma + d}
\end{pmatrix} \ ,
\end{equation}
respectively. From these expressions it follows that the diagonal locus $z=0$ is preserved under the two embedded subgroups, where they operate componentwise on $\tau \in \IH_1$ and $\sigma\in \IH_1$, respectively. Another symplectic transformation preserving the diagonal locus is given by \eqref{eq:SympEx2} with $U=\begin{psmallmatrix}
0 & 1 \\ 1 & 0
\end{psmallmatrix}$, which exchanges the diagonal entries of $Z$.

Now let $f : \IH_2 \rightarrow \IC$ be a holomorphic function, $k$ be an integer and $\Gamma \subset \Sp_4(\IZ)$ be a congruence subgroup (or a discrete subgroup $\Gamma \subset \Sp_4(\IR)$ with finite covolume~\cite{Ibukiyama1997,MR1082834}).
If
\begin{equation}
f(M Z) = \det(C Z + D)^k f(Z) 
\end{equation}
for all $M = \binom{A\ B}{C\ D} \in \Gamma$, then $f$ is called a Siegel modular form of weight $k$ for $\Gamma$.
As in~\cite{Bryan:2018nlv} denote by $\Mod_k^{(2)}(\Gamma)$ the space of Siegel modular forms of weight $k$ for $\Gamma$ and by
\begin{equation}
\Mod^{(2)}(\Gamma) = \bigoplus_{k} \Mod_k^{(2)}(\Gamma)
\end{equation}
the $\IC$-algebra of Siegel modular forms for $\Gamma$. Also introduce the Petersson slash operator for a function $f : \IH_2 \to \IC$, an element $M \in \Sp_4(\IR)$ and an integer $k$ via
\begin{equation}
(f\big|_k M) (Z) = \det(C Z + D)^{-k}\,  f((AZ+B) (CZ + D)^{-1}) \ .
\end{equation}
Then $f \in \Mod_k^{(2)}(\Gamma)$ is equivalent to $f\big|_k M = f $ for all $M\in \Gamma$.\footnote{Here we only deal with the case of a trivial multiplier system. 
 } One often simply writes $f\big| M$. If \eqref{eq:SympEx2} lies in $\Gamma$ for $U=\begin{psmallmatrix}
0 & 1 \\ 1 & 0
\end{psmallmatrix}$, such $f(Z)$ is invariant under exchange of the diagonal entries of $Z$ (possibly up to a root of unity).

\paragraph{Modular forms for level two subgroups.} 
Generators for rings of modular forms can often be expressed in terms of genus two theta constants (german Thetanullwerte), which we introduce now. For column vectors $m'=a=\begin{psmallmatrix}
a_1 \\ a_2
\end{psmallmatrix},\ m''=b=\begin{psmallmatrix}
b_1 \\ b_2
\end{psmallmatrix} \in \IZ^2$ and $m=\binom{m'}{m''}$ consider the genus two theta constant of characteristic $m$
\begin{equation}\label{eq:defThetaConst}
\theta_m(Z)
=
\sum_{x \in \IZ^2} e\left( \frac{1}{2} \left(x + \frac{1}{2} m'\right)^\tp Z \left(x + \frac{1}{2} m'\right)
+ \left(x + \frac{1}{2} m'\right)^\tp \frac{m''}{2} \right)
\end{equation}
with shorthand $e(z) = \exp(2 \pi i z)$ for $z \in \IC$. This is also written as $\theta\begin{bsmallmatrix} a \\ b \end{bsmallmatrix} = \theta_{a_1 a_2 b_1 b_2}$. The theta constants vanish identically iff $a^\tp b \mod 2$ is odd. For genus two there are precisely ten ``even'' non-trivial theta constants. There is a useful transformation formula under $M\in \Sp_4(\IZ)$,
\begin{equation}
\theta_{M \{m \} }(MZ)= v(M,m) \det(CZ+D)^{1/2} \theta_{m}(Z)\ ,
\label{genus2trafo}
\end{equation}
where $ v(M,m)$ is an eigth root of unity and, denoting by $(...)_0$ the diagonal as a column vector,
\begin{equation}
M \{m \} =  M \{ \begin{psmallmatrix}
 a \\ b
 \end{psmallmatrix} \} = M^\mtp \begin{psmallmatrix}
 a \\ b
 \end{psmallmatrix}  + \begin{psmallmatrix}
 (CD^\tp)_0 \\ (AB^\tp)_0
 \end{psmallmatrix}  \ \mod 2 \ .
\end{equation}
 As special cases we have for the elements in \eqref{eq:SympEx1} and \eqref{eq:SympEx2} simplified formulas
\begin{align}
\theta\begin{bsmallmatrix} a \\ b \end{bsmallmatrix} (Z+S) &=  \theta\begin{bsmallmatrix} a \\ b+ S a + S_0 \end{bsmallmatrix} (Z)\cdot e^{\frac{i \pi}{4} a^\tp S a} \label{eq:thetatransf1}\\ \text{and} \qquad 
\theta\begin{bsmallmatrix} a \\ b \end{bsmallmatrix} (U^\tp Z U) &=  \theta\begin{bsmallmatrix} U a \\ U^\mtp b \end{bsmallmatrix} (Z) \label{eq:thetatransf2} \ .
\end{align}
On the diagonal $z=0$ the theta constants factorize as
\begin{equation}
\theta\begin{bsmallmatrix} a \\ b \end{bsmallmatrix} (\begin{psmallmatrix}
\tau & 0 \\ 0 & \sigma
\end{psmallmatrix}) = \theta\begin{bsmallmatrix} a_1 \\ b_1 \end{bsmallmatrix} (
\tau ) \, \theta\begin{bsmallmatrix} a_2 \\ b_2 \end{bsmallmatrix} (
\sigma ) \ .
\label{limit1}
\end{equation}
For $\begin{bsmallmatrix}  1& 1 \\ 1& 1 \end{bsmallmatrix}$ this vanishes linearly in $z \rightarrow 0$, more precisely
\begin{equation}
\theta \begin{bsmallmatrix}  1& 1 \\ 1& 1 \end{bsmallmatrix} (\begin{psmallmatrix}
\tau & z \\ z & \sigma
\end{psmallmatrix}) \rightarrow \frac{z}{2 \pi i} \, \theta' \begin{bsmallmatrix} 1 \\ 1 \end{bsmallmatrix} (
\tau ) \, \theta' \begin{bsmallmatrix} 1 \\ 1 \end{bsmallmatrix} (
\sigma ) \ , \qquad  \text{with} \quad \theta' \begin{bsmallmatrix} 1 \\ 1 \end{bsmallmatrix} = 2\pi \eta^3 \ .
\label{limit2}
\end{equation}
In these expressions we used standard genus one theta constants defined in complete analogy to \eqref{eq:defThetaConst} (read: sum over $x \in \IZ$, $Z\in \IH_1$, $m', m'' \in \IZ$). Special instances, labelled by
\begin{equation}\label{eq:thetagenus1}
\theta\begin{bsmallmatrix} 1 \\ 0 \end{bsmallmatrix} = \theta_2, \qquad 
\theta\begin{bsmallmatrix} 0 \\ 0 \end{bsmallmatrix} = \theta_3, \qquad \text{and} \quad
\theta\begin{bsmallmatrix} 0 \\ 1 \end{bsmallmatrix} = \theta_4  ,
\end{equation}
relate to the eta-products of the $\IZ_2$ CHL orbifold partition function via
\begin{equation}
\theta_2^4(\tau)  \eta^4(\tau) = 2^4 \eta^8(2\tau) \ , \ \ \
\theta_3^4(\tau)  \eta^4(\tau) = -e^{2\pi i / 3} \eta^8( \tfrac{\tau +1}{2}) \ \ \text{and} \ \ \
\theta_4^4(\tau)  \eta^4(\tau) = \eta^8(\tfrac{\tau}{2}) \ .
\label{etaproducts}
\end{equation}
and satisfy
\begin{align}\label{eq:theta234idsA}
\theta_2^4(\tau)- \theta_3^4(\tau) + \theta_4^4(\tau)&=0    \qquad \text{(Riemann identity)}\ , \\
\theta_2(\tau) \theta_3(\tau)  \theta_4(\tau)-2 \eta^3(\tau)  &= 0 \qquad \text{(Jacobi triple product identity)} \label{eq:theta234idsB} \ .
\end{align}

Now set\footnote{It should be clear from the context whether the symbol $Z=\begin{psmallmatrix}
\tau & z \\ z & \sigma
\end{psmallmatrix}$ is referring to a coordinate for $\IH_2$ or the Siegel modular  form $Z$ defined in \eqref{eq:defZSMF}.}
\begin{align}
X & = 2^{-2}\left( \theta_{0000}^4 + \theta_{0001}^4 + \theta_{0010}^4 + \theta_{0011}^4 \right) \label{eq:SMFY1}\\
Y & = \left( \theta_{0000} \theta_{0001} \theta_{0010} \theta_{0011} \right)^2 \label{eq:SMFY} 
\\
Z & = 2^{-14} (\theta_{0100}^4 - \theta_{0110}^4)^2 \label{eq:defZSMF}
\\
W & = 2^{-12} (\theta_{0100} \theta_{0110} \theta_{1000} \theta_{1001} \theta_{1100} \theta_{1111})^2  \\
T & = 2^{-8} \, (\theta_{0100} \theta_{0110})^4 \ . \label{eq: defTSMF} 
\end{align}
As was proven in~\cite{MR1082834} (see also~\cite{aokiibuki2005}), the functions $X,Y,Z,W$ are Siegel modular forms for $\Gamma_0^{(2)}(2)$ of respective weight $2,4,4$ and $6$ and they generate the ring of even weight modular forms for $\Gamma_0^{(2)}(2)$, i.e.,
\begin{equation}
\Mod_{\text{even}}^{(2)}(\Gamma_0^{(2)}(2)) = \IC[ X, Y, Z,W] \ .
\end{equation}
The function $W$ agrees with the function ``$K$'' defined in~\cite{MR2806099}. On the other hand, the function $T$ is a weight four modular form for the Iwahori subgroup $B(2)$ and by~\cite{MR1082834} the structure of the ring of even weight modular forms for $B(2)$ is known to be
\begin{equation}\label{eq:B2ring}
\Mod^{(2)}_{\mathrm{even}}(B(2)) = \IC[X,Y,Z,W,T] \cong \IC[x,y,z,w,t] / j \ ,
\end{equation}
where $x,y,z,w,t$ are five algebraically independent variables and $j$ is the ideal of $\IC[x,y,z,w,t]$ spanned by
\begin{equation}
64 w^2 + 16 xtw + t (-16yz+t [x^2 - y - 1024 z - 64 t] ) \ .
\end{equation}
For the structure of the ring of modular forms for $K(2) \supset B(2)$ we refer to the results given in~\cite{Ibukiyama1997} and just mention that the function $F_4(Z)$ appearing in the untwisted sector quarter-BPS partition function is the unique weight four Siegel modular form for $K(2)$, which may be defined as
\begin{align}
 F_4(Z) &= \frac{1}{960} ( X^2 + 3 Y + 3072 Z + 960 T ) \label{eq:defF4} \ .
\end{align} 
Also the Siegel modular form $G_4(Z)$ appears in the untwisted sector partition function, which satisfies
\begin{align}\label{eq:defG4}
G_4(Z) &= \frac{1}{120} X^2 - \frac{3}{80} Y - \frac{12}{5} Z \quad \in \Mod_4^{(2)}(\Gamma_0^{(2)}(2)) \ .
\end{align}

As in the genus one case, the theta function $\Theta_{E_8}^{(2)}$ for the $E_8$ root lattice yields a (Siegel) Eisenstein series and we have the following expressions in terms of theta constants:
\begin{align}
E_4^{(2)}(Z) & = 4 X^2 - 3Y + 12288 Z \quad  \in \Mod_{4}^{(2)}(\Sp_4(\IZ)) \label{eq:defE4Z}\\
E_4^{(2)}(2Z) & = \frac{1}{4} X^2 + \frac{3}{4} Y - 192 Z  \quad \ \, \in \Mod_{4}^{(2)}(\Gamma_0^{(2)}(2))  \label{eq:defE42Z} \ .
\end{align}
These both appear in section \ref{sec:genustwo}, along with closely related functions $(E_4^{(2)}(2Z))|_{M}$ for appropriate $ M\in \Sp_4(\IZ)\backslash \Gamma_0^{(2)}(2)$, which we give in the form
\begin{align}
 \Theta_{E_8}^{(2)}( 2\tau ,z ,  \tfrac{\sigma}{2})=&\ 2^{-4}\sum_{\substack{(Q_1,Q_2)\in\\ E_8(2) \oplus E_8(2)^*}}\,e^{i\pi Q^r \Omega_{rs} Q^s} \label{eq:ThetaE8halfdouble}
 \\
 \Theta_{E_8}^{(2)}(2\tau ,z ,  \tfrac{\sigma +1}{2})=&\ 2^{-4}\sum_{\substack{(Q_1,Q_2)\in\\ E_8(2) \oplus E_8(2)^*}}(-1)^{Q_2^2}\,e^{i\pi Q^r \Omega_{rs} Q^s} \ . \label{eq:ThetaE8halfdoubleB}
\end{align}

All of these may again be expressed in terms of theta constants. We note that in the limit $z=0$ these reduce to products of the genus one theta series for the $E_8$ root lattice or related functions, which we list here for convenience:
\begin{align}
\theta_{E_8(1)}(\tau)& = \frac{1}{2}\left( \theta_2^8 + \theta_3^8 + \theta_4^8 \right)\label{eq:thetaE8theta}\\
\theta_{E_8(1)}(2\tau)& = \frac{1}{2^4}\left( \theta_3^8 + \theta_4^8 + 14 \theta_3^4 \theta_4^4 \right) \\
\theta_{E_8(1)}(\tfrac{\tau}{2})& =\theta_2^8 +\theta_3^8 +  14 \theta_2^4 \theta_3^4\\
\theta_{E_8(1)}(\tfrac{\tau + 1}{2})& =\theta_2^8 +\theta_4^8 -  14 \theta_2^4 \theta_3^4 \ .
\end{align}
Besides those, of interest are also
\begin{align}
\theta_{E_8(2),1}(\tau) &= \frac{1}{2^4}\left( \theta_3^8 + \theta_4^8 + 14 \theta_3^4 \theta_4^4 \right)&  &\!\!\!\! = \frac{1}{2^4}\left( \theta_2^4 \theta_3^4 + 16 \, \theta_3^4 \theta_4^4 - \theta_2^4 \theta_4^4  \right)& & \\
\theta_{E_8(2),248}(\tau)& = \frac{1}{2^4}\left( \theta_3^8 -  \theta_4^8 \right) & &  \!\!\!\! = \frac{1}{2^4}\left( \theta_2^4 \theta_3^4 + \theta_2^4 \theta_4^4 \right) & & \\
\theta_{E_8(2),3875}(\tau)& = \frac{1}{2^4} \ \ \theta_2^8  & & \!\!\!\! = \frac{1}{2^4}\left( \theta_2^4 \theta_3^4 - \theta_2^4 \theta_4^4 \right) & & \ ,
\end{align}
with the notation of eq. \eqref{eq:defthetaE8P}, and the two sets are related via
\begin{align}
\theta_{E_8(1)}(2\tau)& = \theta_{E_8(2),1}(\tau) \\
\theta_{E_8(1)}(\tfrac{\tau}{2})& = \theta_{E_8(2),1}(\tau)+ 120 \, \theta_{E_8(2),248}(\tau) + 135\, \theta_{E_8(2),3875}(\tau)\\
\theta_{E_8(1)}(\tfrac{\tau + 1}{2})& = \theta_{E_8(2),1}(\tau) - 120 \, \theta_{E_8(2),248}(\tau) + 135\, \theta_{E_8(2),3875}(\tau) \ . \label{eq:thetaE8taup1half}
\end{align}

Coming back to Siegel modular forms, the Igusa cusp form $\chi_{10} \in \Mod_{10}^{(2)}(\Sp_4(\IZ))$, whose reciprocal counts unit-torsion dyons in Het[$T^6$], is given by the well-known product of the squares of all even genus two theta constants
\begin{equation}
\chi_{10}(Z)  = Y W \label{eq:defchi10} \ .
\end{equation}
In the $\IZ_2$ orbifold we also encounter the $\Gamma_0^{(2)}(2)$ cusp form $\Phi_{6,0}=W$ and its modular images 
\begin{equation}\label{eq:DefPhi6i}
\Phi_{6,i}\coloneqq \Phi_{6,0}|_{M_i}
\end{equation} 
 under $M_i\in \Sp_4(\IZ)\backslash\Gamma_0^{(2)}(2)$ (the $M_i$ being specified in \eqref{eq:slashmatrices1} and \eqref{eq:slashmatrices2}):
\begin{align}
\frac{1}{\Phi_{6,0}} & = \ \, \,  \frac{\theta_{0000}^2 \theta_{0001}^2 \theta_{0010}^2 \theta_{0011}^2}{\chi_{10}} = \frac{1}{W} \label{eq:defPhi60}  \\
\frac{1}{\Phi_{6,1}} & = \ \, \,  \frac{\theta_{0000}^2 \theta_{0001}^2 \theta_{1000}^2 \theta_{1001}^2}{\chi_{10}}  \\
\frac{1}{\Phi_{6,2}} & = - \frac{\theta_{1000}^2 \theta_{1001}^2 \theta_{0010}^2 \theta_{0011}^2}{\chi_{10}} \label{eq:defPhi62}  \\
\frac{1}{\Phi_{6,3}} & =   \ \, \, \frac{ \theta_{0000}^2 \theta_{0010}^2 \theta_{0100}^2 \theta_{0110}^2}{\chi_{10}} = \frac{Y'}{YW} \\
\frac{1}{\Phi_{6,4}} & = - \frac{ \theta_{0001}^2 \theta_{0011}^2 \theta_{0100}^2 \theta_{0110}^2}{\chi_{10}} = \frac{Y''}{YW}  \ .\label{eq:defPhi64}
\end{align} 
Here we have kept the notation of~\cite{Bossard_2019} and introduced
\begin{equation}\label{eq:SMFY2}
Y' = (\theta_{0000} \theta_{0010}\theta_{0100}\theta_{0110})^2 \quad \text{and} \quad Y'' = -(\theta_{0001} \theta_{0011}\theta_{0100}\theta_{0110})^2 \ .
\end{equation}
With the help of \eqref{eq:thetatransf1} one easily checks that%
\footnote{The minus sign in \eqref{eq:defPhi62} and \eqref{eq:defPhi64} is imporant for reproducing the result for the orbifold block $\mathcal{Z}_8\begin{bsmallmatrix}0 & 0 \\ 0 & 1 \end{bsmallmatrix}$ obtained in~\cite[eq. (4.38)]{Dabholkar_2007}, c. f. the relative signs between the terms in \eqref{eq:Z80001}.\label{fn:minusY3Y4}}
\begin{equation}\label{eq:SMFY3}
Y'(Z+\begin{psmallmatrix}
0 & 0 \\ 0 & 1
\end{psmallmatrix}) = Y''(Z) \quad \Rightarrow \quad \Phi_{6,3}(Z+\begin{psmallmatrix}
0 & 0 \\ 0 & 1
\end{psmallmatrix}) = \Phi_{6,4}(Z) \ .
\end{equation}
The corresponding elements $M_i \in \Sp_4(\IZ)\backslash\Gamma_0^{(2)}(2)$ are in the notation of \eqref{eq:SL2embedded1} and \eqref{eq:SL2embedded2}
\begin{equation}\label{eq:slashmatrices1}
M_1 = \begin{psmallmatrix}
0 & -1 \\ 1 & 0
\end{psmallmatrix}_\tau \ ,  \ \ M_2 = \begin{psmallmatrix}
1 & -1 \\ 1 & 0
\end{psmallmatrix}_\tau \ , \ \ M_3 = \begin{psmallmatrix}
0 & -1 \\ 1 & 0
\end{psmallmatrix}_\sigma \ ,  \ \text{and} \ \ M_4 = \begin{psmallmatrix}
1 & -1 \\ 1 & 0
\end{psmallmatrix}_\sigma  \ .
\end{equation}
Indeed $\Phi_{6,1/2}$ and $\Phi_{6,3/4}$ map to each other under exchange of the diagonal elements of $Z\in \IH_2$, for instance, $M_1$ is conjugate to $M_3$ by the element \eqref{eq:SympEx2} with $U=\begin{psmallmatrix}
0 & 1 \\ 1 & 0
\end{psmallmatrix}$. For the other Siegel modular forms $\Phi_{6,k}$, with $k\in \lbrace 5, 6 , 10 ,11 \rbrace$, that appear in section \ref{sec:genustwo} we do not need explicit expressions and just give
\begin{equation}\label{eq:slashmatrices2}
M_5 = \begin{psmallmatrix}
0 & -1 \\ 1 & 0
\end{psmallmatrix}_\tau \begin{psmallmatrix}
0 & -1 \\ 1 & 0
\end{psmallmatrix}_\sigma \ , \ \ M_6 = \begin{psmallmatrix}
1 & -1 \\ 1 & 0
\end{psmallmatrix}_\tau \begin{psmallmatrix}
0 & -1 \\ 1 & 0
\end{psmallmatrix}_\sigma \ , \ \ M_{10} = \begin{psmallmatrix}
1 & 0 & 0 & 1\\
0 & 1 & 1 & 0\\
0 & 0 & 1 & 0\\
0 & 0 & 0 & 1
\end{psmallmatrix}M_5 \ , \ \ M_{11} = \begin{psmallmatrix}
1 & 0 & 1 & 1\\
0 & 1 & 1 & 0\\
0 & 0 & 1 & 0\\
0 & 0 & 0 & 1
\end{psmallmatrix}M_5 \ \! .
\end{equation}

There are many quadratic relations that the squares of the theta constants satisfy and which have, for instance, been reviewed  in~\cite{Cl_ry_2015}. 
One particular identity important for our untwisted partition functions is the relation
~\cite[eq. (5.1)]{Cl_ry_2015}
\begin{equation}\label{eq:SiegelThetaId}
\theta_{0100}^2 \theta_{0110}^2 = \theta_{0000}^2 \theta_{0010}^2 - \theta_{0001}^2 \theta_{0011}^2 \ ,
\end{equation}
which implies for the above Siegel modular forms
\begin{equation} \label{eq:TYYrelation}
16^2 \, T = Y'+ Y'' \ .
\end{equation}

Finally, we remark that the quadratic divisors, on which $\chi_{10}$ and its orbifold analog $\Phi_{6,1}$ (or $\Phi_{6,3}$ with the roles of the diagonal entries swapped) vanish quadratically, can, for instance, be found in~\cite[section 4]{David_2006} or~\cite[appendix D]{Sen:2007qy}. By an appropriate $\Sp_4(\IZ)$-transformation they can be mapped to the standard diagonal divisor, as was used in~\cite{Murthy:2009dq}.


\section{Comments on the $\mathbb{Z}_2$-twisted BPS index}
\label{sec:twistedIndex}

In \cite{Sen:2009Atwist,Sen:2010Discrete,Govindarajan:2010fu} so-called ``twisted'' helicity supertraces were introduced as helicity supertraces with a discrete $\IZ_N$ symmetry generator inserted into the trace.
As some of the BPS states considered so far are in an untwisted sector of a $\IZ_2$ orbifold, one could naively expect that their BPS index can be obtained by projecting the index of the original theory onto the subspace of $\IZ_2$ invariant states, which could practically be realized by linearly combining (un-)twisted helicity supertraces.\footnote{We thank the anonymous referee for suggesting us to comment on this idea.} 
This can be ruled out, as we will see now.

First recall that the type of BPS index we have considered so far is the ($2n$)th helicity supertrace \cite{Bachas_1997,Gregori_1998,Kiritsis:1997hj}, which can be defined as\footnote{This is denoted by $B_{2n}$ in \cite{Sen:2009Atwist,Sen:2010Discrete,Govindarajan:2010fu}.}
\begin{equation}\label{eq:defHST}
\Omega_{2n}(Q,P; \cdot) = \frac{1}{(2n)!} \operatorname{Tr}_{(Q,P)}\left[(-1)^{2h} (2h)^{2n} \right]\, ,
\end{equation}
where $h$ is the third component of the angular momentum of a state in the rest frame and the trace is taken over all states carrying charge $(Q,P)$. For $2n=4$  and for $2n=6$ these traces were called half-BPS or quarter-BPS index, respectively, as they only get non-zero contributions from states with the indicated amount of broken supersymmetries.

On special moduli subspaces of the $\mathcal{N}=4$ compactification under consideration --- here, in the type IIA picture, a $\IZ_{N_{\text{o}}}$ CHL orbifold by an order $N_{\text{o}}$ translation along a circle $S^1\subset T^2$ of K3${}\times T^2$ combined with an order $N_{\text{o}}$ symplectic K3 symmetry --- the theory might possess a discrete $\IZ_{N_{\text{s}}}$ symmetry, generated by an element $g_{\text{s}}$ that commutes with the orbifold group. The subscripts ``o'' and ``s'' stand for ``orbifold'' or ``symmetry'', respectively. 
For dyons of charge $(Q,P)$, which must be a $g_{\text{s}}$-\textit{invariant} charge, one defines a twisted index via
\begin{equation}\label{eq:defTwHST}
\Omega_{2n}^{g_{\text{s}}}(Q,P ;\cdot) = \frac{1}{(2n)!} \operatorname{Tr}_{(Q,P)}\left[ g_{\text{s}} \,(-1)^{2h} (2h)^{2n} \right]\, ,
\end{equation} where the trace is taken over states with the respective charge in the $\IZ_{N_{\text{o}}}$ orbifold theory. The index \eqref{eq:defTwHST} receives contributions from quarter-BPS states \cite{Sen:2010Discrete}. As with the standard non-twisted index \eqref{eq:defHST} (equivalently the trivially twisted index), the dot denotes potential dependence on the moduli, and indeed, the twisted sixth helicity supertrace shares the wall-crossing features of $\Omega_6$, see \cite{Sen:2009Atwist,Sen:2010Discrete,Govindarajan:2010fu} for details.

As suggested by Sen \cite[eq. (1.15)]{Sen:2010Discrete}, one can compute the non-twisted index for states carrying a definite $g_{\text{s}}$-eigenvalue $\lambda_{\text{s}}=\exp(2\pi i\, a/N_{\text{s}})$ and $g_{\text{s}}$-invariant charge $(Q,P)$ using the combination
\begin{align}\label{eq:HSTprojection}
\frac{1}{N_{\text{s}}} \sum_{b=0}^{N_{\text{s}}-1} e^{2\pi i\, ab/N_{\text{s}}}\, \operatorname{Tr}_{(Q,P)}\left[ (g_{\text{s}})^b \,(-1)^{2h} (2h)^{2n} \right] 
\overset{N_{\text{s}}=2}{=} \  \operatorname{Tr}_{(Q,P)}\left[ \frac{1\pm g_{\text{s}}}{2} \,(-1)^{2h} (2h)^{2n} \right] \ .
\end{align}

\pagebreak[3] Following \cite{Sen:2009Atwist,Sen:2010Discrete,Govindarajan:2010fu} we first consider a symplectic K3 automorphism $g_{\text{s}}$ that commutes with the orbifolding by $g_{\text{o}}$ in the construction of the CHL model. Examples of such $\IZ_{N_{\text{o}}}\times \IZ_{N_{\text{s}}}$ groups are given in \cite{Chaudhuri:1995dj,Aspinwall_1996}. Note, however, that the $g_{\text{s}}$ symmetry is \textit{not} accompanied by any shift along a circle (as is the case for $\IZ_{N_{\text{o}}}$). Generating functions for these twisted quarter-BPS indices have been derived from the D1-D5 system in \cite{Sen:2009Atwist,Sen:2010Discrete}, leading to product formulae very similar to the Borcherds lifts giving rise to the (standard) quarter-BPS counting functions considered earlier in  \cite{David_2006}. Another construction of such twisted-index generating functions was given in \cite{Govindarajan:2010fu}, namely via an additive lift of suitable Jacobi forms, in turn similar to the construction in \cite{Jatkar:2005bh} for the standard quarter-BPS counting functions in the $\IZ_{N_{\text{o}}}$ orbifold. 

We are interested in the $\IZ_{N_{\text{s}}}=\IZ_2$ symmetry, acting as Nikulin involution on K3 or swapping the $E_8$'s in the heterotic picture, and we first recall relevant results \cite{Sen:2009Atwist,Sen:2010Discrete,Govindarajan:2010fu}. 
The generating function for the $\IZ_2$-twisted quarter-BPS index of unit-torsion dyons in the $\IZ_1$-orbifold is given  by the weight -6 Siegel modular form $1/\Phi_{6,0}(Z)$. 
In contrast, 
the generating function for the $\IZ_1$-twisted quarter-BPS index of unit-torsion dyons in the $\IZ_2$ CHL orbifold   is given by the weight -6 Siegel modular form $1/(2^4 \, \Phi_{6,3}(Z))$, if the electric charge lies in $\Lambda_e \backslash \Lambda_m$.
 Also recall that the $\IZ_1$-twisted quarter-BPS indices for unit-torsion dyons in the $\IZ_1$-orbifold are generated by the original DVV-result $1/\chi_{10}$.

Now a $g_{\text{s}}$-invariant charge $(Q,P)\in \Lambda_{22,6}\oplus \Lambda_{22,6}$ in the trivial orbifold necessarily has the same components along the two $E_8$ root lattices in each $\Lambda_{22,6}$, i.e., the respective components of $Q$  must lie on the diagonal $E_8(-2) \subset E_8(-1) \oplus  E_8(-1) \subset \Lambda_{22,6}$. This implies that for these $Q\in\Lambda_{22,6}$ the coset element $\mathcal{P}=0$ vanishes when writing the two $E_8$ components in terms of their sum and their difference.
The momentum-winding charges along $T^6$ are not affected by this $g_{\text{s}}$, hence $Q \in U^{\oplus 6} \oplus E_8(-2)$ for a generic $g_{\text{s}}$-invariant charge. By S-duality $(Q,P)\mapsto (-P, Q)$ in this theory, the same is expected to hold for the invariant magnetic charges $P$.\footnote{Note that the $\IZ_2$-twisted index generating function $\Phi_{6,0}$ is symmetric under $\sigma \leftrightarrow \tau$ and the half-BPS counting functions on the diagonal divisor $z=0$ are both given by $\eta^8(\cdot)\eta^8(2\, \cdot)$.} So a $g_{\text{s}}$-invariant charge $(Q,P)$ is subject to
\begin{align}
Q &\in \ \, U \oplus U^{\oplus 5 } \oplus E_8(-2) \ \, \subset \Lambda_{22,6}=\Lambda_e\\
P &\in \ \, U \oplus U^{\oplus 5 } \oplus E_8(-2) \ \, \subset \Lambda_{22,6}=\Lambda_m \ ,
\end{align}
which includes as a special case $P$-charges in the intersection $(U(2)\oplus U^{\oplus 5} \oplus E_8(-2) )\cap \Lambda_{22,6}$.

Considering $2n=6$, by the standard contour prescription \eqref{eq:contourint} the projected index \eqref{eq:HSTprojection} for a $g_{\text{s}}$-invariant unit-torsion charge can thus be obtained as a Fourier coefficient of the generating function
\begin{equation}\label{eq:projSMF}
\frac{1}{2} \left( \frac{1}{\chi_{10}} \pm \frac{1}{\Phi_{6,0}} \right) \ .
\end{equation}
 However, \eqref{eq:projSMF} differs from the untwisted sector (unit-torsion, $\mathcal{P}=0$) quarter-BPS partition functions in the $\IZ_2$ orbifold in \eqref{eq:g2untwTerms2}, though we can formally compare quarter-BPS indices for charges $(Q,P)$ lying in the same charge sublattices. 
 As the orbifolding projects out 8 gauge bosons odd under $\IZ_2$ the rank of $\Lambda_e$ and $\Lambda_m=\Lambda_e^*$ is reduced from 28 to 20. After orbifolding the physical electric charge coming from $(P_1,P_2)\in E_8(-1)^{\oplus 2}$ is their sum $P_1 + P_2$. Internal momentum vectors $P_1 - P_2$, odd under $\IZ_2$, distinguish electric charges in the parent theory, but not in the $\IZ_2$ CHL orbifold theory. In the latter we thus have states with the same electric charge (lying in a rank 20 electric lattice), but with different values for $P_1-P_2$ (see also \cite[section 3.3]{Dabholkar:2005dt}). 
 This means when taking the helicity supertrace over states of fixed electric-magnetic charge $(Q,P)$ \textit{in the $\IZ_2$ orbifold}, the lattice vector $P_1-P_2$ is \textit{not} fixed by $(Q,P)$, i.e., various states of different $P_1-P_2$ might (and do) contribute to the BPS-index. For instance, in eq. \eqref{eq:B4_Z2chl_untwEps} or eq. \eqref{eq:electricHBPSeps} this gave rise to the $E_8(2)$-theta function in the numerator of the half-BPS counting function. 
  This is different for a helicity supertrace of $g_{\text{s}}$-invariant charge in the parent theory, where the $E_8$ charges are completely fixed $(Q,P)$ and must be equal (hence $P_1 - P_2 = 0$).   
 The half-BPS index for electric DH states of $g_{\text{s}}$-invariant charge $(Q,0)\in  U^{\oplus 6} \oplus E_8(-2)\subset \Lambda_{22,6}$ are now counted by
\begin{equation}\label{eq:electricHBPSeps2}
f_{\pm}(\sigma) \coloneqq \frac{1}{2} \left( \frac{1}{ \eta^{24}(\sigma)} \pm \frac{1}{ \eta^{8}(\sigma)\eta^{8}(2\sigma)} \right) \ ,
\end{equation}
where the sign is that picked up by the monomial in the 24 bosonic oscillators, (2+6+8=16) of them even under $g_{\text{s}}$ and 8 of them odd under $g_{\text{s}}$ after diagonalization. 
Note that the $E_8$ theta series associated with $P_1-P_2 = 2P_- $ in the anti-diagonal $E_8(-2) \subset E_8(-1)^{\oplus 2} \subset \Lambda_{22,6}$ is in \eqref{eq:electricHBPSeps2} absent in comparison with eqs.  \eqref{eq:B4_Z2chl_untwEps} and \eqref{eq:electricHBPSeps}.
And indeed, the quadratic pole of \eqref{eq:projSMF} at $z=0$ reads (dropping an overall factor)
\begin{equation}
\frac{1}{2}\left( \frac{1}{\eta^{24}(\sigma)\eta^{24}(\tau)} \pm \frac{1}{\eta^{8}(\sigma)\eta^{8}(2\sigma)\eta^{8}(\tau)\eta^{8}(2\tau)}   \right) = \begin{cases}  f_+(\sigma)f_+(\tau) + f_-(\sigma)f_-(\tau)  &: \ ``+''\\ f_+(\sigma)f_-(\tau) + f_-(\sigma)f_+(\tau) & : \ ``-'' \end{cases} \ ,
\end{equation}
which admits the interpretation that the $g_{\text{s}}$-eigenvalue of the decaying quarter-BPS dyon is the product of the half-BPS decay products with charges $(Q,0)$ and $(0,P)$. 

In fact, although the $\IZ_2$ symmetry just discussed is not quite the symmetry used in construction of the $\IZ_2$ CHL model as it did not include the simultaneous order-two translation along a circle of $T^6$ (or the type II elliptic curve), the argument concerning the different roles played by the off-diagonal vectors $P_1-P_2$ still goes through. In other words, one should not expect that the generating function for projected BPS indices \eqref{eq:HSTprojection} (of dyons carrying a $g_{\text{s}}$ invariant charge) in the parent theory with  $g_{\text{s}}$ now including the CHL translation matches the corresponding untwisted sector quarter-BPS counting functions in the $\IZ_2$ orbifold.
This mismatch can explicitly be seen for the fourth helicity supertrace. For perturbative DH states of charge $(Q,0)\in U^{\oplus 6} \oplus E_8(-2)\subset \Lambda_{22,6}$, the $g_{\text{s}}$ eigenvalue is given by the sign picked up by the oscillator monomial times $(-1)^{\delta \cdot Q}$, where $(-1)^{\delta \cdot Q}=+1$ for an even momentum quantum number along the CHL cirle (i.e., if $(Q,0)\in U(2) \oplus U^{\oplus 5} \oplus E_8(-2)\subset \Lambda_{22,6}$) and $(-1)^{\delta \cdot Q}=-1$ otherwise. This implies that fixing $(Q,0)\in U(2) \oplus U^{\oplus 5} \oplus E_8(-2)$ (i.e., $(-1)^{\delta \cdot Q}=+1$) the projected half-BPS index for $\lambda_{\text{s}}=+1$ is given by $f_+(\sigma)$, while that for $\lambda_{\text{s}}=-1$ is given by $f_-(\sigma)$. In turn taking charges $(Q,0)\in U^{\oplus 6} \oplus E_8(-2)\subset \Lambda_{22,6}$ with $(-1)^{\delta \cdot Q}=-1$ switches the roles of $f_+$ and $f_-$. For neither choice of the invariant charge\footnote{Of course, the states surviving the orbifolding are the invariant states with eigenvalue $\lambda_{\text{s}}=+1$.} $(Q,0)\in U^{\oplus 6} \oplus E_8(-2)\subset \Lambda_{22,6}$  the untwisted sector DH partition function \eqref{eq:electricHBPSeps} from the $\IZ_2$ CHL orbifold is recovered, as the $\theta$-contribution is missing. By the same logic, we expect a mismatch for the Siegel modular forms generating the corresponding (projected) sixth helicity supertraces.



\bibliographystyle{JHEP}
\bibliography{References}


\end{document}